\documentclass[journal,9pt]{IEEEtran}

\usepackage{amsmath,amsfonts}
\usepackage{algorithm} 
\usepackage{algpseudocode}
\usepackage{setspace}
\usepackage{array}

\usepackage[caption=false,font=normalsize,labelfont=sf,textfont=sf]{subfig}

\usepackage{textcomp}
\usepackage{stfloats}
\usepackage{url}
\usepackage{verbatim}
\usepackage{graphicx}
\usepackage{cite}
\usepackage[hidelinks]{hyperref}
\usepackage{booktabs} 
\usepackage{makecell}
\usepackage{xcolor}
\graphicspath{{figures/}}

\begin{document}

\title{\huge{Alternating Optimization Framework with Cross-Correlated Loss and Implicit Neural Representation for Electromagnetic Inverse Scattering}}

\author{\Large{Shilong~Sun}
\thanks{Manuscript received on January XX; revised on XX. This work was supported by the National Natural Science Foundation of China under Grant 62471476 and 62231026.}
\thanks{S. Sun (e-mail: sunshilong@nudt.edu.cn) is with the College of Electronic Science and Technology, National University of Defense Technology, Changsha, China.}}

\markboth{IEEE Transactions on Antennas and Propagation,~Vol.~xx, No. xx, 2026}%
{Sun: Alternating Optimization Framework with Cross-Correlated Loss and Implicit Neural Representation for Electromagnetic Inverse Scattering}

\maketitle

\begin{abstract}
Electromagnetic inverse scattering imaging is a crucial computational technique in microwave sensing systems. However, quantitative reconstruction of high-contrast targets remains a formidable challenge due to severe multiple scattering effects. To overcome this bottleneck, this communication proposes an alternating optimization framework based on a cross-correlated implicit neural representation (Alt-CC-INR) as a preliminary study towards the robust solution of high-contrast targets reconstruction. Unlike conventional purely neural architectures, this framework introduces a heterogeneous optimization paradigm. Specifically, the method employs an analytical Polak-Ribi\`{e}re conjugate gradient (PR-CG) algorithm, driven by a cross-correlated loss and accelerated by 2D-FFT, to optimally resolve the dynamic contrast sources. Subsequently, a stochastic deep learning optimizer updates the continuous neural representation to infer the static medium profile. Validations on simulated and measured data demonstrate that Alt-CC-INR effectively overcomes local minima when inverting high-contrast objects, providing a highly robust computational electromagnetic solver.
\end{abstract}


\section{Introduction} \label{sec:intro}

    \IEEEPARstart{E}{lectromagnetic} inverse scattering imaging aims to quantitatively reconstruct the spatial distribution of the complex permittivity of unknown targets. It has demonstrated irreplaceable value in applications such as microwave nondestructive evaluation \cite{amineh2020nondestructive,ghattas2024review} and microwave medical imaging \cite{meaney2000clinical}. The core challenge of inverse scattering imaging lies in the highly nonlinear and ill-posed nature of the multiple scattering effects. Traditional iterative optimization algorithms, such as the Contrast Source Inversion (CSI) method \cite{van1997contrast,sun2017cross} and the Distorted Born Iterative Method \cite{wang1989iterative}, achieve promising results but are prone to local minima when confronted with strong, high-contrast microwave scatterers \cite{11106328}.

    Recently, the integration of deep learning has introduced new pathways for inverse scattering. State-of-the-art supervised methods, including convolutional neural networks (CNNs) \cite{8565987,DastfalTGRSL2025}, diffusion models \cite{9897092,bi2024diffusionemis}, and Transformers \cite{du2024inhomogeneous}, have established new benchmarks for imaging quality and offer instantaneous inference once trained. However, establishing a robust direct mapping from scattered fields to target contrast is extremely difficult, making these methods heavily reliant on massive offline training datasets. Consequently, supervised networks suffer from a severe ``simulation-to-reality gap,'' where generalizing to real-world measured data with unforeseen non-idealities remains a formidable challenge. As a powerful alternative to overcome this generalization bottleneck, unsupervised architectures based on Physics-Informed Neural Networks (PINNs) \cite{RAISSI2019686} optimize the medium profile on-the-fly, acting as zero-shot solvers \cite{du2025physics,ISPNet2025}. In the previous research, a cross-correlated PINN (CC-PINN) \cite{sun2026CC_PINN} was proposed, which uniquely introduces a cross-correlated term forcefully coupling the predicted contrast and the contrast source to the far-field observations. Although CC-PINN greatly enhances the ability to escape local minima, its convergence stability encounters a bottleneck for extreme high-contrast targets. This ceiling stems from the limitations of empirical step-size gradient descent. In the original CC-PINN, spatial contrast (network weights) and internal contrast sources are jointly updated by the Adam optimizer. For extreme high-contrast targets, such empirical step-size updates are more likely to lose the global optimal solution.

    To resolve this specific computational bottleneck, this communication proposes an Alternating Optimization INR framework (Alt-CC-INR). The core contribution lies in abandoning the end-to-end joint optimization and replacing it with a heterogeneous alternating optimization paradigm. Unlike CC-CSI's purely discrete numerical iterations or CC-PINN's purely stochastic neural updates, this architecture optimally maps distinct mathematical variables to their most suitable solvers: an analytical PR-CG update strictly resolves the dynamic, highly oscillatory physical fields, while a stochastic Adam optimizer alongside INR infers the static geometric medium profile. This heterogeneous iterative framework perfectly aligns with the physical algebraic structure of microwave field evolution, profoundly improving reconstruction fidelity and robustness for high-contrast targets. Since the solution for contrast sources (Phase 1) is delegated to PR-CG, while the network is responsible only for the remaining optimization steps, I adopt the term Implicit Neural Representation (INR) \cite{sitzmann2020implicit} to describe the architecture.

\section{Algorithm Architecture and Implementation} \label{sec:architecture}

\subsection{Problem Modeling and Cross-Correlated Loss} \label{sec:modeling}

    For a multi-frequency inverse scattering scenario, the complex contrast of the target at spatial position $\mathbf{r} \in \mathcal{D}$ under frequency $f \in \mathcal{F}_s$ is $\chi_f(\mathbf{r}) = \varepsilon_\text{r}(\mathbf{r}) - 1 - \mathrm{j}\sigma(\mathbf{r}) / (\omega_f \varepsilon_0)$, where $\varepsilon_\text{r}$ and $\sigma$ denote the relative permittivity and conductivity, $\omega_f$ is the angular frequency, and $\varepsilon_0$ is the vacuum permittivity. An implicit $\exp(\mathrm{j}\omega t)$ time dependence is assumed throughout this communication. By introducing the equivalent contrast source variable $\mathbf{J}_f = \chi_f \odot \mathbf{E}_{\text{tot}, f}$ (with $\odot$ denoting the Hadamard product and $\mathbf{E}_{\text{tot}, f}$ the internal total electric field), the forward process is governed by the state equation $\mathbf{E}_{\text{tot}, f} = \mathbf{E}_{\text{inc}, f} + \mathcal{G}_{\mathcal{D}, f} \mathbf{J}_f$ and the data equation $\mathbf{E}_{\text{meas},f} = \mathcal{G}_{\mathcal{S}, f} \mathbf{J}_f$. Here, $\mathbf{E}_{\text{inc}, f}$ and $\mathbf{E}_{\text{meas},f}$ are the incident and measured scattered fields, respectively, while $\mathcal{G}_{\mathcal{D}, f}$ and $\mathcal{G}_{\mathcal{S}, f}$ represent the domain and observation Green's function integral operators.

    The total loss function for active frequency set $\mathcal{F}_s$ consists of the superposition of the data term, state term, and a cross-correlated term:
    \begin{equation}
    \begin{split}
    L_{\text{total}}(\mathbf{J}, \theta) = \sum_{f \in \mathcal{F}_s} \Bigl[ & L_{\text{data},f}(\mathbf{J}_f) + L_{\text{state},f}(\mathbf{J}_f, \theta) \\
    & + \beta_{\text{Lc},s}^{(k)} L_{\text{cross},f}(\mathbf{J}_f, \theta) \Bigr].
    \end{split}
    \end{equation}
    Here, the cross-correlated error acts as a robust global regularizer. To prevent manual weight tuning, the exponential decay factor $\beta_{\text{Lc},s}^{(k)} = \exp(-10 \cdot k / K_s)$ automates the balancing of the cross-correlated term across $K_s$ epochs, where $k$ is the current iteration index. Using $\|\cdot\|_F$ to denote the Frobenius norm, the physical residual terms are defined as:
    $L_{\text{data}, f} = \| \mathcal{G}_{\mathcal{S}, f} \mathbf{J}_f - \mathbf{E}_{\text{meas},f} \|_F^2 / \| \mathbf{E}_{\text{meas},f} \|_F^2$, 
    $L_{\text{state}, f} = \| \boldsymbol{\chi}_f(\theta) \odot (\mathbf{E}_{\text{inc}, f} + \mathcal{G}_{\mathcal{D}, f} \mathbf{J}_f) - \mathbf{J}_f \|_F^2 / \| \mathbf{E}_{\text{inc}, f} \|_F^2$, and 
    $L_{\text{cross}, f} = \| \mathcal{G}_{\mathcal{S}, f} [ \boldsymbol{\chi}_f(\theta) \odot (\mathbf{E}_{\text{inc}, f} + \mathcal{G}_{\mathcal{D}, f} \mathbf{J}_f) ] - \mathbf{E}_{\text{meas},f} \|_F^2 / \| \mathbf{E}_{\text{meas},f} \|_F^2$.

\subsection{Alternating Optimization Engine}

    Alt-CC-INR decomposes the complex optimization into two well-posed subproblems, fully exploiting the quadratic nature of the physical states for exact analytical descent.
    
    \textbf{Phase 1: Batched PR-CG for Contrast Sources Update.} In the PyTorch implementation, $\mathbf{J}_f$ is defined as an explicit complex-valued leaf tensor. To significantly accelerate computation and bypass the massive memory overhead of automatic differentiation (autograd), $\mathbf{J}_f$ is detached from the computational graph (\texttt{.detach()} operation) and calculate its complex gradient $\mathbf{g}_{\mathbf{J}}^{(k)}$ analytically. With the network parameters $\theta$ frozen, the objective function becomes strictly quadratic with respect to $\mathbf{J}_f$. I use the Polak-Ribi\`{e}re conjugate gradient method to determine the descent direction $\mathbf{v}^{(k)}$. The update rule is $\mathbf{J}_f^{(k)} = \mathbf{J}_f^{(k-1)} + \alpha^{(k)} \mathbf{v}^{(k)}$, where the optimal step size $\alpha^{(k)}$ is analytically determined via a 1D line search independently for each active frequency. By substituting the update into the loss function and setting its derivative to zero, $\alpha^{(k)} = -\Re\{ \langle \mathbf{g}_{\mathbf{J}}^{(k)}, \mathbf{v}^{(k)} \rangle \} / (2 \cdot \text{Denom}^{(k)})$ is obtained, where $\Re\{\cdot\}$ and $\langle \cdot, \cdot \rangle$ denote the real part and inner product, respectively. The denominator $\text{Denom}^{(k)}$ is exactly derived as:
    \begin{equation}
    \begin{split}
       \text{Denom}^{(k)} = &\frac{\left\| \mathcal{G}_{\mathcal{S}, f} \mathbf{v}^{(k)} \right\|_F^2}{\left\| \mathbf{E}_{\text{meas},f} \right\|_F^2} + \frac{\left\| \mathbf{v}^{(k)} - \boldsymbol{\chi}_f(\theta) \odot \mathcal{G}_{\mathcal{D}, f} \mathbf{v}^{(k)} \right\|_F^2}{\left\| \mathbf{E}_{\text{inc}, f} \right\|_F^2} \\
       &+ \beta_{\text{Lc},s}^{(k)} \frac{\left\| \mathcal{G}_{\mathcal{S}, f} \left[ \boldsymbol{\chi}_f(\theta) \odot \mathcal{G}_{\mathcal{D}, f} \mathbf{v}^{(k)} \right] \right\|_F^2}{\left\| \mathbf{E}_{\text{meas},f} \right\|_F^2},
    \end{split}
    \end{equation}
    which directly encapsulates the respective quadratic variations of the data, state, and cross-correlated residuals. To rigorously preserve descent properties under gradient clipping, dynamically annealing weights, and frequency hopping, an automatic restart safeguard is strictly enforced (Algorithm 1, lines 21--23): if the conjugacy breaks and $\mathbf{v}^{(k)}$ ceases to be a descent direction ($\Re\{\langle \mathbf{g}_{\mathbf{J}}^{(k)}, \mathbf{v}^{(k)} \rangle\} \ge 0$), or when a new frequency stage begins, the search direction automatically resets to the steepest descent direction $-\mathbf{g}_{\mathbf{J}}^{(k)}$.
    
    \textbf{Phase 2: Neural Network Update.} After accomplishing the precise analytical advancement of the contrast sources, $\mathbf{J}_f^{(k)}$ is treated as a constant tensor via the \texttt{.detach()} operation. The neural network $\theta$ is unfrozen, and the algorithm enters the inner loop. At this point, utilizing the updated and fixed internal total field $\mathbf{E}_{\text{tot},f}^{(k)} = \mathbf{E}_{\text{inc},f} + \mathcal{G}_{\mathcal{D}, f} \mathbf{J}_f^{(k)}$, and considering that the data residual $L_{\text{data},f}$ is completely independent of $\theta$, the target loss function for network optimization degenerates to:
    \begin{equation}
        L_\text{NN}(\theta) = \sum_{f \in \mathcal{F}_s} \left[ L_{\text{state}, f}(\theta) + \beta_{\text{Lc},s}^{(k)} L_{\text{cross}, f}(\theta) \right].
    \end{equation}
    The Adam optimizer is employed to minimize this degenerated loss via automatic differentiation, compelling the network-output contrast $\boldsymbol{\chi}_f(\theta)$ to accurately match the current optimal physical field distribution. Before the parameter update, a maximum norm clipping (e.g., restricted to $1.0$) is applied to $\nabla_{\theta} L_\text{NN}$ to further ensure stable network convergence:
    \begin{equation}
    \theta^{(k)} = \theta^{(k-1)} - \eta \cdot \text{Adam}\left\{ \nabla_{\theta} L_\text{NN}\left(\theta^{(k-1)}\right) \right\},
    \end{equation}
    where $\eta$ is the learning rate. For every single update of the contrast sources, the contrast is updated only once. 

\subsection{Network Architecture and Operator Acceleration}

    \begin{figure}[!t]
        \centering
        \includegraphics[width=0.95\linewidth]{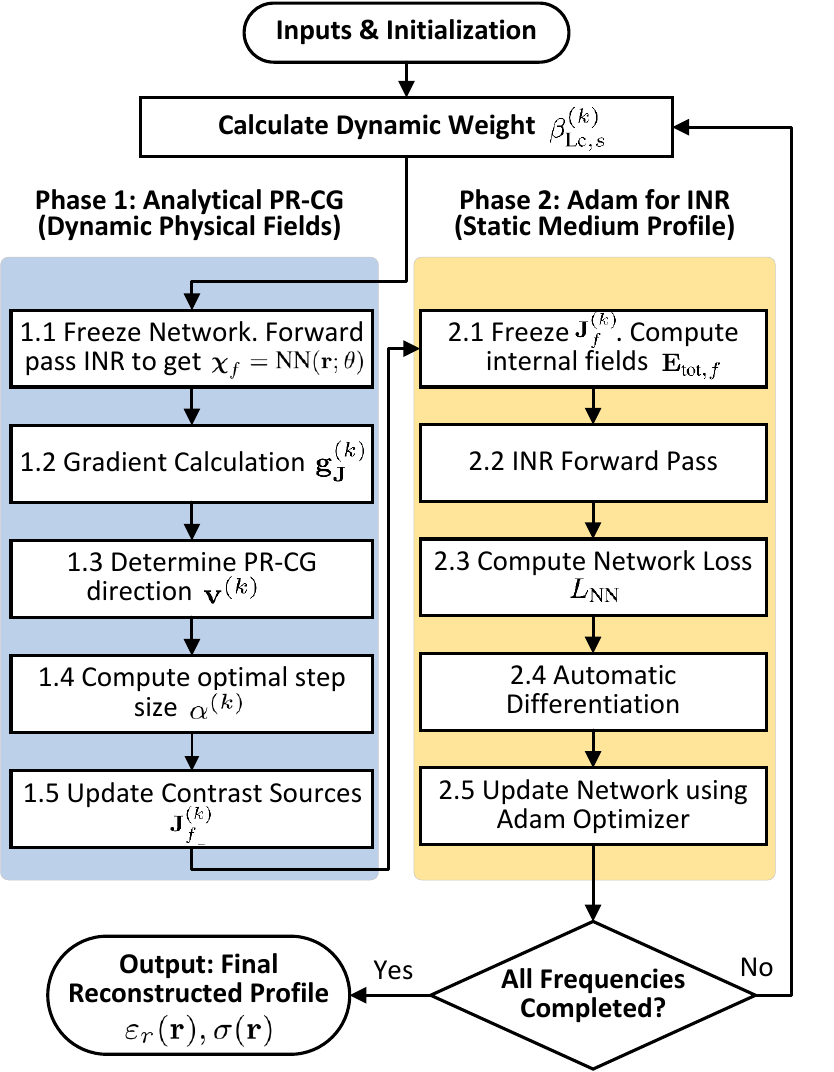}
        \caption{Flowchart of the proposed Alt-CC-INR framework.}
        \label{fig:Flowchart_Alt-CC-INR}
    \end{figure}

    Unlike recent architectures that parameterize the highly oscillatory contrast sources with neural fields \cite{sun2026differentiable}, I implicitly parameterize the target contrast using a continuous Fourier feature Multi-Layer Perceptron (MLP). Specifically, the normalized 2D spatial coordinates $\mathbf{r} = (x, y)^\text{T}$ are first projected into a 128-dimensional periodic feature space via $\gamma(\mathbf{r}) = \left[\sin(2\pi\mathbf{B}\mathbf{r})^\text{T}, \cos(2\pi\mathbf{B}\mathbf{r})^\text{T}\right]^\text{T}$. Here, $\mathbf{B} \in \mathbb{R}^{64 \times 2}$ is a fixed, non-trainable Gaussian random projection matrix with a standard deviation scale of $1.0$, ensuring the network overcomes spectral bias without overfitting. The subsequent MLP backbone comprises sequential linear layers with dimensions $[128, 256, 256, 128, 2]$. To maintain smooth optimization trajectories without inducing the granular spatial artifacts typical of LayerNorm, Weight Normalization and SiLU activation functions are exclusively applied to all hidden layers. The final layer outputs are passed through a Sigmoid activation, strictly bounding the reconstructed relative permittivity $\varepsilon_\text{r}$ and conductivity $\sigma$ within physically reasonable intervals of $[1.0, 80.0]$ and $[0.0, 1.0]$~S/m, respectively. To guarantee stable early-stage convergence, the final layer biases are explicitly initialized to $-3.0$, forcing the initial output to closely match the free-space background medium. More details about the Fourier feature MLP can be found in \cite{sun2026CC_PINN}. To ensure strict fairness in the subsequent comparative experiments, the network topologies and initialization parameters utilized across all evaluated algorithms are kept completely identical. Additionally, to eliminate the computational bottleneck introduced by the domain Green's function integration $\mathcal{G}_{\mathcal{D}}$, batched zero-padded 2D-FFT is comprehensively deployed. This guarantees circular convolution equivalence while reducing complexity to $\mathcal{O}(N^2 \log N)$. 

    The complete procedure of the proposed Alt-CC-INR, executing a frequency-hopping progression strategy, is summarized in Algorithm 1. A critical parameter governing the optimization dynamics is the exponentially decaying weight $\beta_{\text{Lc},s}^{(k)}$, whose theoretical rationale was initially established in \cite{sun2026CC_PINN}. By dynamically annealing across the $K_s$ epochs within each frequency stage, it exerts a stringent cross-correlated global regularization to forcefully bypass local minima during the early iterations. Subsequently, it smoothly phases out, allowing the network to perform high-precision fine-tuning driven purely by standard data and state residuals. Fig.~\ref{fig:Flowchart_Alt-CC-INR} gives the flowchart of the proposed Alt-CC-INR framework. 

    \begingroup
    \small
    \noindent\rule{\linewidth}{1pt}
    \noindent \textbf{Algorithm 1} Fast Batched Alt-CC-INR with Frequency Hopping \\
    \noindent\rule{\linewidth}{0.5pt}
    \begin{algorithmic}[1] 
        \Require Multi-frequency measured scattered data $\mathbf{E}_{\text{meas},f}$, background incident fields $\mathbf{E}_{\text{inc},f}$, domains $\mathcal{D}$ and $\mathcal{S}_\text{obs}$, and frequency stages configuration $\left\{ \mathcal{F}_s \right\}_{s=1}^S$.
        \Ensure Reconstructed relative permittivity $\varepsilon_\text{r}(\mathbf{r})$ and conductivity $\sigma(\mathbf{r})$.
        
        \State \textbf{Initialization:}
        \State Initialize neural network parameters $\theta$ (Weight Norm \& SiLU) with specific biases (e.g., bias $= -3.0$ for background fitting)
        \State Initialize batched contrast sources $\mathbf{J}_f^{(0)}$ via data equation back-propagation
        \State Pre-compute zero-padded 2D-FFT Green's kernels for fast domain integration
        \State Initialize PR-CG variables: $\mathbf{g}_{\mathbf{J}}^{(0)} = \mathbf{0}$, $\mathbf{v}^{(0)} = \mathbf{0}$

        \Statex \textit{// Outer Loop: Frequency Hopping Strategy}
        \For{stage $s = 1 \to S$}
            \State Determine active frequencies $\mathcal{F}_s$ and allocate epochs $K_s$ for current stage
            
            \Statex \indent \textit{// Inner Loop: Optimization within Current Stage}
            \For{epoch $k = 1 \to K_s$}
                \State Calculate dynamic cross-term weight based on stage progress: $\beta_{\text{Lc},s}^{(k)} = \exp\left(-10 \cdot \frac{k}{K_s}\right)$
                
                \Statex \indent \indent \textit{// Phase 1: Batched PR-CG}

                \State \textbf{Freeze} network parameters $\theta$
                \State Forward pass INR to obtain current contrast: 
                \Statex \hskip7em $\boldsymbol{\chi}_f = \text{NN}(\mathbf{r}; \theta)$
                \State Compute batched internal total fields with FFT: 
                \Statex \hskip5em $\mathbf{E}_{\text{tot},f} = \mathbf{E}_{\text{inc},f} + \mathcal{G}_{\mathcal{D}, f} \mathbf{J}_f^{(k-1)}$
                
                \State Compute complex gradient $\mathbf{g}_{\mathbf{J}}^{(k)}$ using sum of losses over active frequencies: 
                \Statex \hskip3em $\mathbf{g}_{\mathbf{J}}^{(k)} = \nabla_{\mathbf{J}_f} \sum_{f \in \mathcal{F}_s} \left[ L_{\text{data},f} + L_{\text{state},f} + \beta_{\text{Lc},s}^{(k)} L_{\text{cross},f} \right]$
                
                \State Apply magnitude clipping to $\mathbf{g}_{\mathbf{J}}^{(k)}$ (max norm $\le 100.0$) to prevent gradient explosion
                
                \If{$k == 1$ \textbf{and} $s == 1$}
                    \State $\gamma^{(k)} = 0$
                \Else
                    \State Compute PR-CG coefficient: 
                    \Statex \hskip6em $\gamma^{(k)} = \max \left( 0, \frac{\Re\left\{ \langle \mathbf{g}_{\mathbf{J}}^{(k)}, \mathbf{g}_{\mathbf{J}}^{(k)} - \mathbf{g}_{\mathbf{J}}^{(k-1)} \rangle \right\}}{\left\|\mathbf{g}_{\mathbf{J}}^{(k-1)}\right\|_2^2} \right)$
                \EndIf
                
                \State Update conjugate direction: 
                \Statex \hskip6em $\mathbf{v}^{(k)} = -\mathbf{g}_{\mathbf{J}}^{(k)} + \gamma^{(k)} \mathbf{v}^{(k-1)}$
                \If{$\Re\left\{ \langle \mathbf{g}_{\mathbf{J}}^{(k)}, \mathbf{v}^{(k)} \rangle \right\} \ge 0$}
                    \State Reset conjugate direction to steepest descent: 
                    \Statex \hskip6em $\mathbf{v}^{(k)} = -\mathbf{g}_{\mathbf{J}}^{(k)}$
                \EndIf
                
                \State Analytically compute optimal step size $\alpha^{(k)}$ via 1D line search on quadratic loss surface
                \State Update contrast sources: $\mathbf{J}_f^{(k)} = \mathbf{J}_f^{(k-1)} + \alpha^{(k)} \mathbf{v}^{(k)}$
                
                \Statex \indent \indent \textit{// Phase 2: Update Neural Network Parameters (Adam)}
                \State \textbf{Freeze} contrast sources $\mathbf{J}_f \leftarrow \mathbf{J}_f^{(k)}$
                \State Re-compute static internal total fields based on updated $\mathbf{J}_f^{(k)}$
                
                \State Forward pass INR to get $\boldsymbol{\chi}_f(\theta)$
                \State Compute regression loss (Data loss is independent of $\theta$ here): 
                \Statex \hskip4.5em $L_\text{NN} = \sum_{f \in \mathcal{F}_s} \left[ L_{\text{state}, f}(\theta) + \beta_{\text{Lc},s}^{(k)} L_{\text{cross}, f}(\theta) \right]$
                \State Compute gradient $\nabla_{\theta} L_\text{NN}$ and apply gradient clipping (max norm $\le 1.0$)
                \State Update $\theta$ using Adam optimizer: 
                \Statex \hskip6em $\theta \leftarrow \theta - \eta \cdot \text{Adam}(\nabla_{\theta} L_\text{NN})$
            \EndFor
            \State Pass optimized $\mathbf{J}_f$ of current active frequencies to the next stage
        \EndFor
        \State \Return $\varepsilon_\text{r}(\mathbf{r}), \sigma(\mathbf{r})$ parameterized by optimal $\theta$
    \end{algorithmic}
    \noindent\rule{\linewidth}{1pt}
    \endgroup

\section{Experimental Validation}

\subsection{Simulation Experiment Analysis}

    The simulated experimental configuration simulates a Fresnel data acquisition setup, where 12 transmitting antennas are positioned uniformly on a 3~m radius circle. To highly replicate physical blind zones in authentic measurements, the received data within a $30^\circ$ sectorial region centered on the active transmitter is excluded during each excitation. In the remaining $300^\circ$ observation arc, receivers are densely sampled at an angular resolution of $3^\circ$, yielding $12 \times 101$ transceiver pairs. The system operates at 0.3, 0.4, and 0.5~GHz under TM polarization. The background is free-space, and complex Gaussian white noise is injected into the scattered field to achieve a signal-to-noise ratio (SNR) of 20~dB. To thoroughly test the algorithmic robustness against severe multiple scattering, I employ two classically challenging profiles: the ``Austria'' target (two disks and one ring) and the ``Bowtie-Cross'' target. The proposed Alt-CC-INR is evaluated against Alt-INR (without the cross-correlated term), CC-PINN (representing the joint optimization structure of the original architecture \cite{sun2026CC_PINN}), its decoupled alternating Adam variant CC-PINN (Alt) (updating variables sequentially via Adam to isolate the fundamental contribution of PR-CG), and the traditional pure iterative method CC-CSI. To avoid statistical bias, results are analyzed using the median and variance over 11 independent Monte Carlo runs. The displayed reconstructed images are selected from the inversion results corresponding to the median PSNR, with the true target contours outlined by dashed lines.

    \begin{figure}[!t]
        \hspace*{\fill}%
        \subfloat[]{\includegraphics[width=0.5\linewidth]{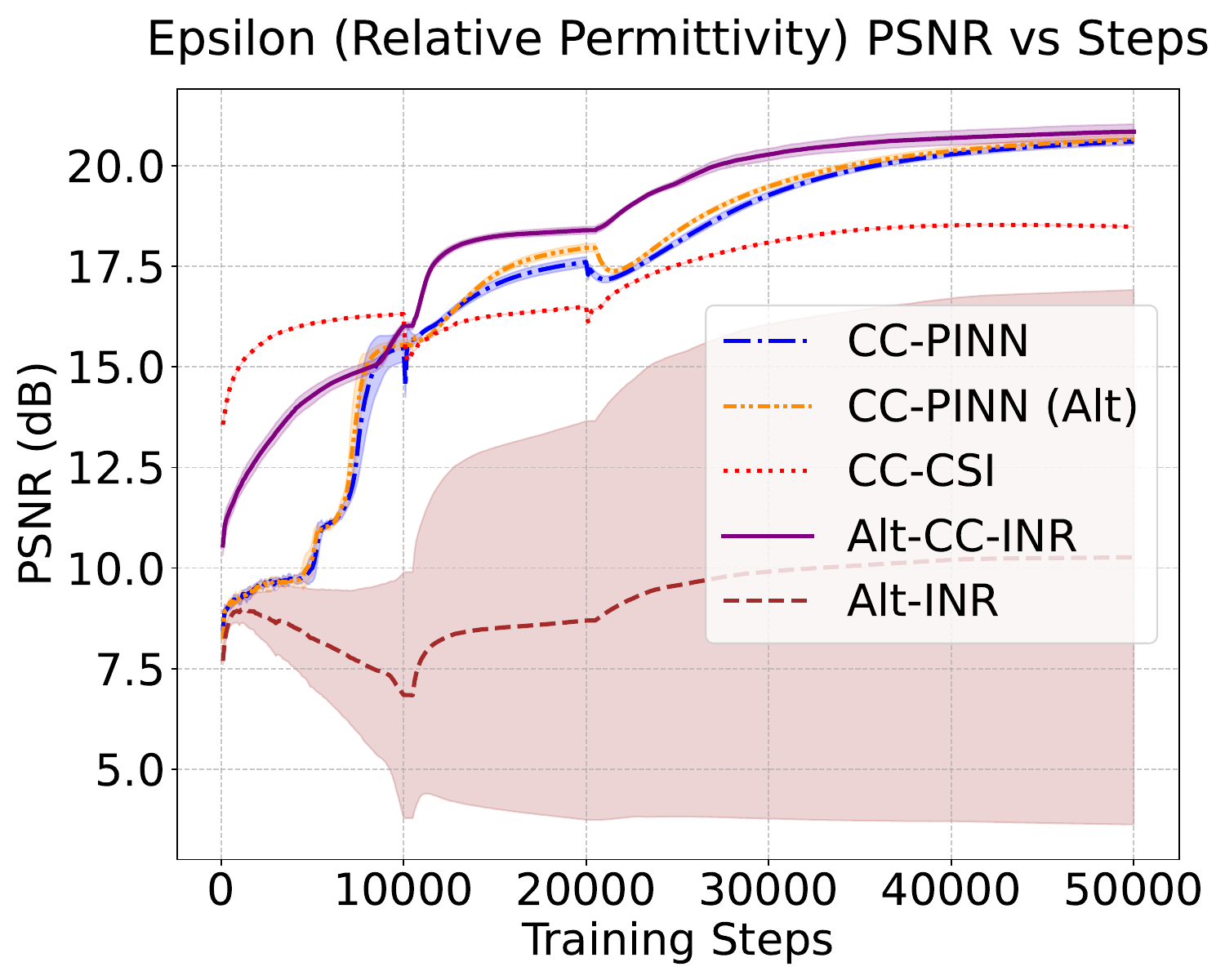}}\hfill
        \subfloat[]{\includegraphics[width=0.5\linewidth]{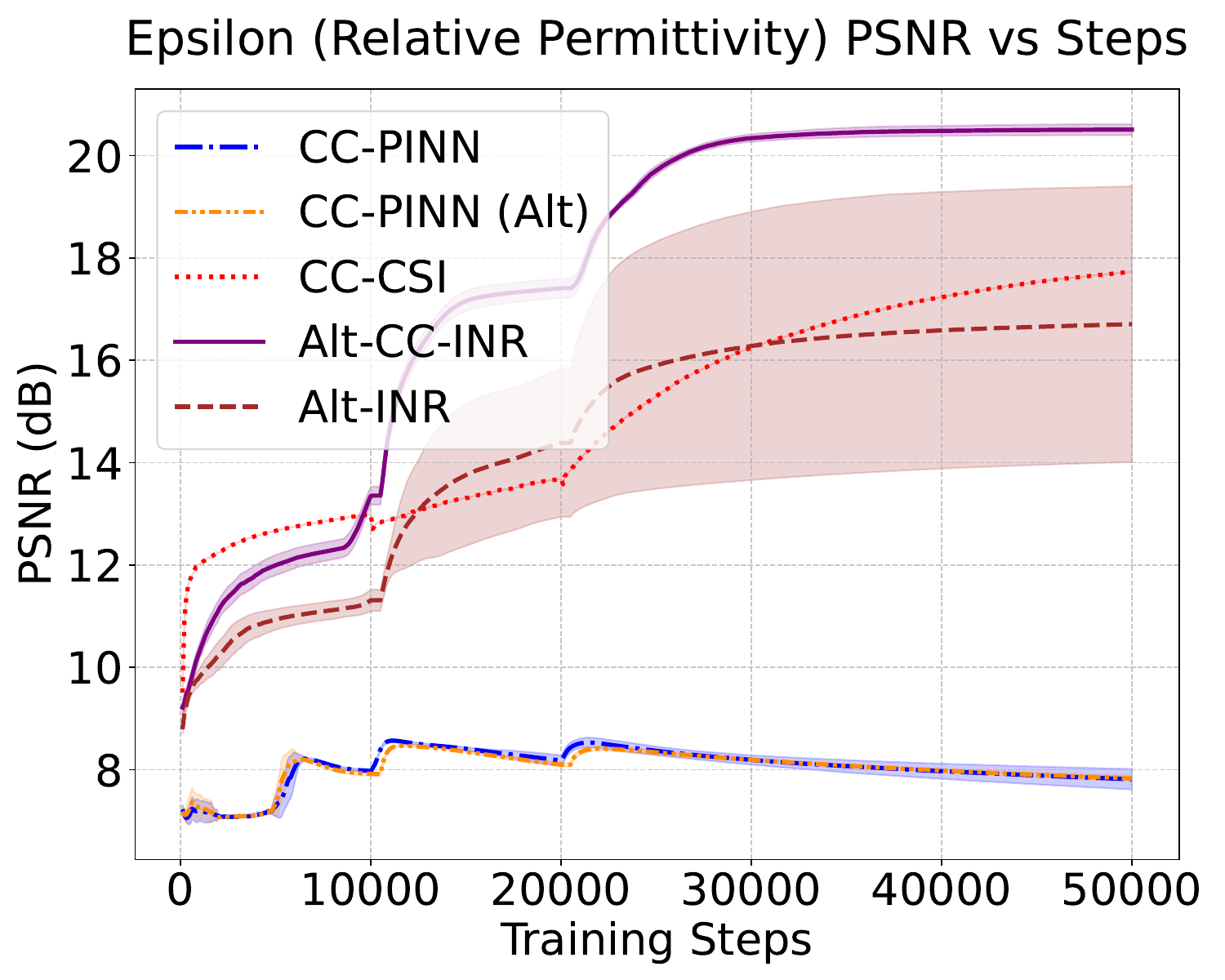}}\hspace*{\fill}

        \vspace{-1.0ex}

        \hspace*{\fill}%
        \subfloat[]{\includegraphics[width=0.5\linewidth]{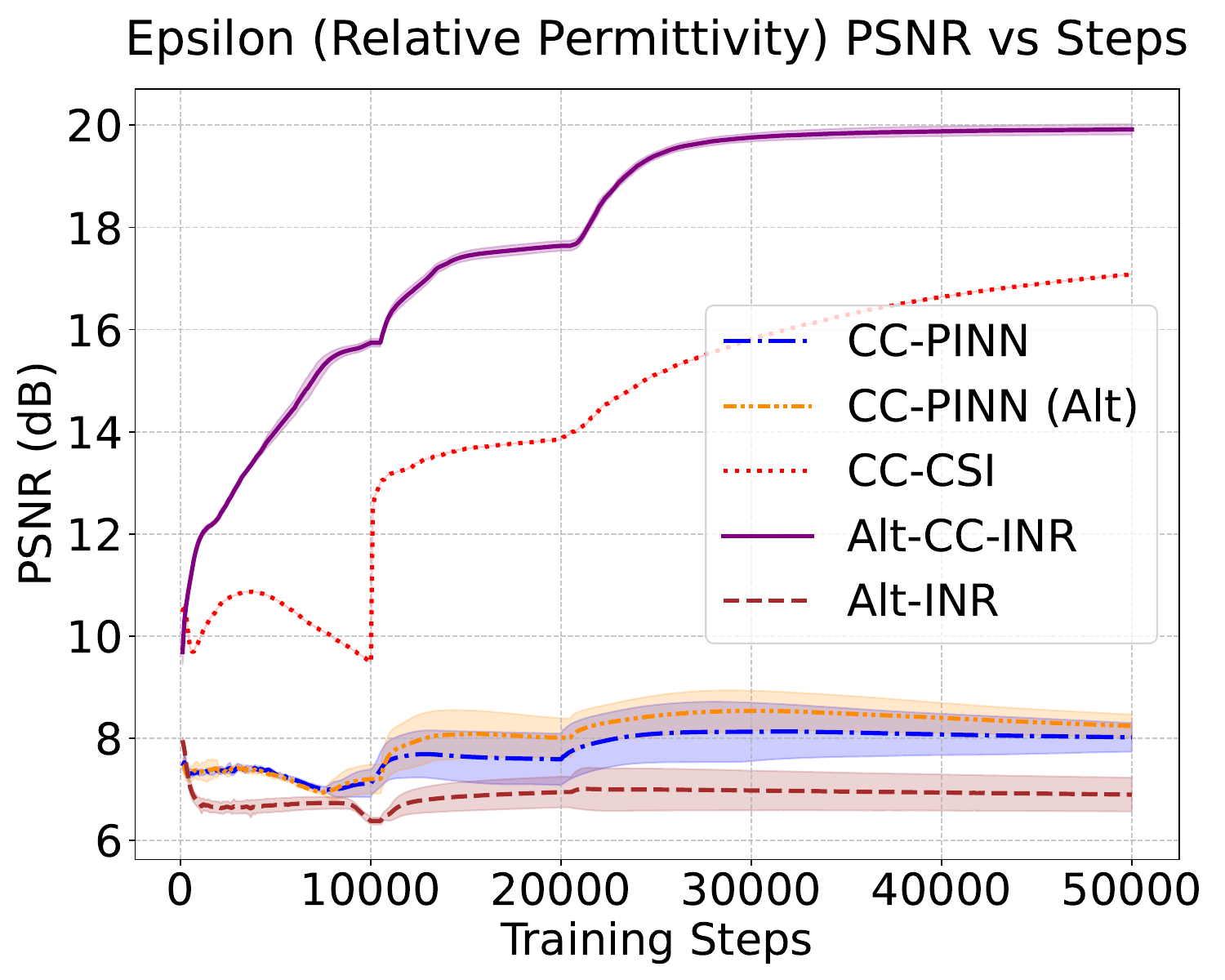}}\hfill
        \subfloat[]{\includegraphics[width=0.5\linewidth]{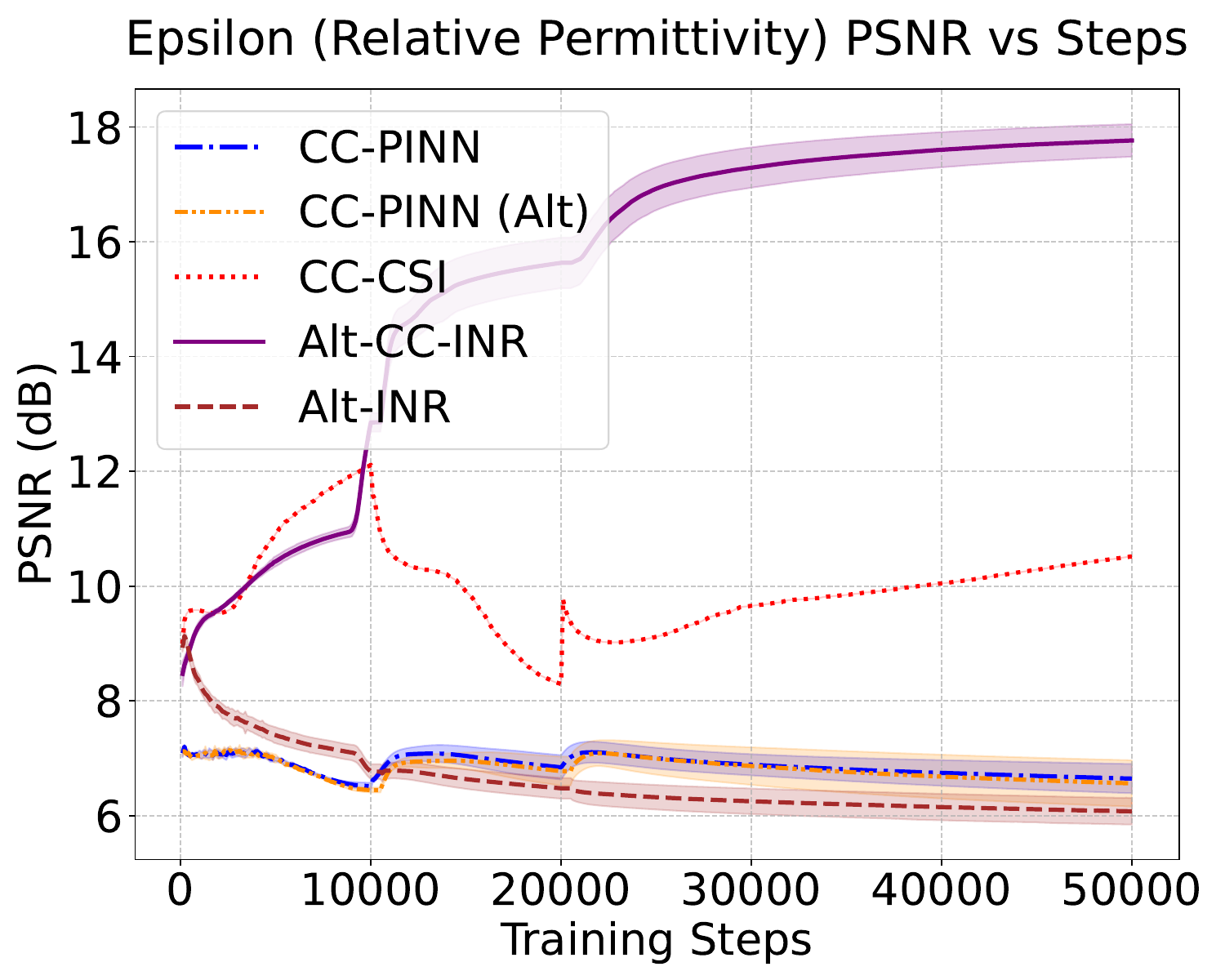}}\hspace*{\fill}

        \caption{Comparison of mean PSNR convergence with standard deviation band for CC-PINN, CC-PINN (Alt), CC-CSI, Alt-CC-INR and Alt-INR using the frequency-hopping strategy to invert ``Austria'' dielectric targets of different contrasts. The relative permittivity of the dielectric targets is $\varepsilon_\text{r} = 6$ (a), $\varepsilon_\text{r} = 7$ (b), $\varepsilon_\text{r} = 8$ (c), and $\varepsilon_\text{r} = 9$ (d).}
        \label{fig:PSNR_eps_AustriaDiel}   
    \end{figure}

    \begin{figure}[!t]
        \hspace*{\fill}%
        \subfloat[]{\includegraphics[width=0.5\linewidth]{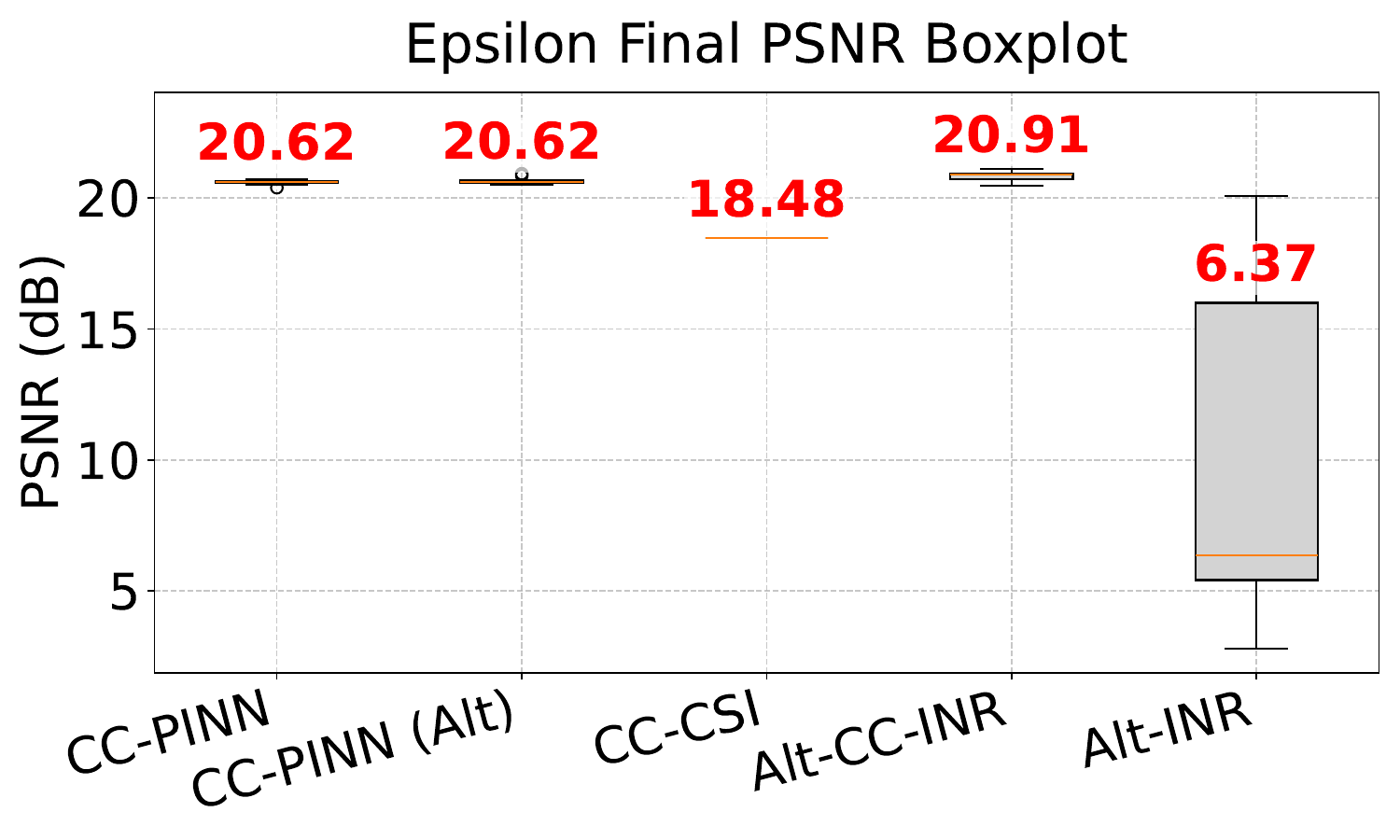}}\hfill
        \subfloat[]{\includegraphics[width=0.5\linewidth]{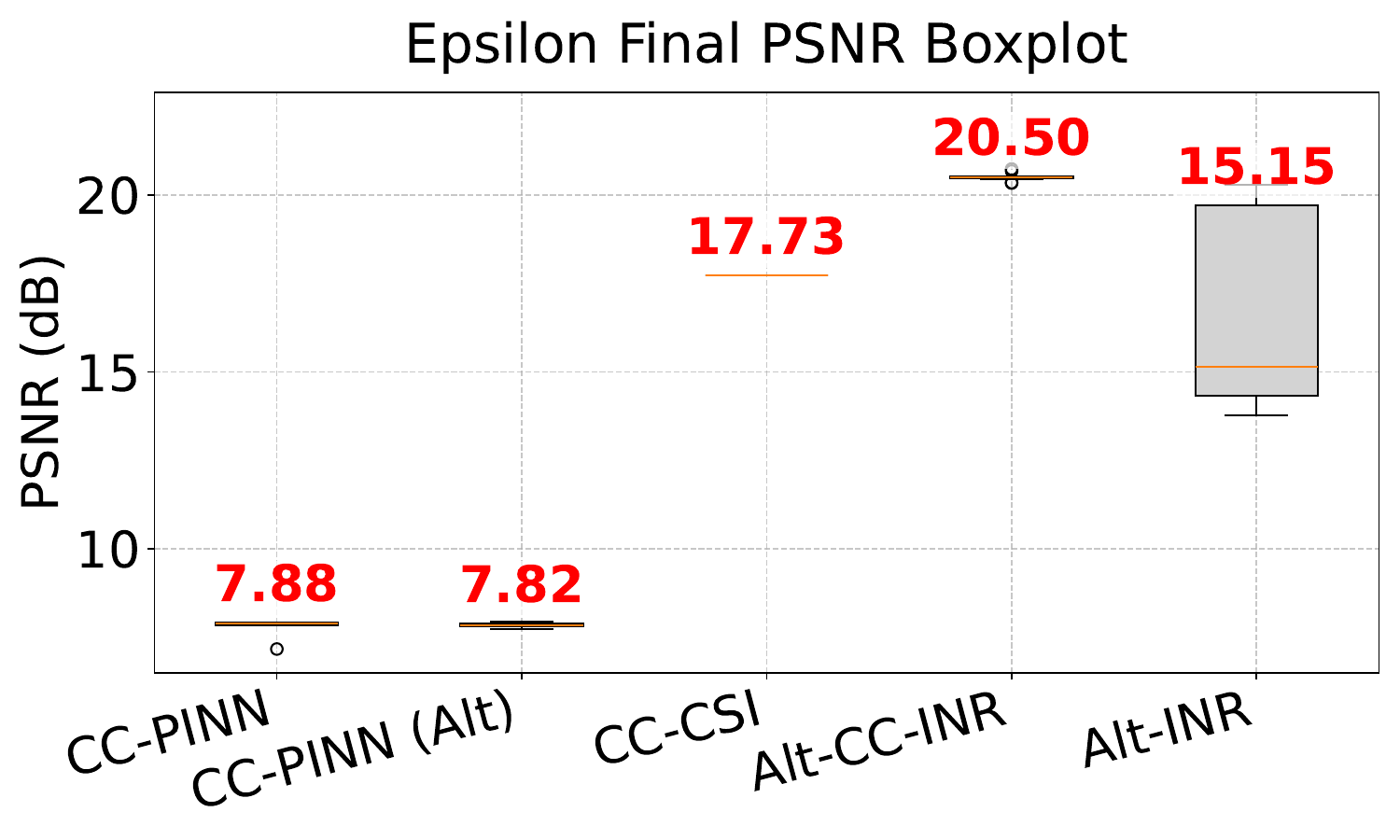}}\hspace*{\fill}

        \vspace{-1.0ex}

        \hspace*{\fill}%
        \subfloat[]{\includegraphics[width=0.5\linewidth]{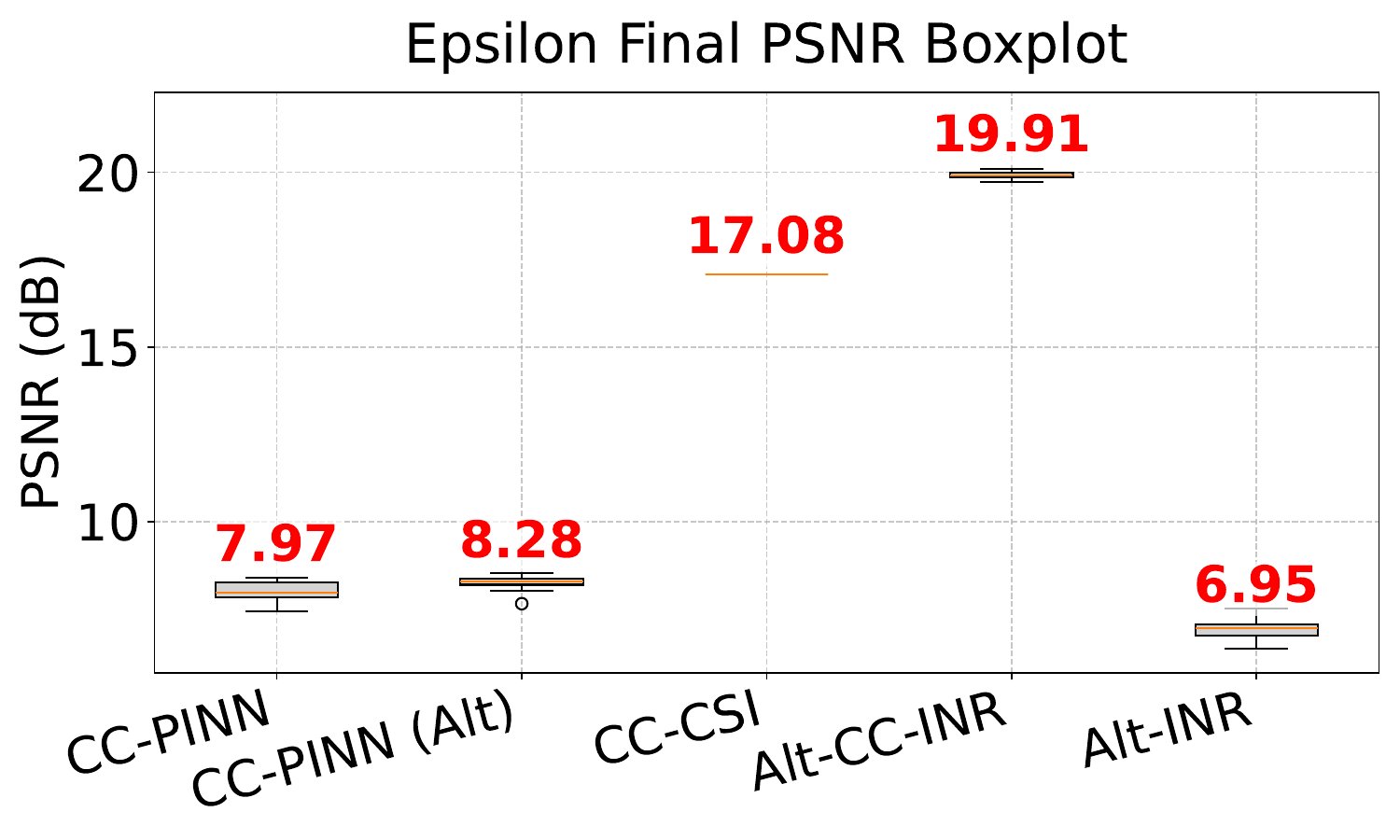}}\hfill
        \subfloat[]{\includegraphics[width=0.5\linewidth]{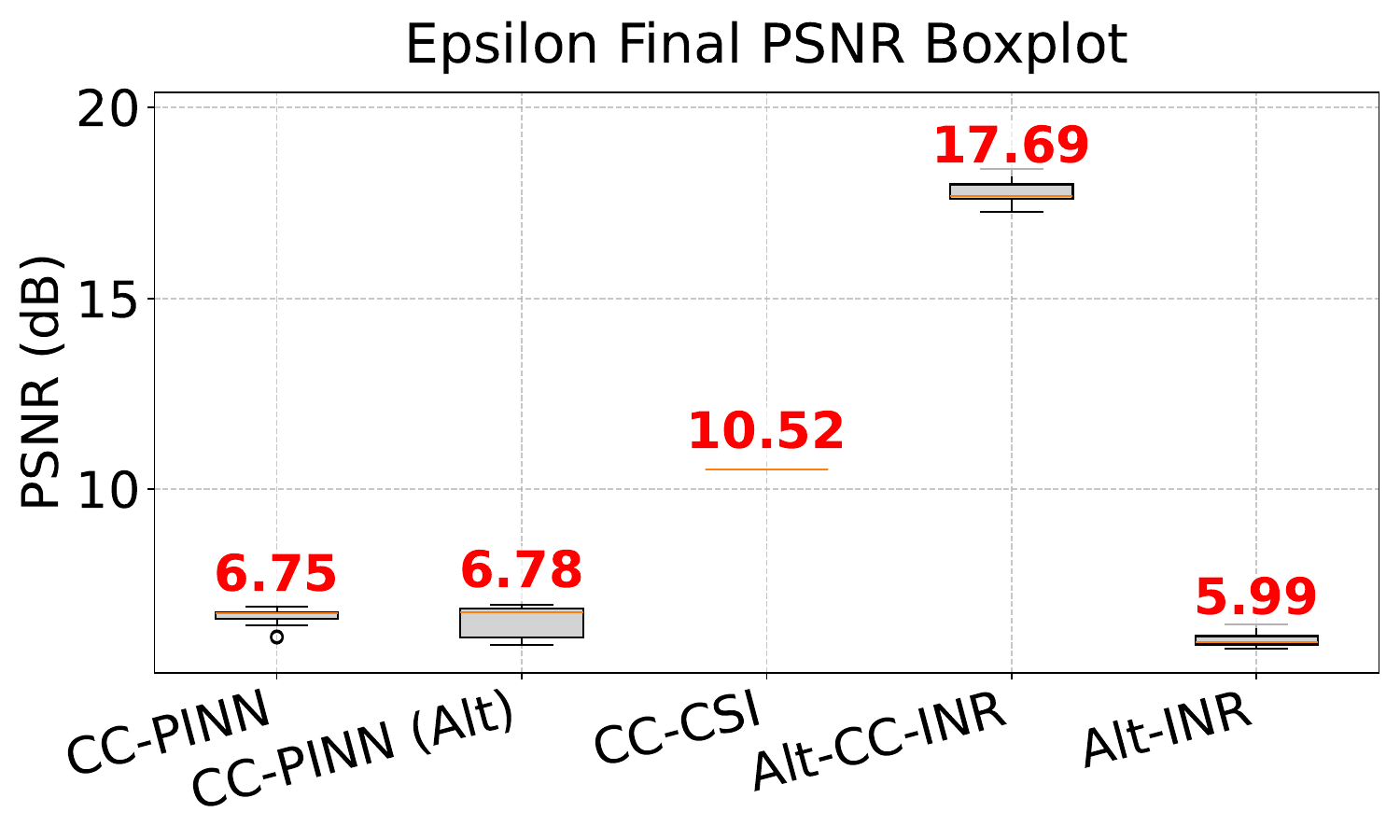}}\hspace*{\fill}

        \caption{Boxplots of final PSNR for CC-PINN, CC-PINN (Alt), CC-CSI, Alt-CC-INR and Alt-INR using the frequency-hopping strategy to invert ``Austria'' dielectric targets of different contrasts. The relative permittivity of the dielectric targets is $\varepsilon_\text{r} = 6$ (a), $\varepsilon_\text{r} = 7$ (b), $\varepsilon_\text{r} = 8$ (c) and $\varepsilon_\text{r} = 9$ (d).}
        \label{fig:PSNR_Boxplots_eps_AustriaDiel}   
    \end{figure}

    \begin{figure}[!t]
        \makebox[\columnwidth][c]{%
            \begin{minipage}{\columnwidth}
                \hspace*{\fill}%
                \subfloat[]{\includegraphics[height=0.235\linewidth]{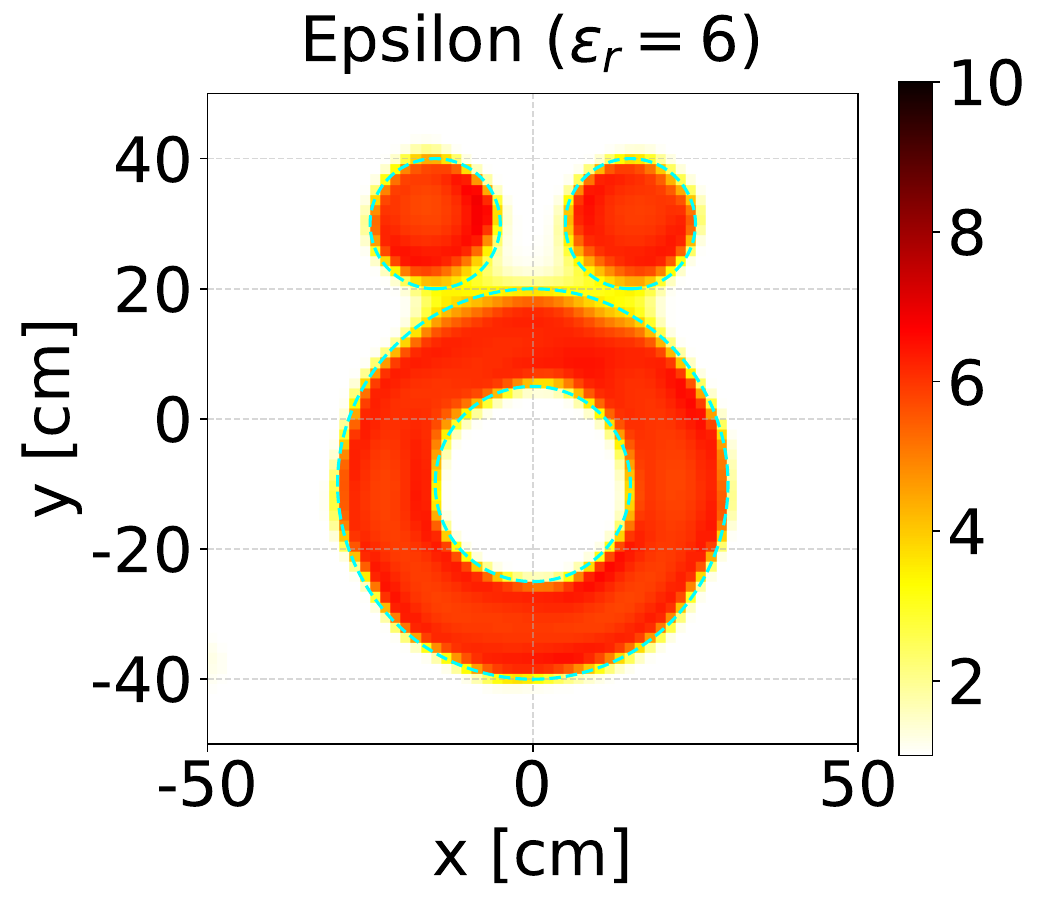}}\hfill
                \subfloat[]{\includegraphics[height=0.235\linewidth]{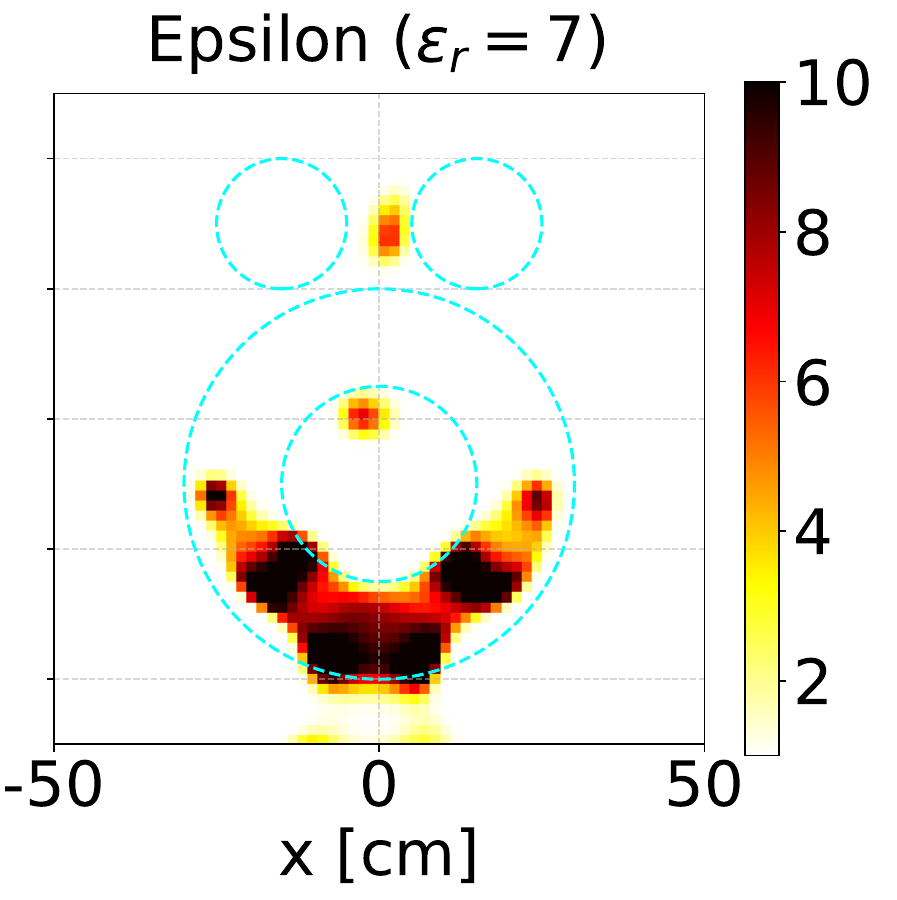}}\hfill
                \subfloat[]{\includegraphics[height=0.235\linewidth]{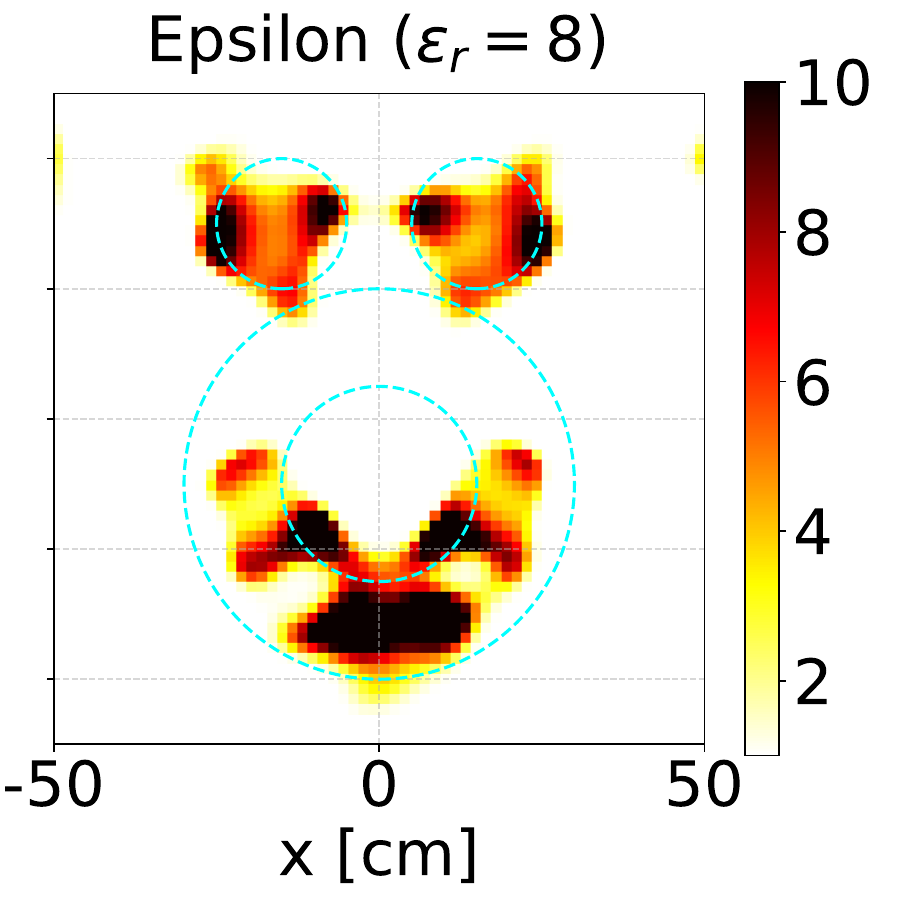}}\hfill
                \subfloat[]{\includegraphics[height=0.235\linewidth]{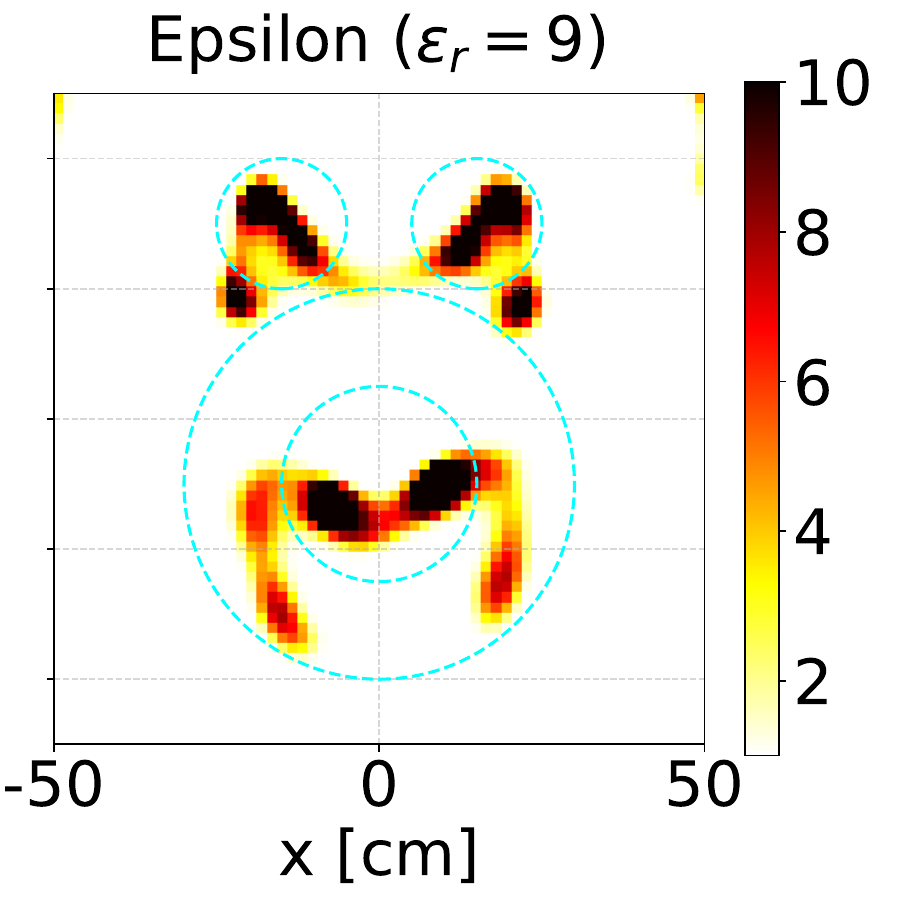}}\hspace*{\fill}

                \vspace{-1.5ex} 

                \hspace*{\fill}
                \subfloat[]{\includegraphics[height=0.235\linewidth]{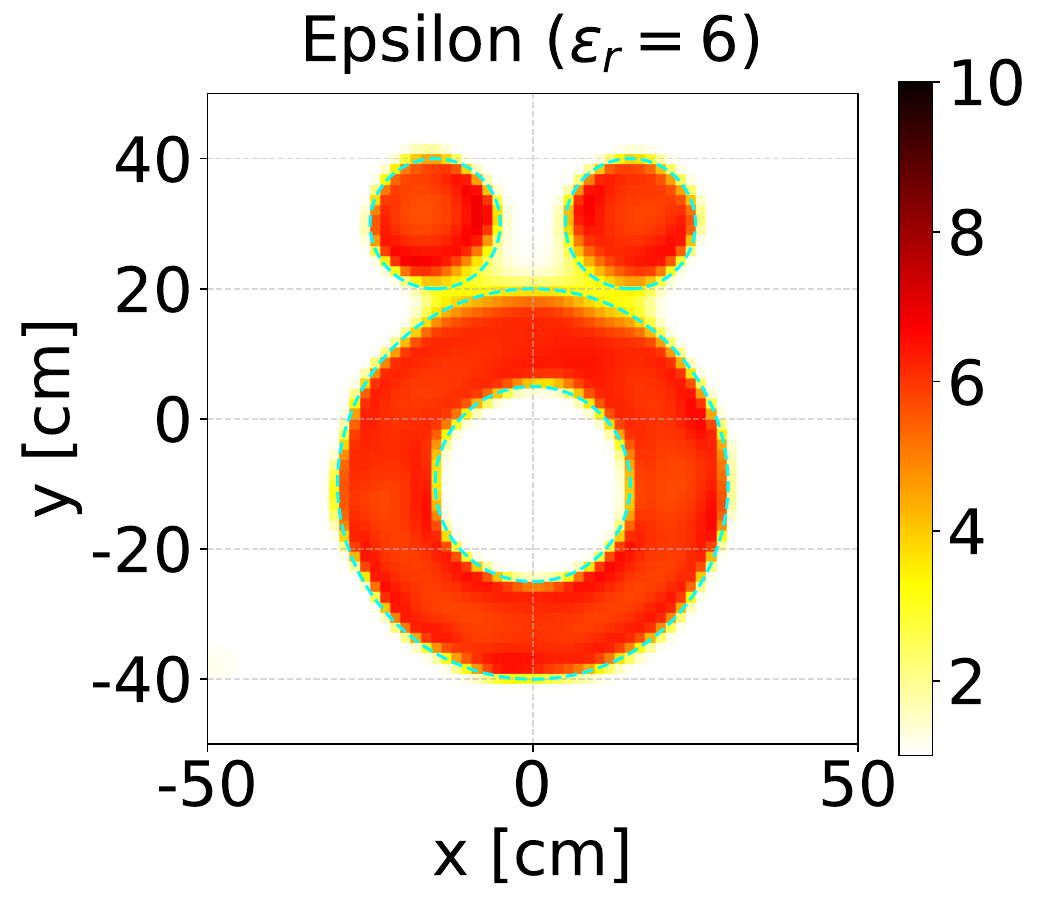}}\hfill
                \subfloat[]{\includegraphics[height=0.235\linewidth]{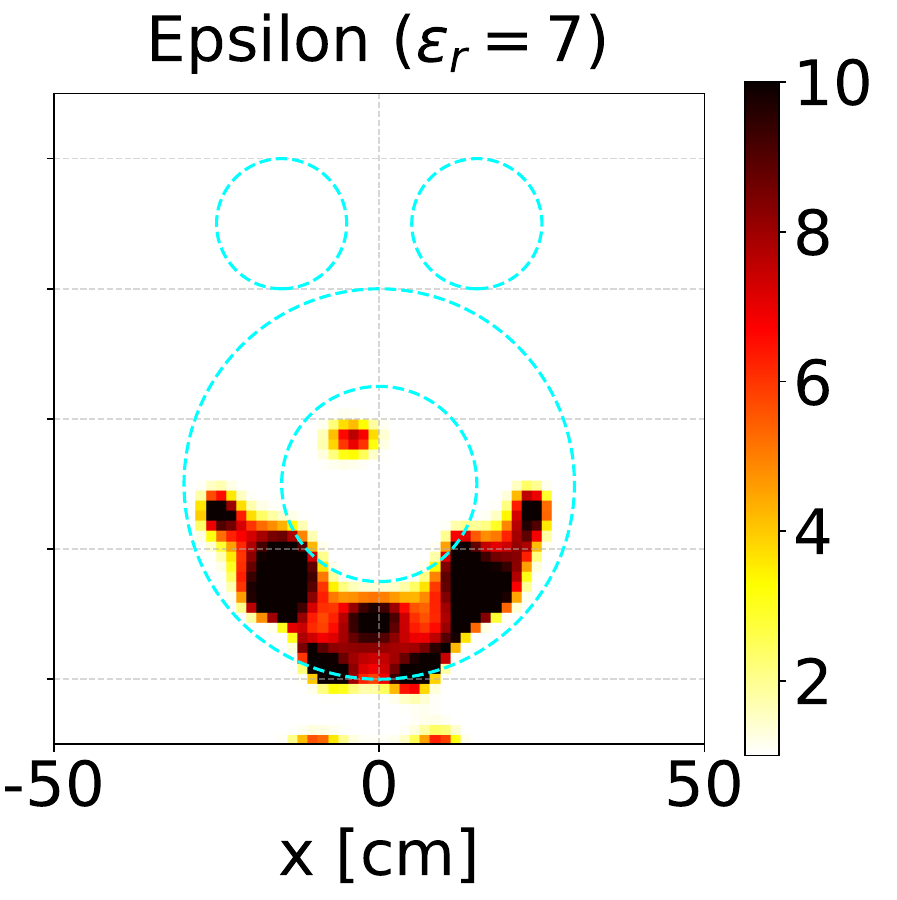}}\hfill
                \subfloat[]{\includegraphics[height=0.235\linewidth]{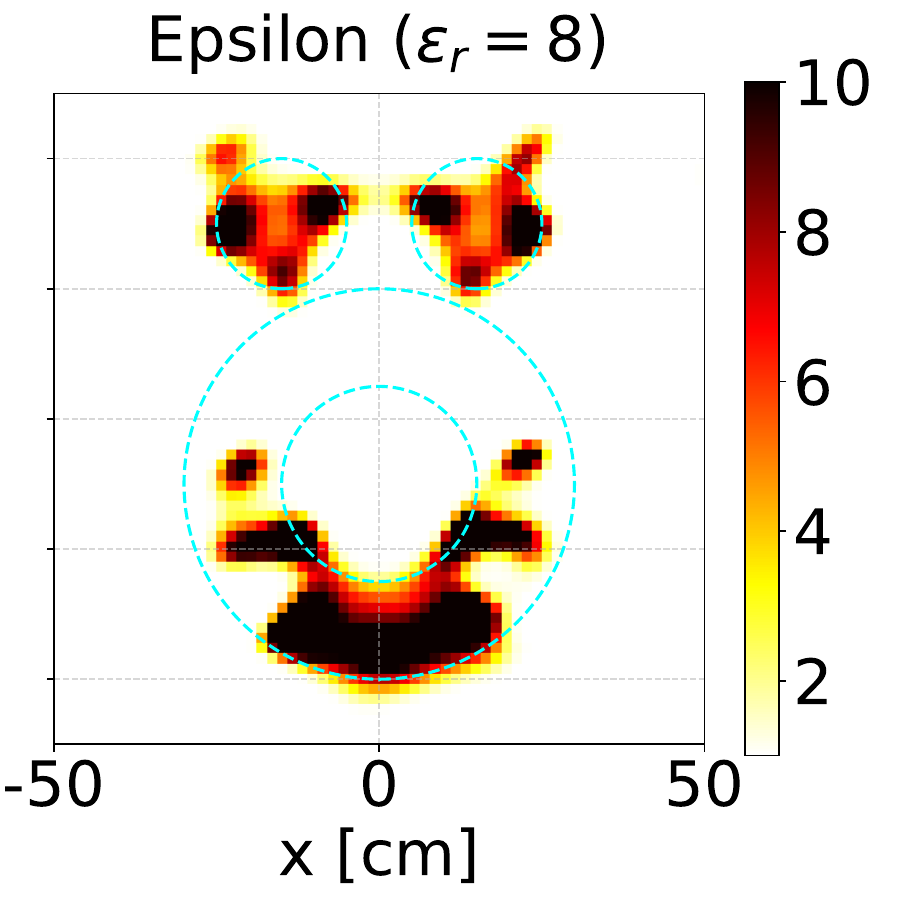}}\hfill
                \subfloat[]{\includegraphics[height=0.235\linewidth]{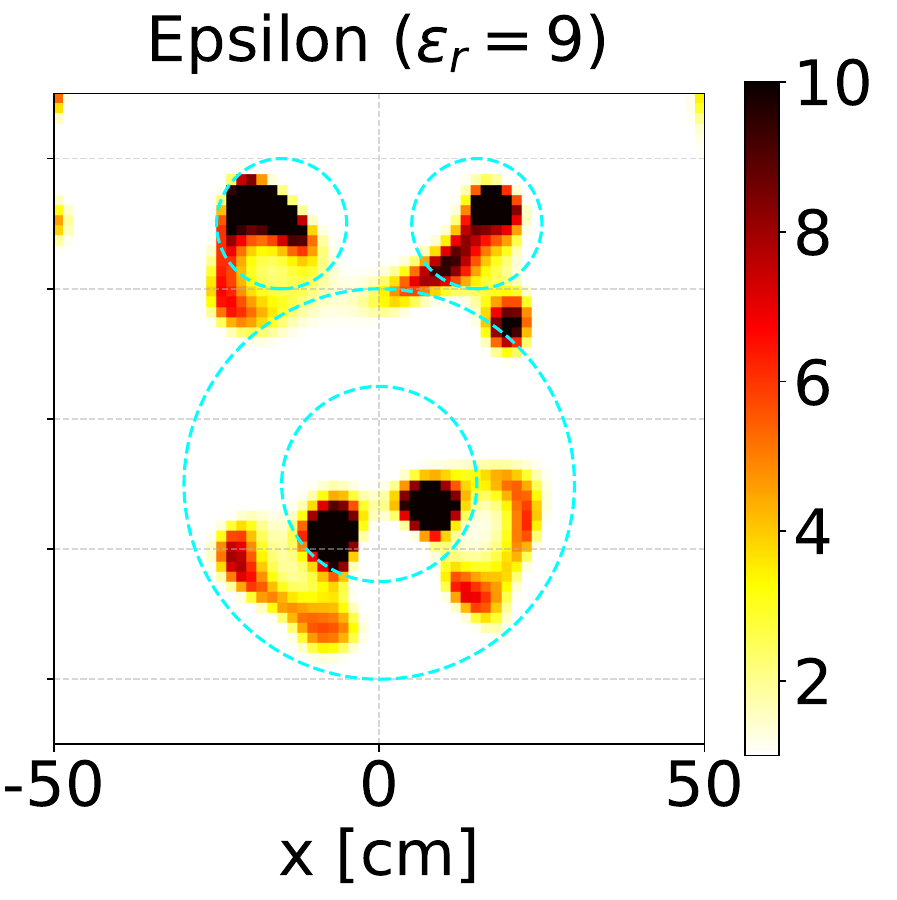}}\hspace*{\fill}

                \vspace{-1.5ex}

                \hspace*{\fill}
                \subfloat[]{\includegraphics[height=0.235\linewidth]{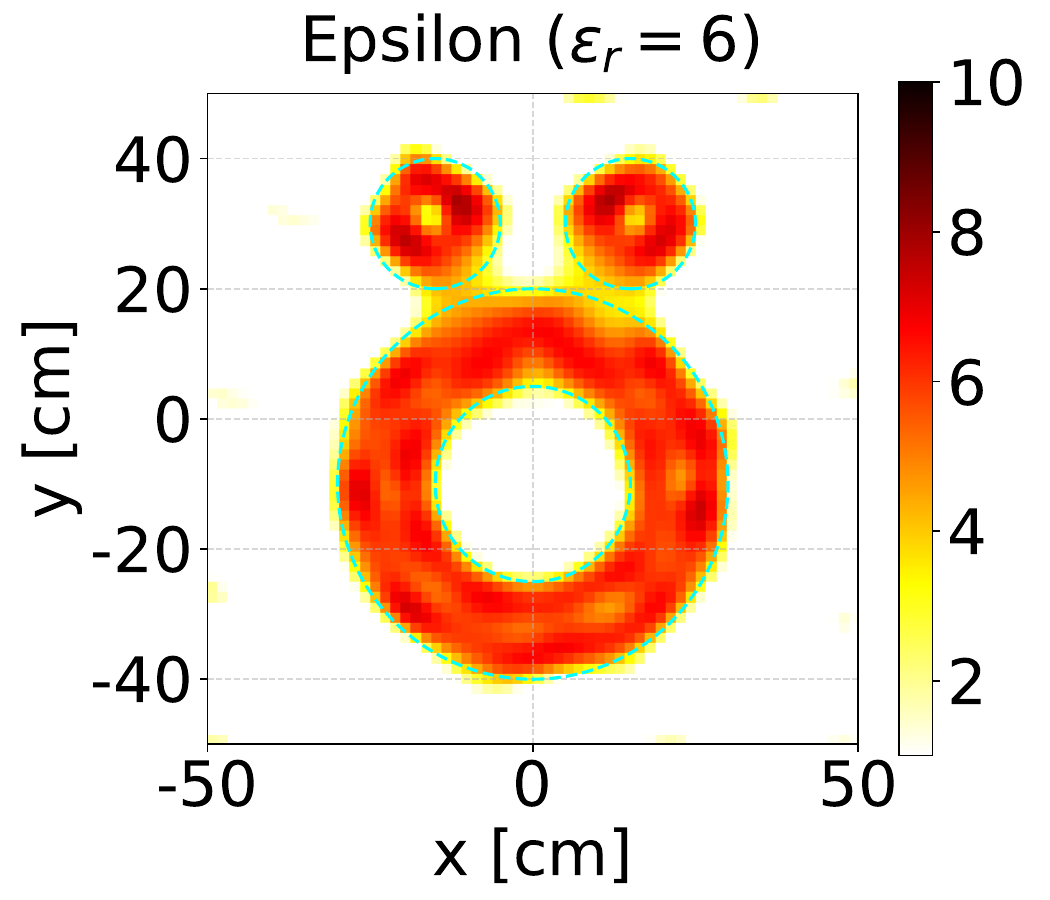}}\hfill
                \subfloat[]{\includegraphics[height=0.235\linewidth]{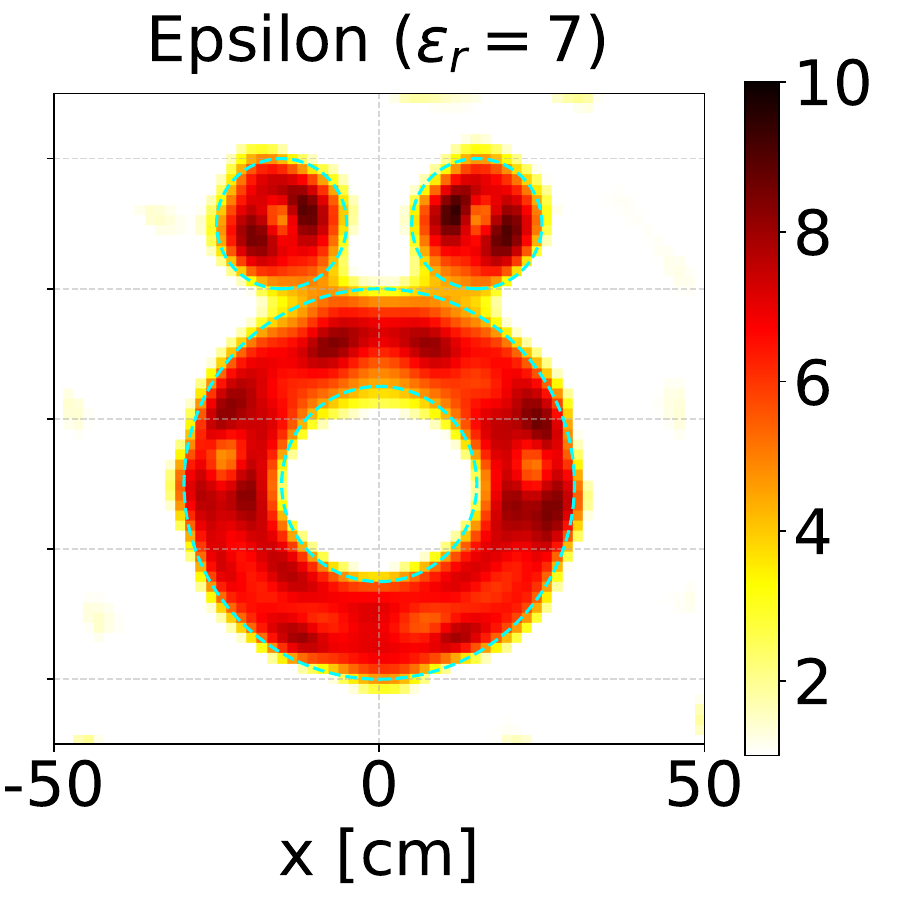}}\hfill
                \subfloat[]{\includegraphics[height=0.235\linewidth]{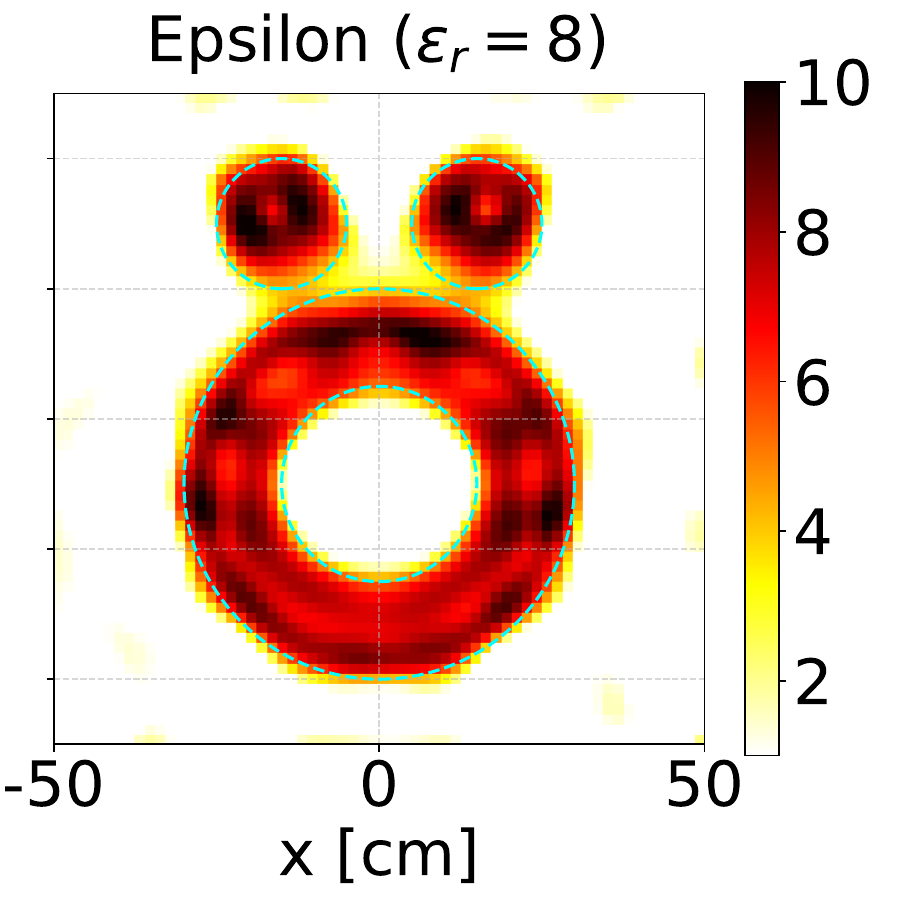}}\hfill
                \subfloat[]{\includegraphics[height=0.235\linewidth]{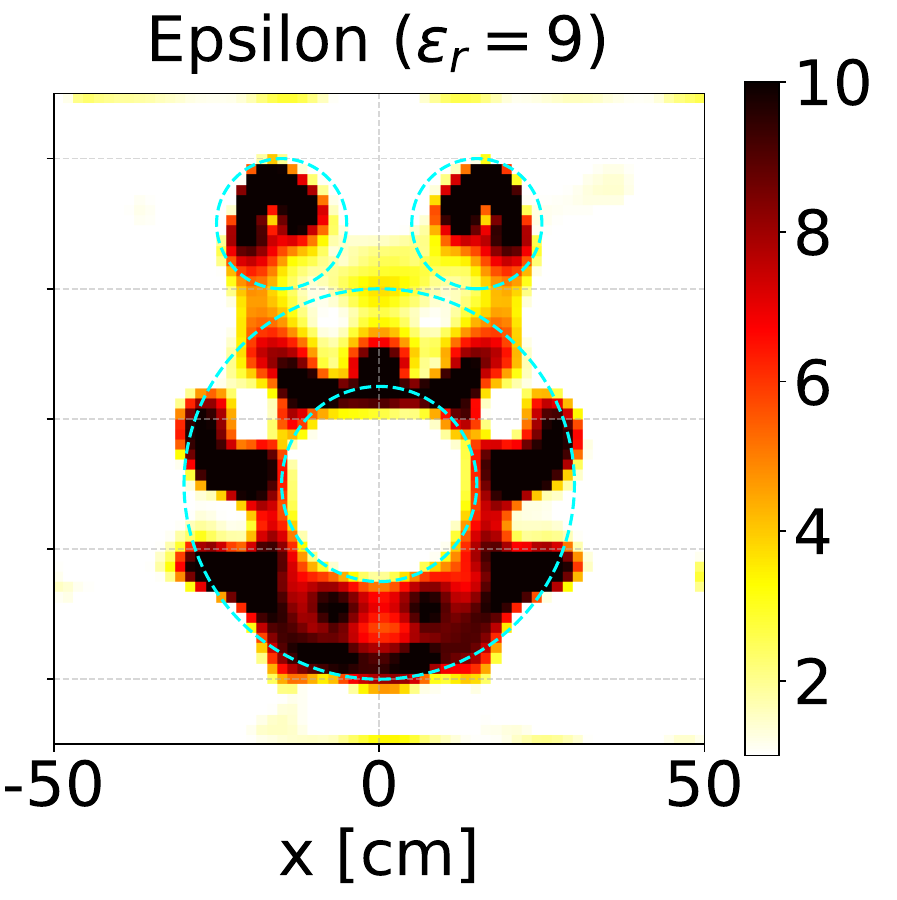}}\hspace*{\fill}

                \vspace{-1.5ex}

                \hspace*{\fill}
                \subfloat[]{\includegraphics[height=0.235\linewidth]{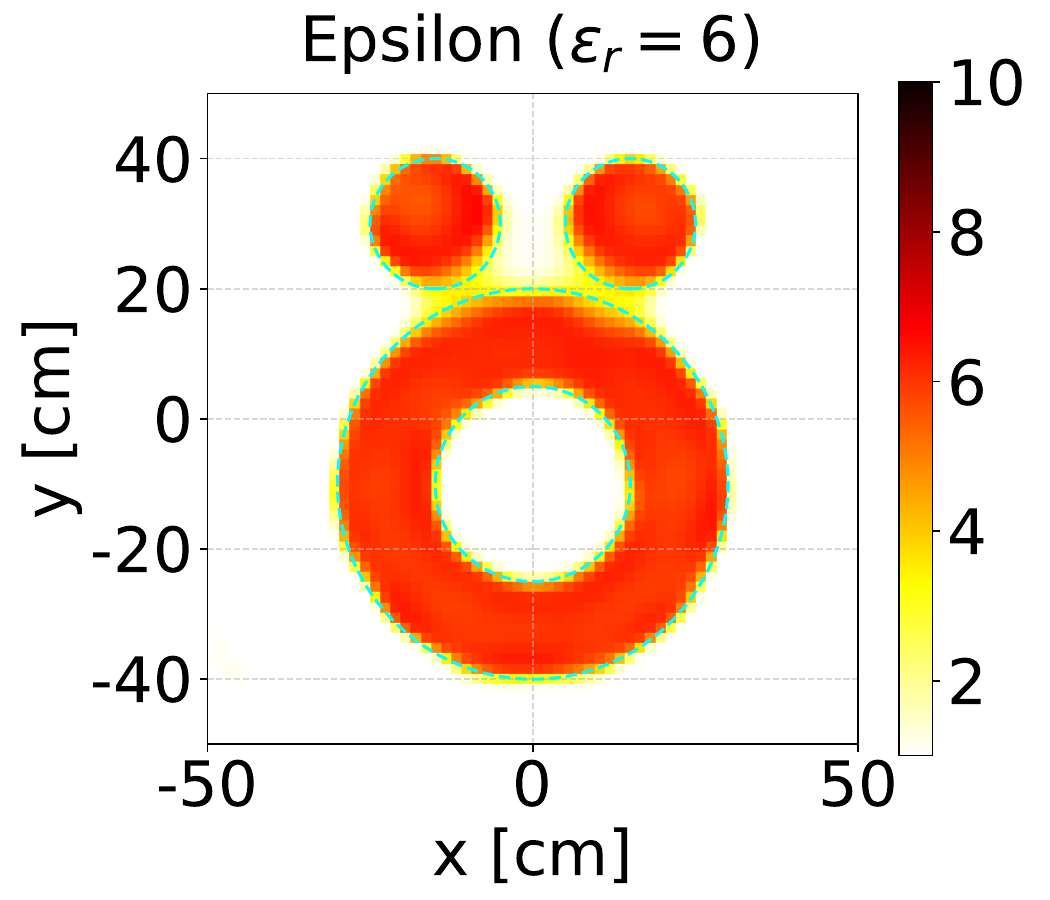}}\hfill
                \subfloat[]{\includegraphics[height=0.235\linewidth]{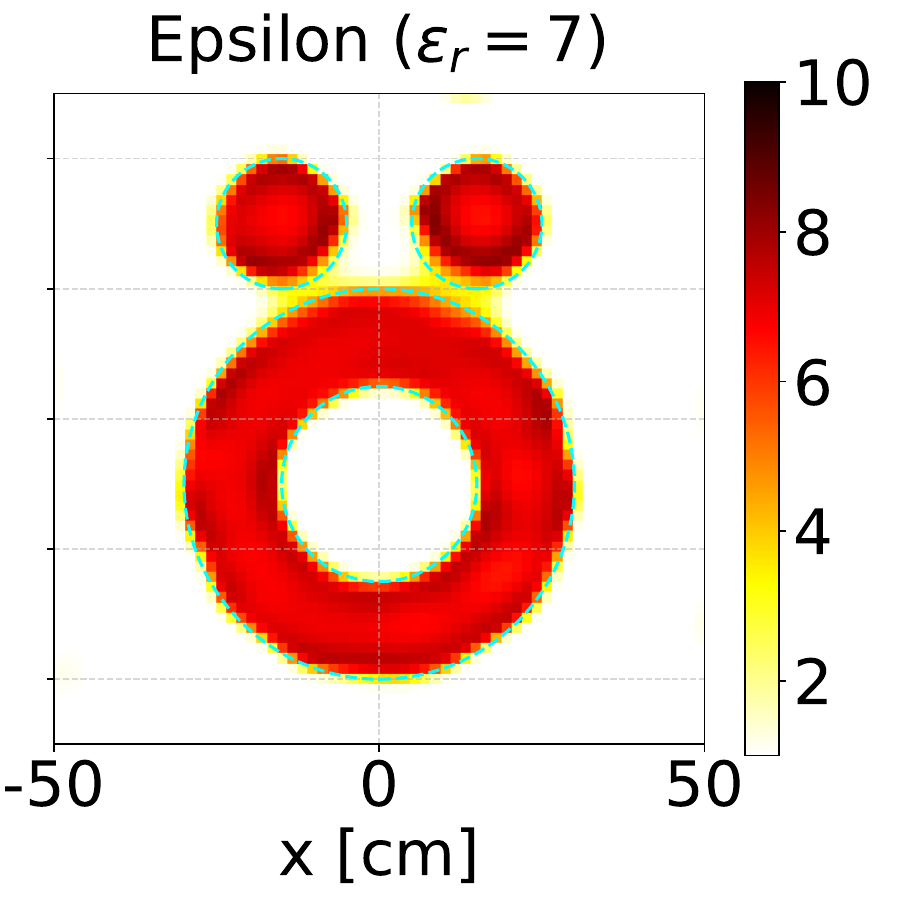}}\hfill
                \subfloat[]{\includegraphics[height=0.235\linewidth]{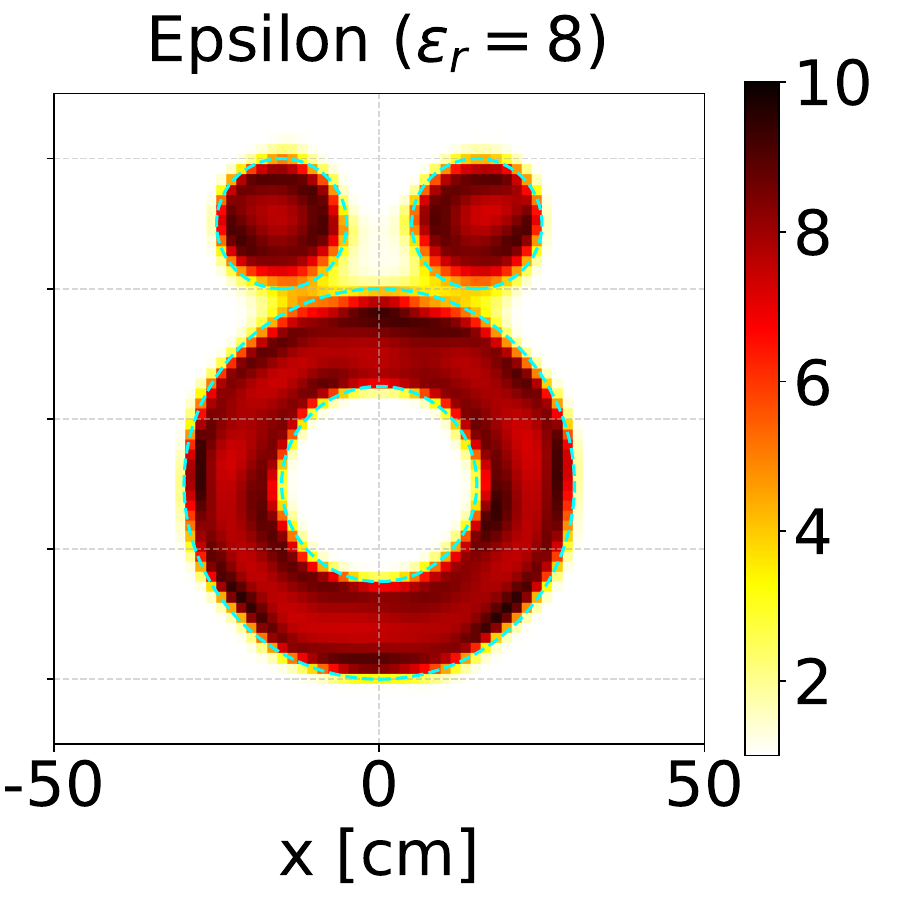}}\hfill
                \subfloat[]{\includegraphics[height=0.235\linewidth]{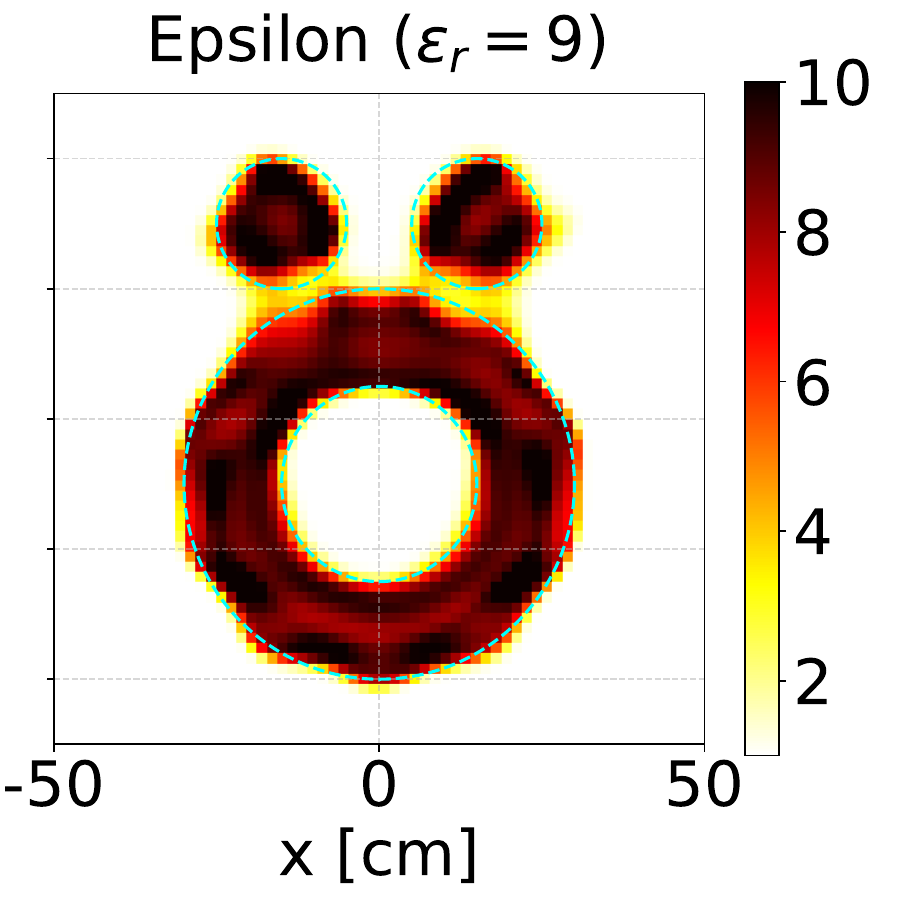}}\hspace*{\fill}

                \vspace{-1.5ex}

                \hspace*{\fill}
                \subfloat[]{\includegraphics[height=0.235\linewidth]{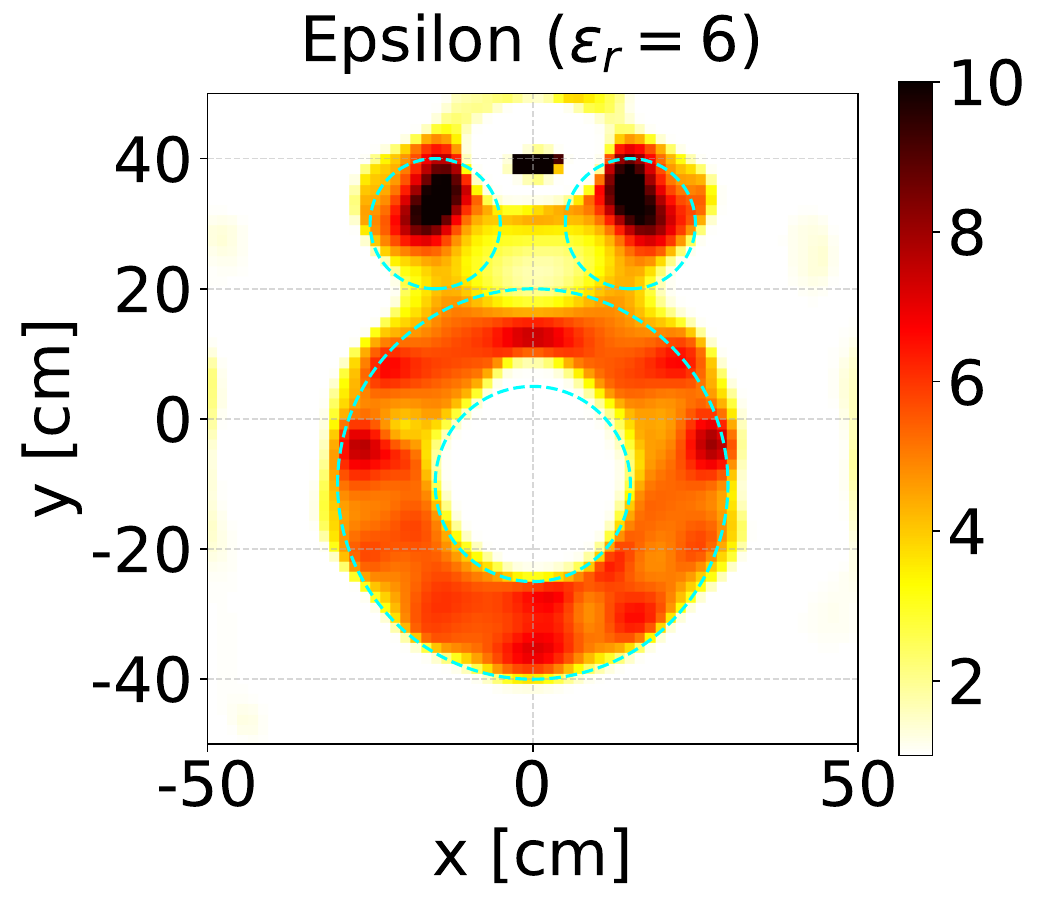}}\hfill
                \subfloat[]{\includegraphics[height=0.235\linewidth]{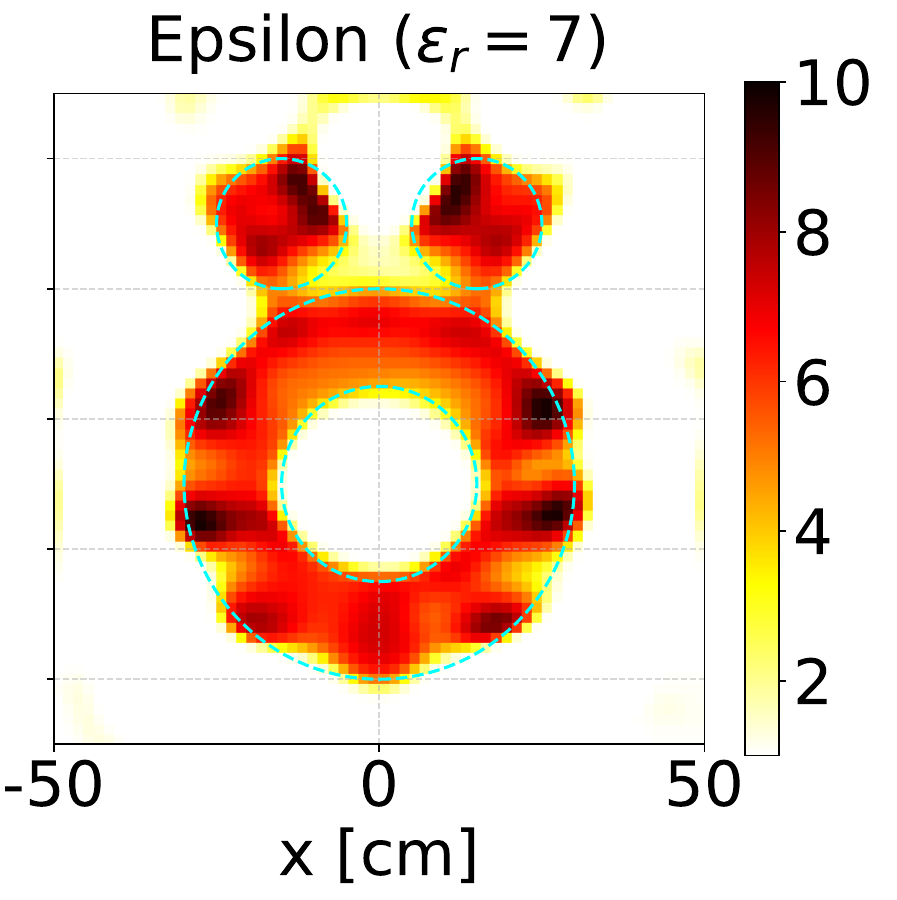}}\hfill
                \subfloat[]{\includegraphics[height=0.235\linewidth]{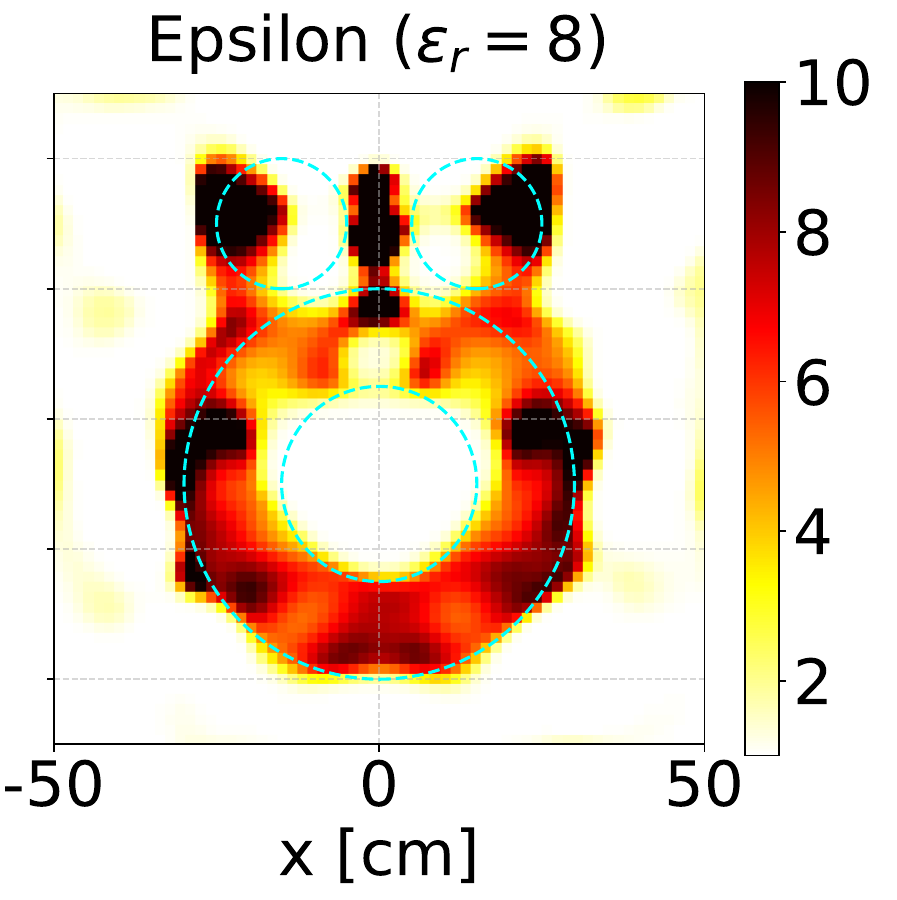}}\hfill
                \subfloat[]{\includegraphics[height=0.235\linewidth]{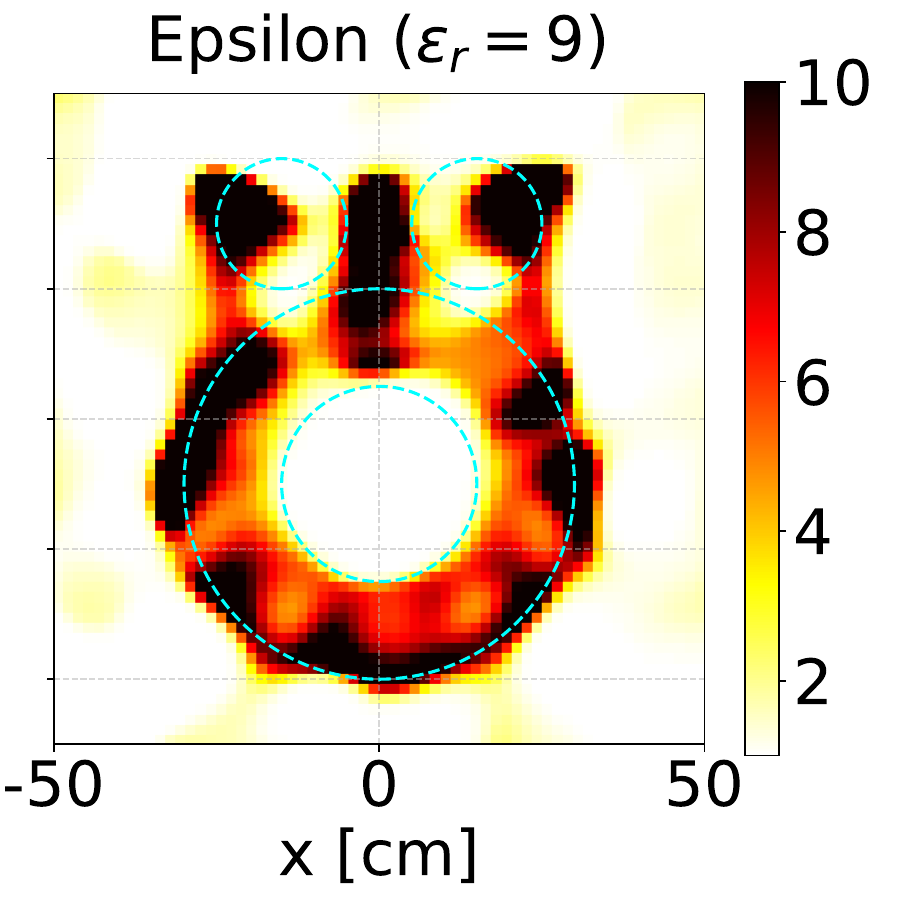}}\hspace*{\fill}
            \end{minipage}%
        }
        \caption{Final reconstructed images for CC-PINN, CC-PINN (Alt), CC-CSI, Alt-CC-INR and Alt-INR using the frequency-hopping strategy to invert ``Austria'' dielectric targets of different contrasts. From top to bottom: CC-PINN (a, b, c, d), CC-PINN (Alt) (e, f, g, h), CC-CSI (i, j, k, l), Alt-CC-INR (m, n, o, p) and Alt-INR (q, r, s, t). From left to right: $\varepsilon_\text{r} = 6$ (a, e, i, m, q), $\varepsilon_\text{r} = 7$ (b, f, j, n, r), $\varepsilon_\text{r} = 8$ (c, g, k, o, s), and $\varepsilon_\text{r} = 9$ (d, h, l, p, t).}
        \label{fig:eps_AustriaDiel}   
    \end{figure}

    \begin{figure}[!t]
        \hspace*{\fill}%
        \subfloat[]{\includegraphics[width=0.5\linewidth]{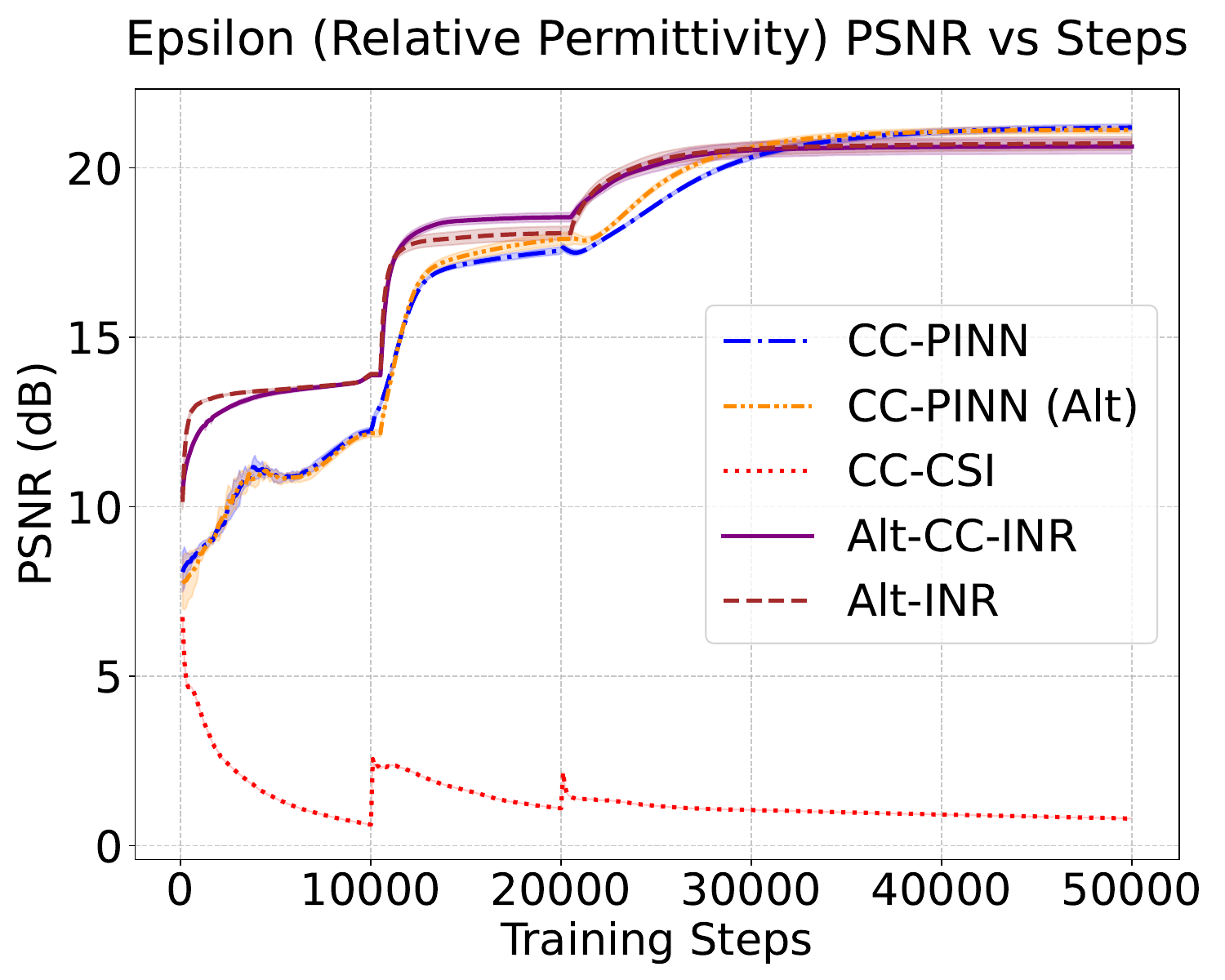}}\hfill
        \subfloat[]{\includegraphics[width=0.5\linewidth]{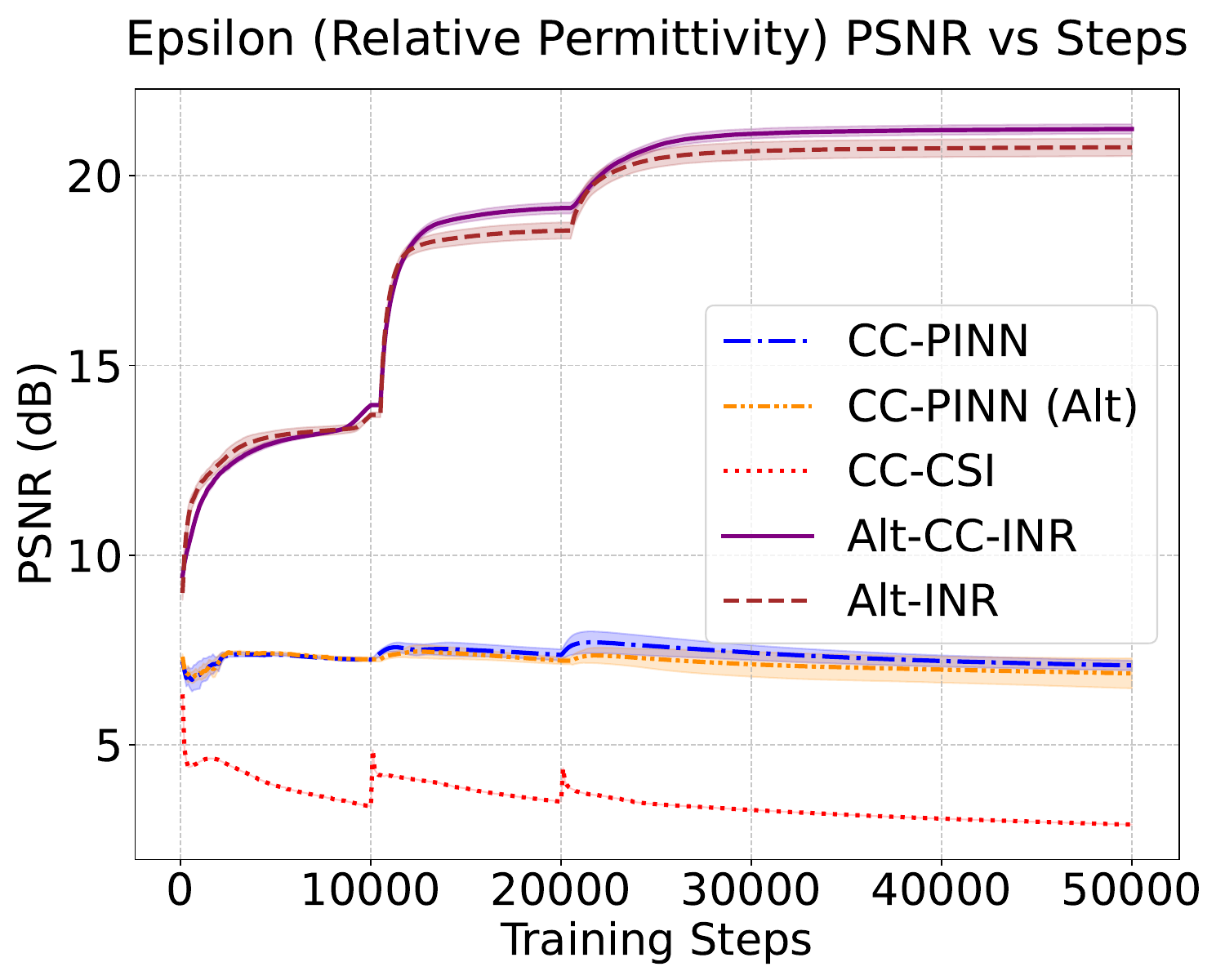}}\hspace*{\fill}

        \vspace{-1.0ex}

        \hspace*{\fill}%
        \subfloat[]{\includegraphics[width=0.5\linewidth]{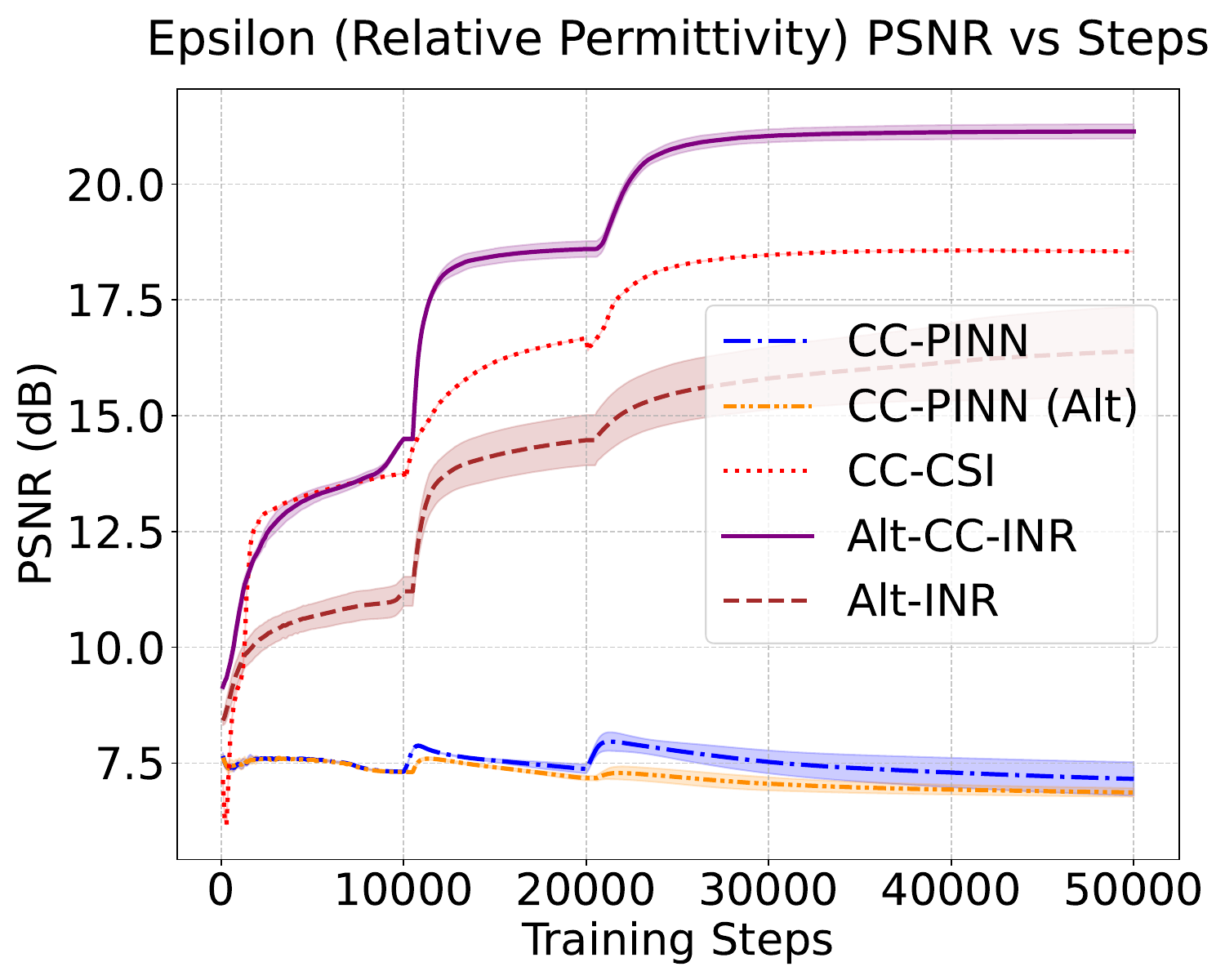}}\hfill
        \subfloat[]{\includegraphics[width=0.5\linewidth]{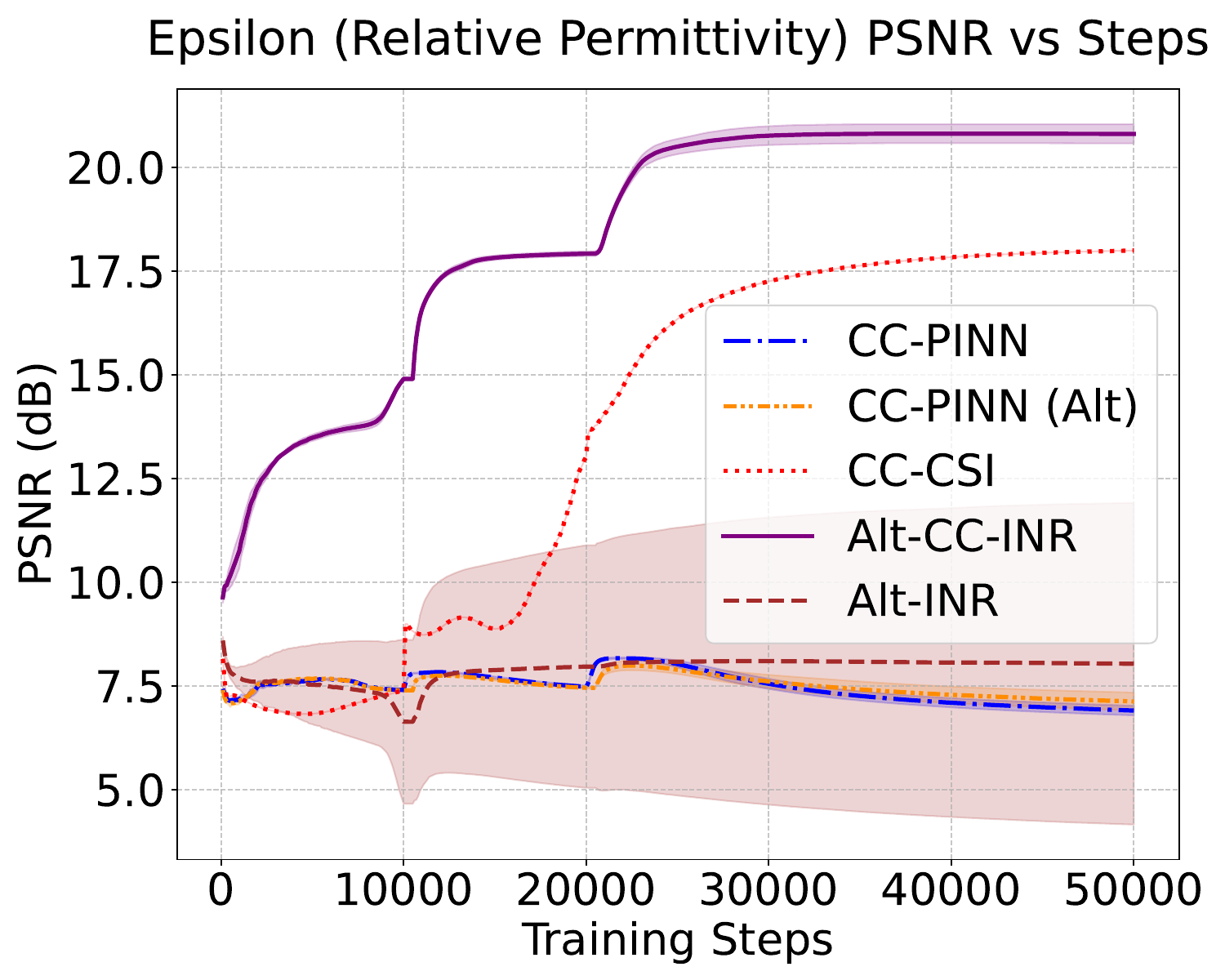}}\hspace*{\fill}

        \caption{Comparison of mean PSNR convergence with standard deviation band for CC-PINN, CC-PINN (Alt), CC-CSI, Alt-CC-INR and Alt-INR using the frequency-hopping strategy to invert ``Bowtie-Cross'' dielectric targets of different contrasts. The relative permittivity of the dielectric targets is $\varepsilon_\text{r} = 6$ (a), $\varepsilon_\text{r} = 6.5$ (b), $\varepsilon_\text{r} = 7$ (c), and $\varepsilon_\text{r} = 7.5$ (d). }
        \label{fig:PSNR_eps_BowtieCrossDiel}   
    \end{figure}

    \begin{figure}[!t]
        \hspace*{\fill}%
        \subfloat[]{\includegraphics[width=0.5\linewidth]{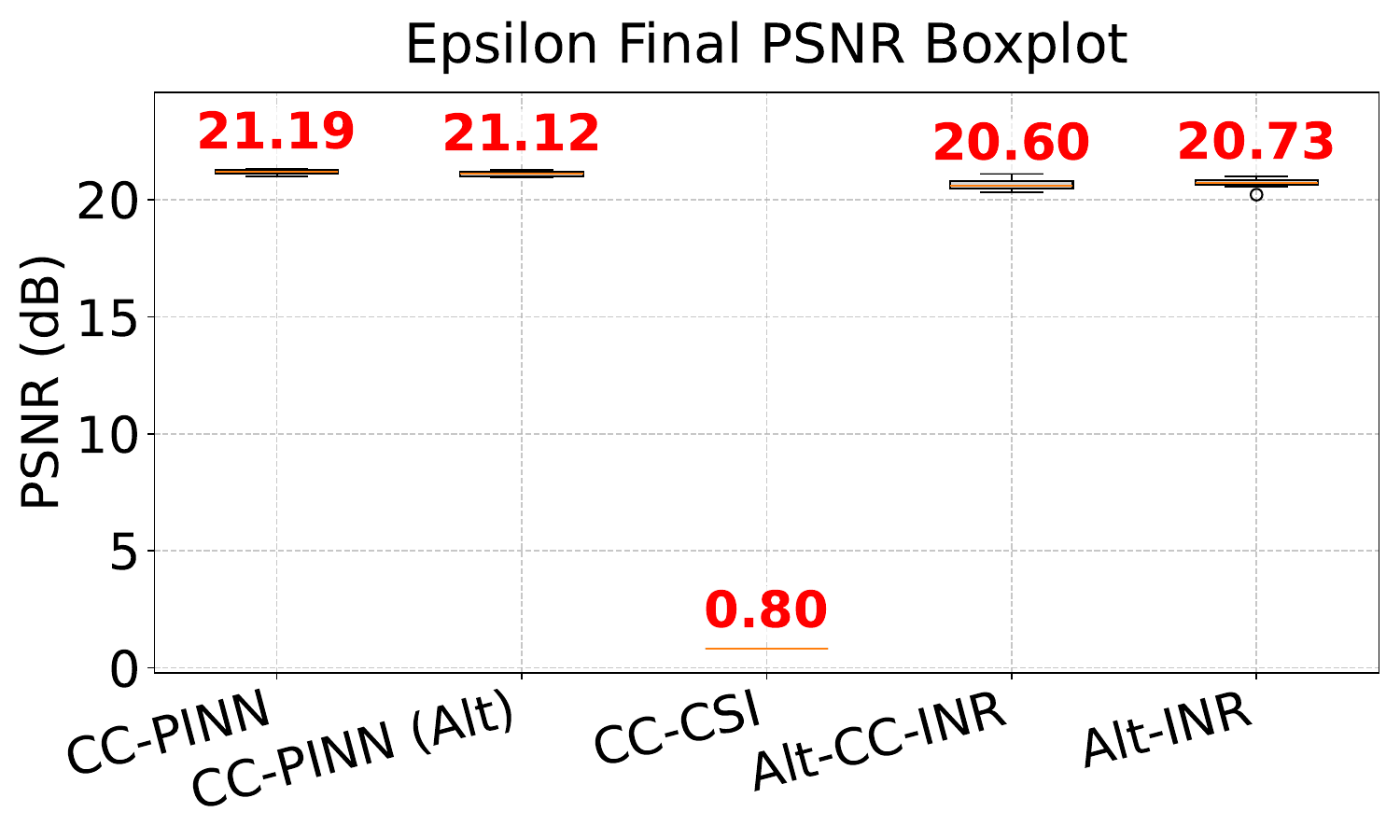}}\hfill
        \subfloat[]{\includegraphics[width=0.5\linewidth]{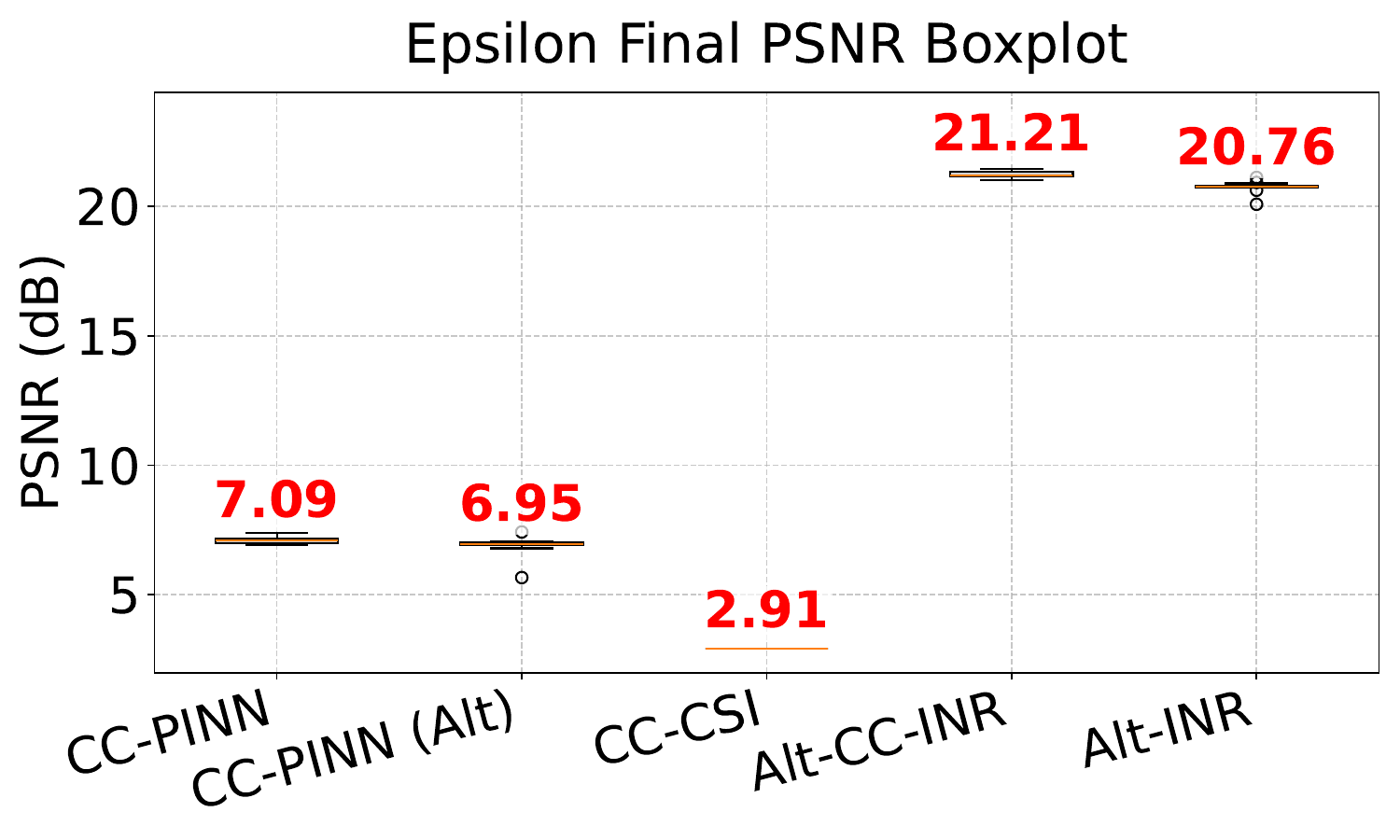}}\hspace*{\fill}

        \vspace{-1.0ex}

        \hspace*{\fill}%
        \subfloat[]{\includegraphics[width=0.5\linewidth]{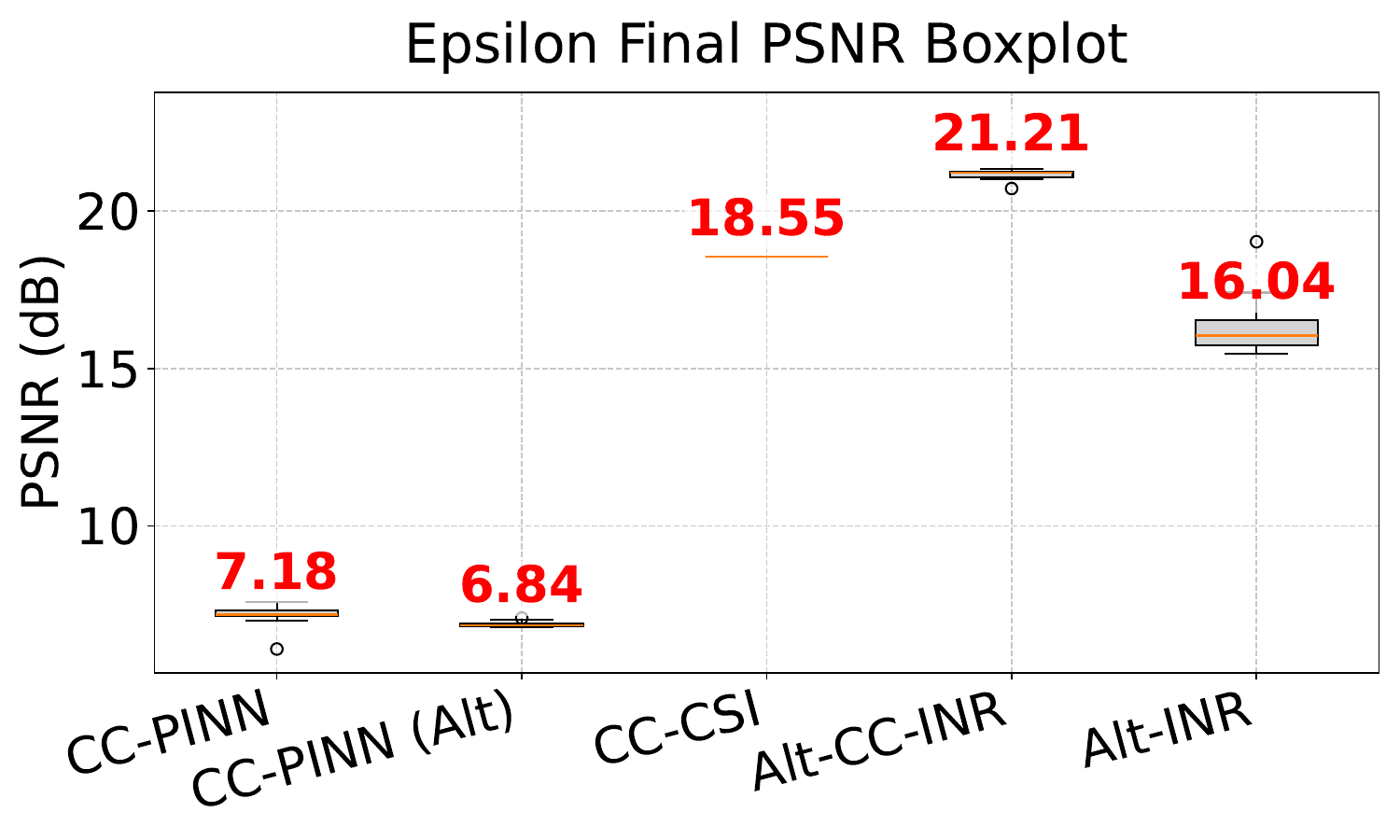}}\hfill
        \subfloat[]{\includegraphics[width=0.5\linewidth]{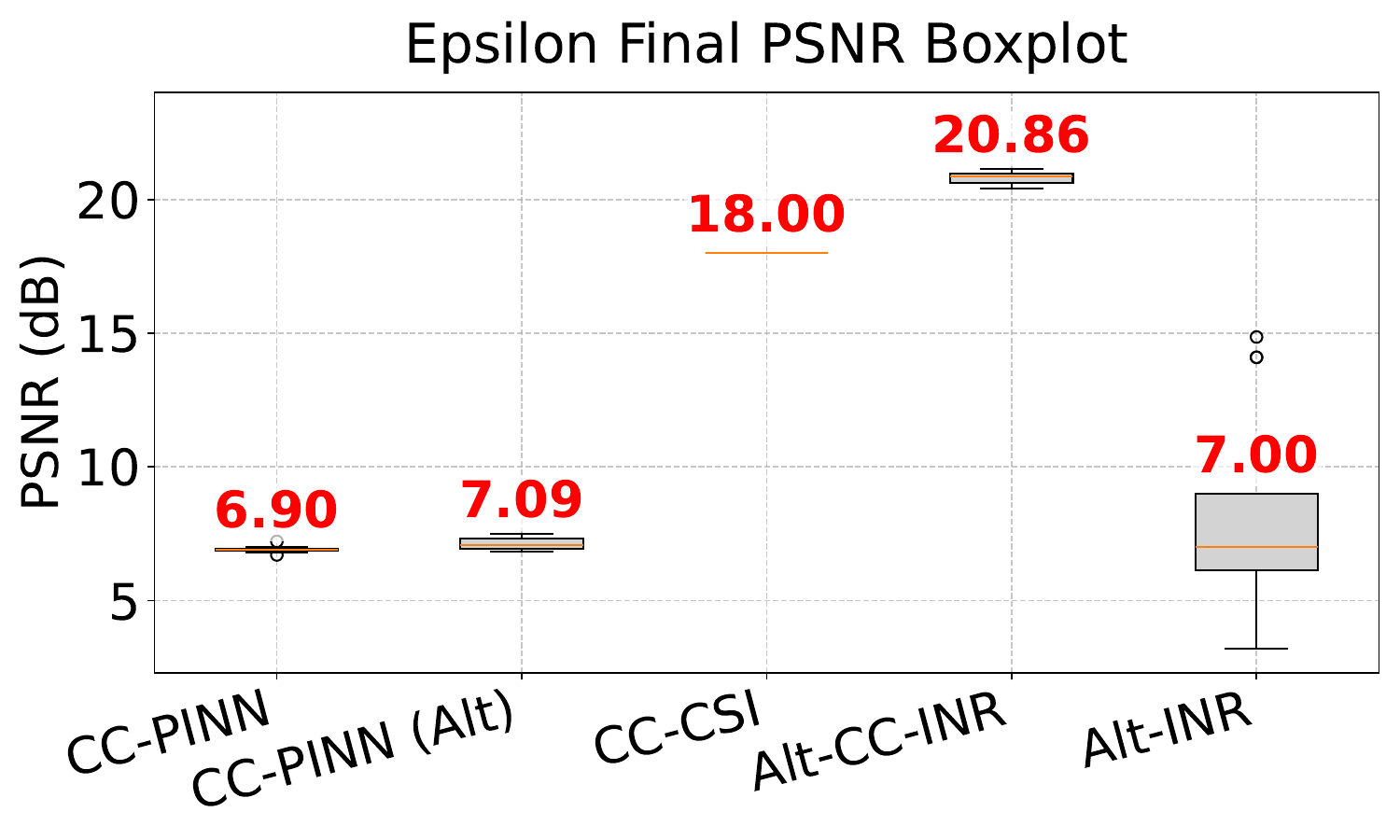}}\hspace*{\fill}

        \caption{Boxplots of final PSNR for CC-PINN, CC-PINN (Alt), CC-CSI, Alt-CC-INR and Alt-INR using the frequency-hopping strategy to invert ``Bowtie-Cross'' dielectric targets of different contrasts. The relative permittivity of the dielectric targets is $\varepsilon_\text{r} = 6$ (a), $\varepsilon_\text{r} = 6.5$ (b), $\varepsilon_\text{r} = 7$ (c) and $\varepsilon_\text{r} = 7.5$ (d).} 
        \label{fig:PSNR_Boxplots_eps_BowtieCrossDiel}   
    \end{figure}

    \begin{figure}[!t]
        \makebox[\columnwidth][c]{%
            \begin{minipage}{\columnwidth}
                \hspace*{\fill}%
                \subfloat[]{\includegraphics[height=0.235\linewidth]{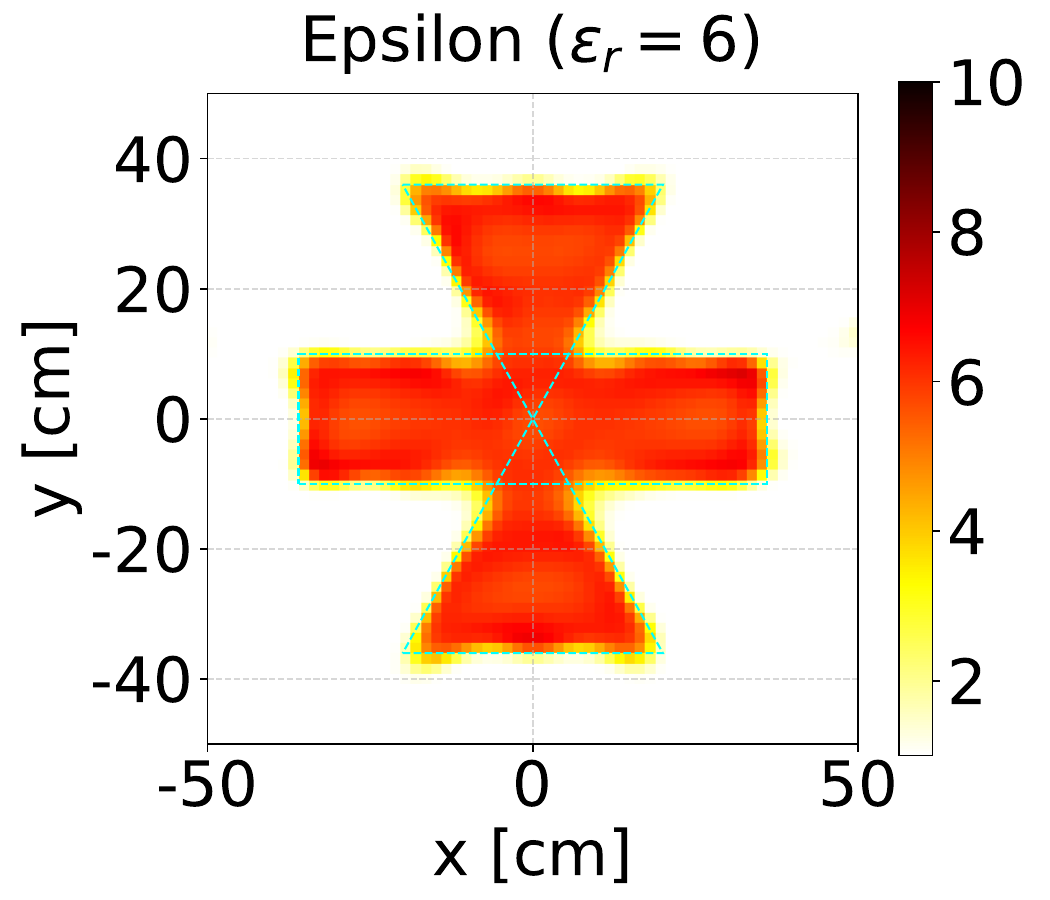}}\hfill
                \subfloat[]{\includegraphics[height=0.235\linewidth]{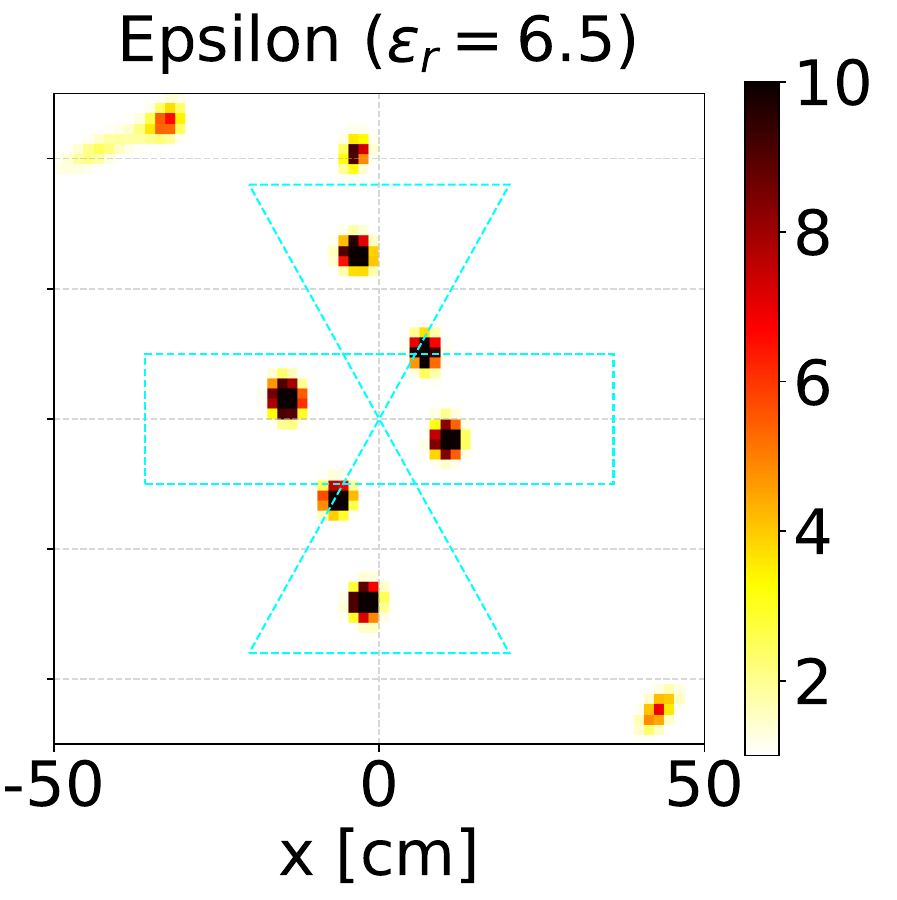}}\hfill
                \subfloat[]{\includegraphics[height=0.235\linewidth]{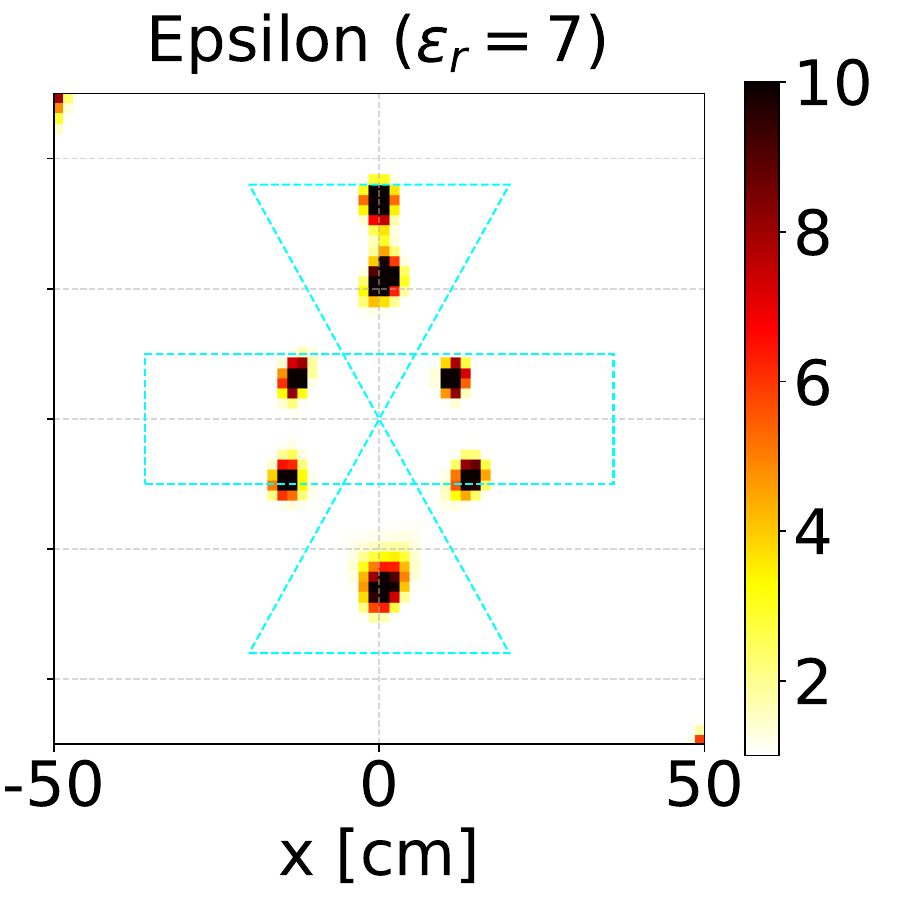}}\hfill
                \subfloat[]{\includegraphics[height=0.235\linewidth]{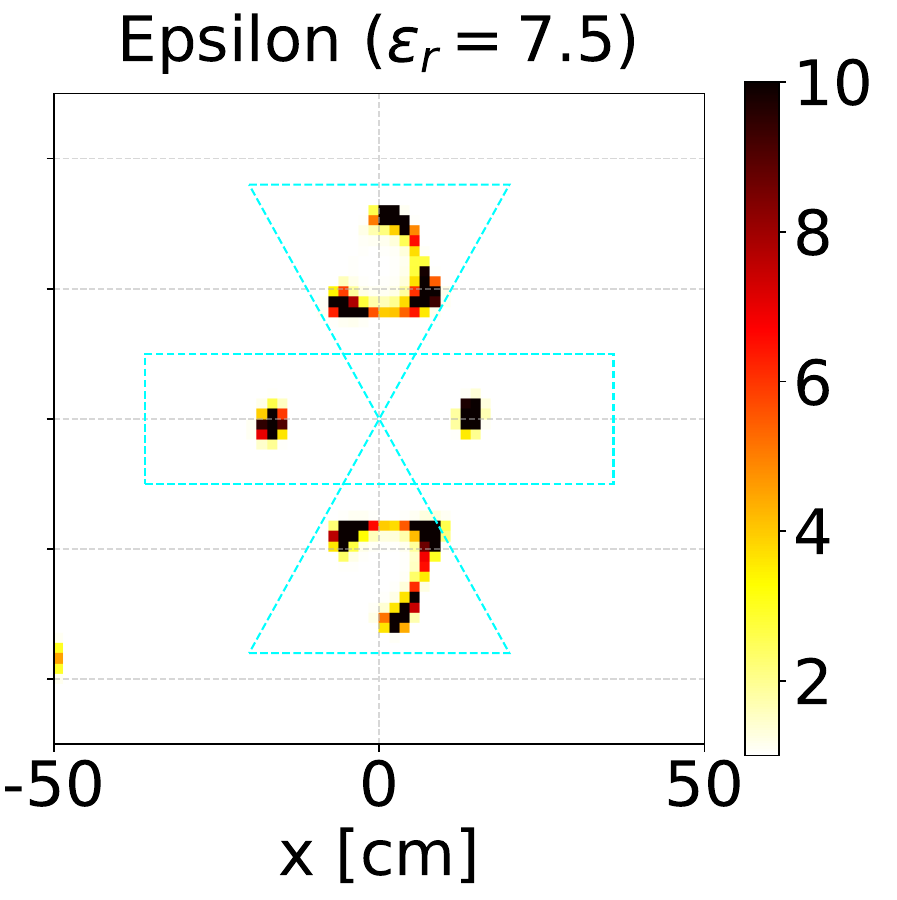}}\hspace*{\fill}

                \vspace{-1.5ex} 

                \hspace*{\fill}%
                \subfloat[]{\includegraphics[height=0.235\linewidth]{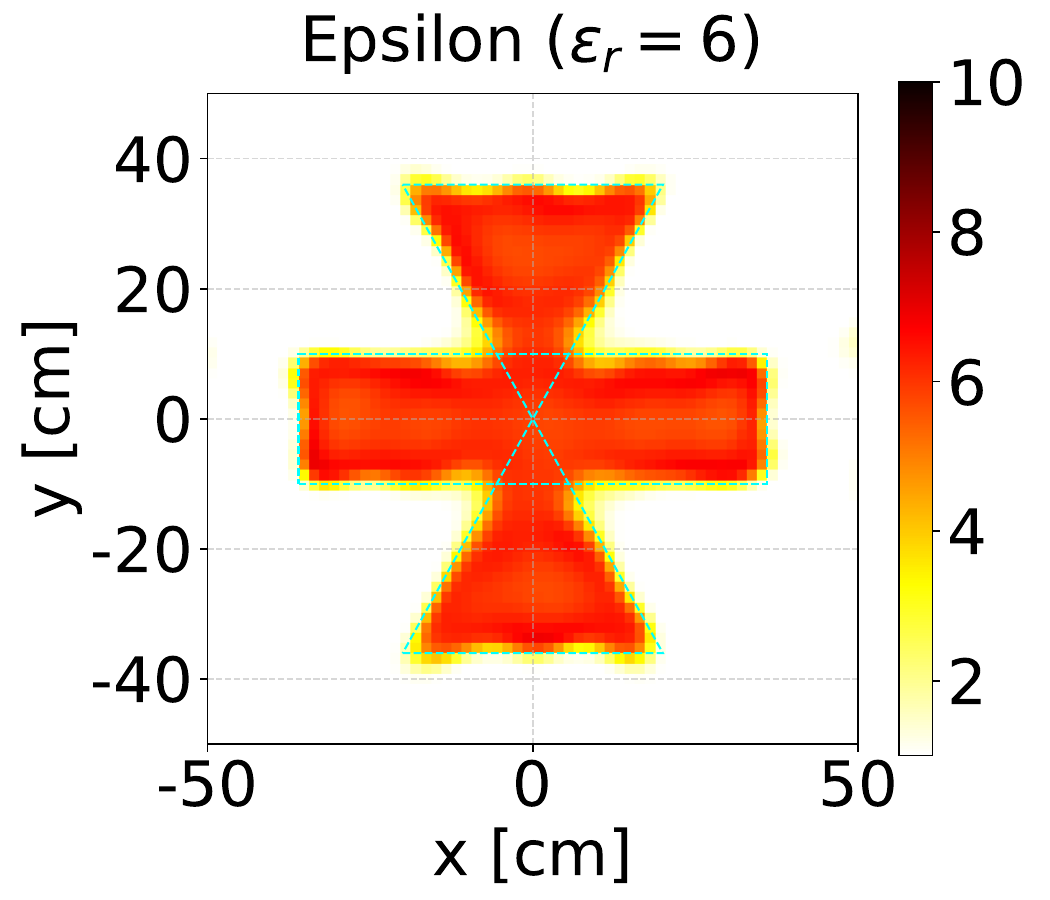}}\hfill
                \subfloat[]{\includegraphics[height=0.235\linewidth]{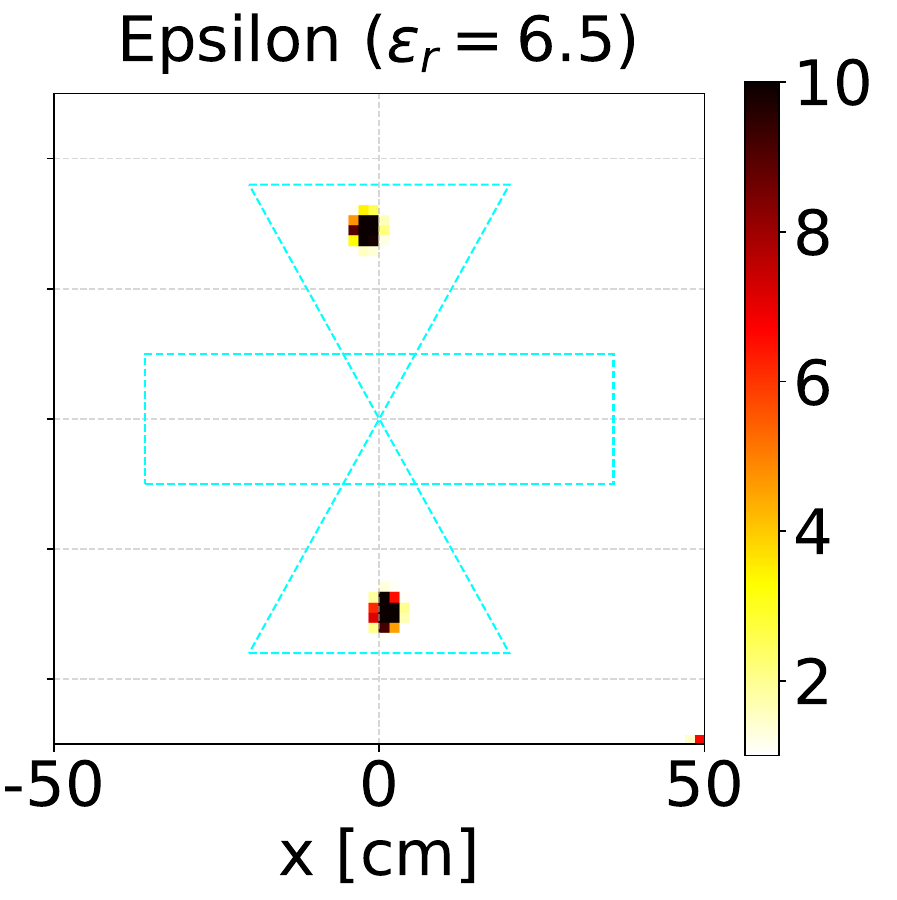}}\hfill
                \subfloat[]{\includegraphics[height=0.235\linewidth]{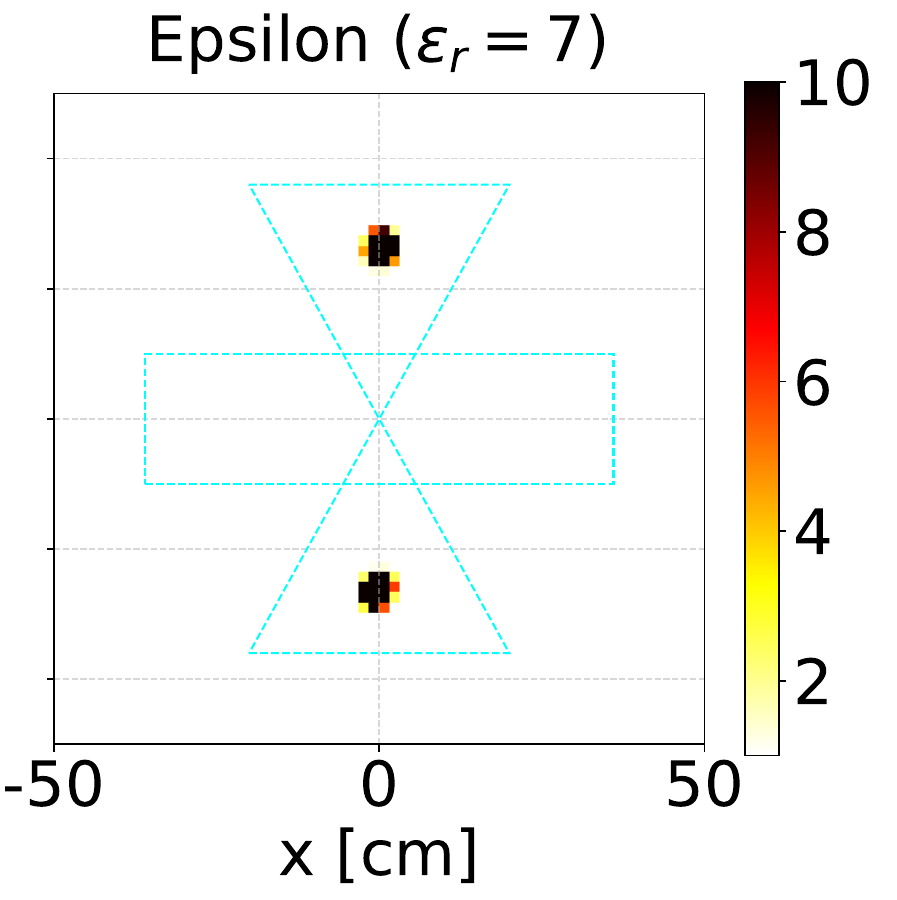}}\hfill
                \subfloat[]{\includegraphics[height=0.235\linewidth]{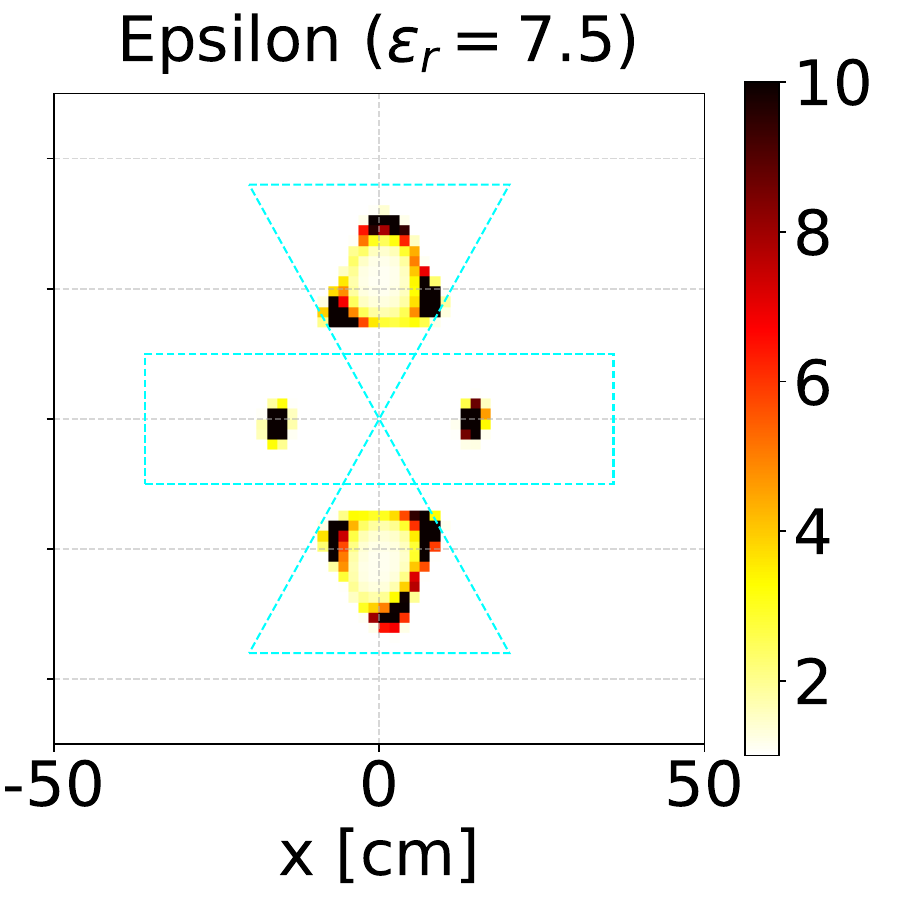}}\hspace*{\fill}

                \vspace{-1.5ex} 

                \hspace*{\fill}
                \subfloat[]{\includegraphics[height=0.235\linewidth]{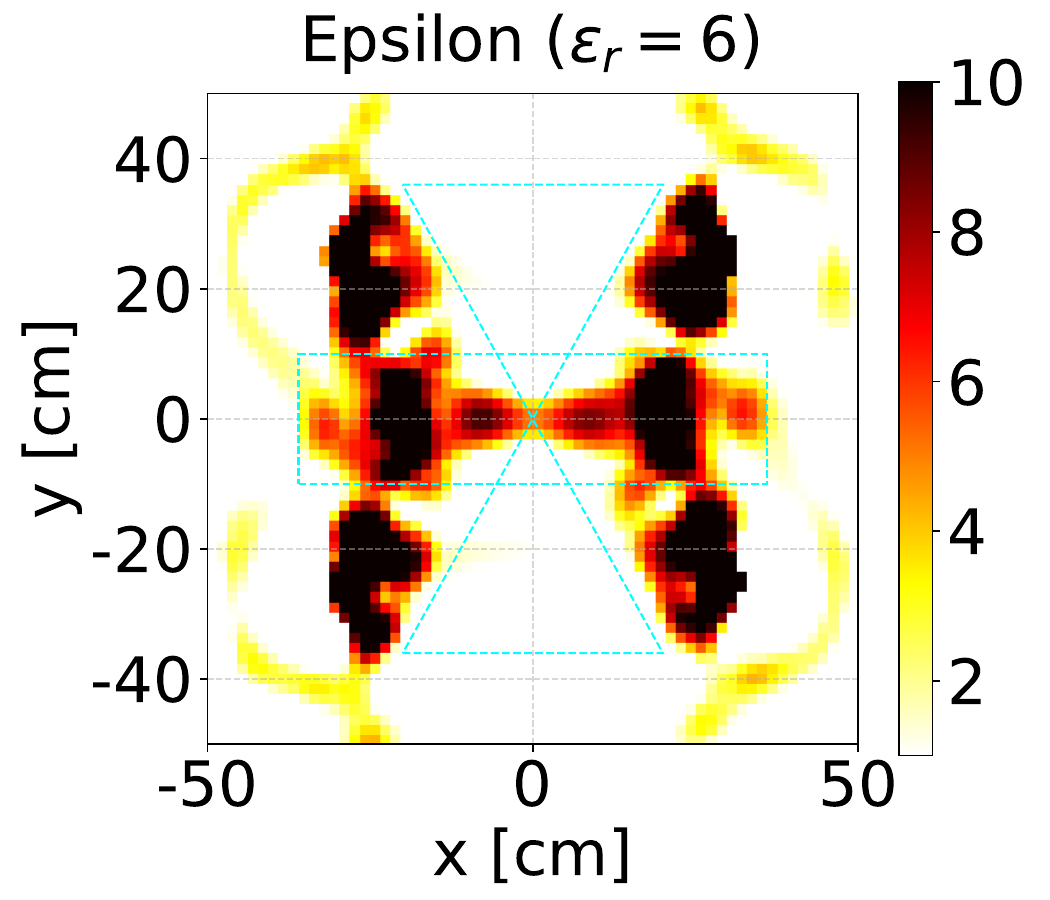}}\hfill
                \subfloat[]{\includegraphics[height=0.235\linewidth]{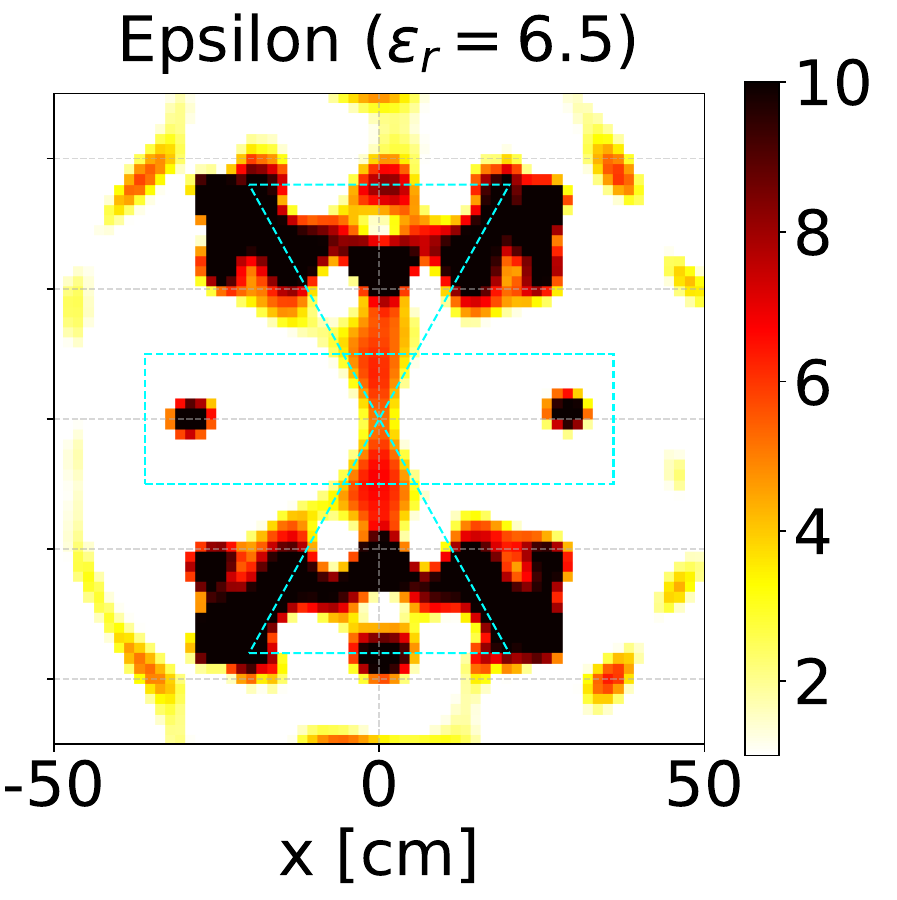}}\hfill
                \subfloat[]{\includegraphics[height=0.235\linewidth]{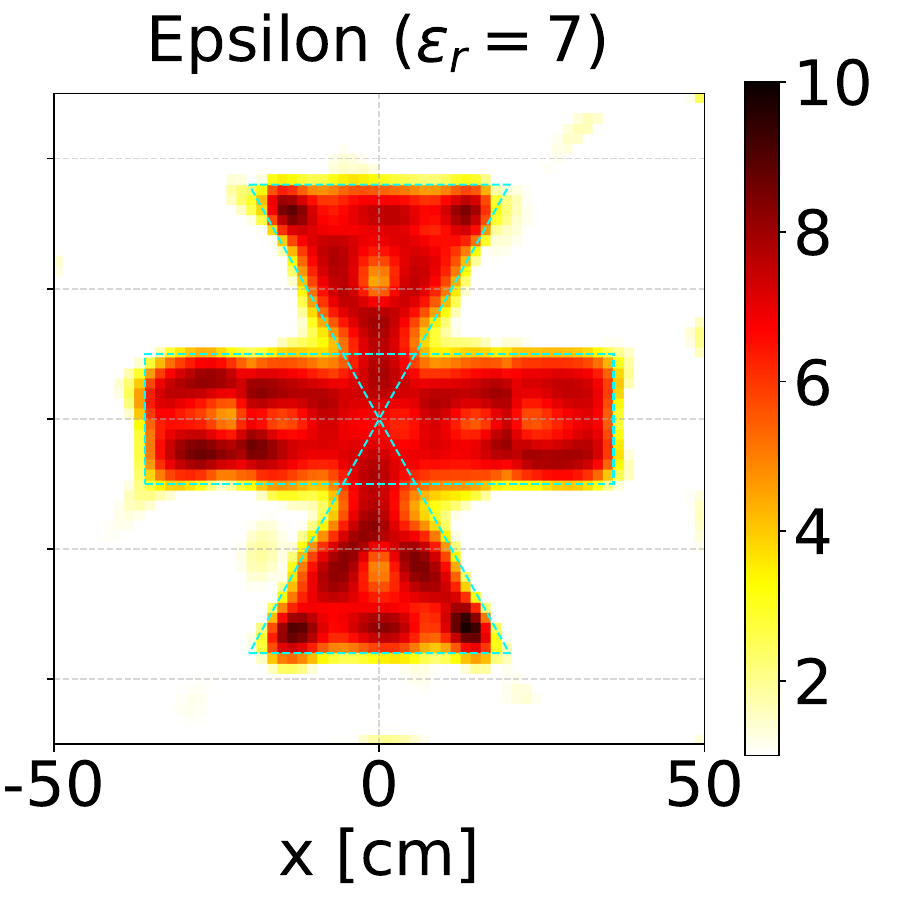}}\hfill
                \subfloat[]{\includegraphics[height=0.235\linewidth]{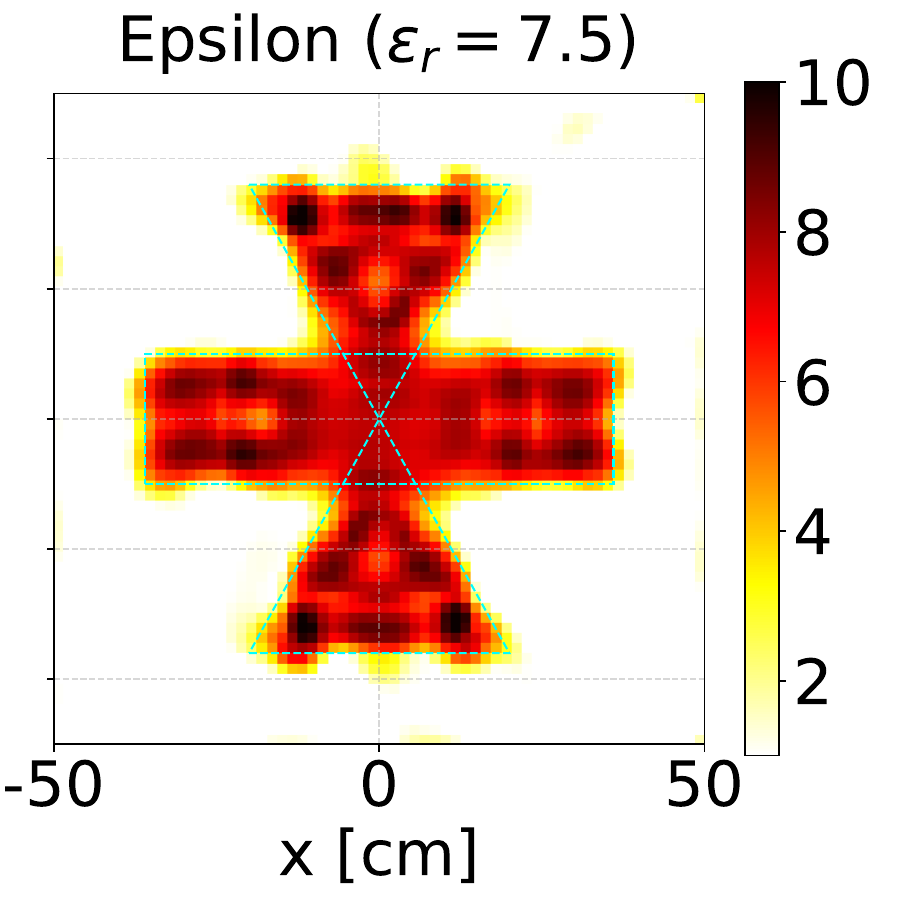}}\hspace*{\fill}

                \vspace{-1.5ex}

                \hspace*{\fill}
                \subfloat[]{\includegraphics[height=0.235\linewidth]{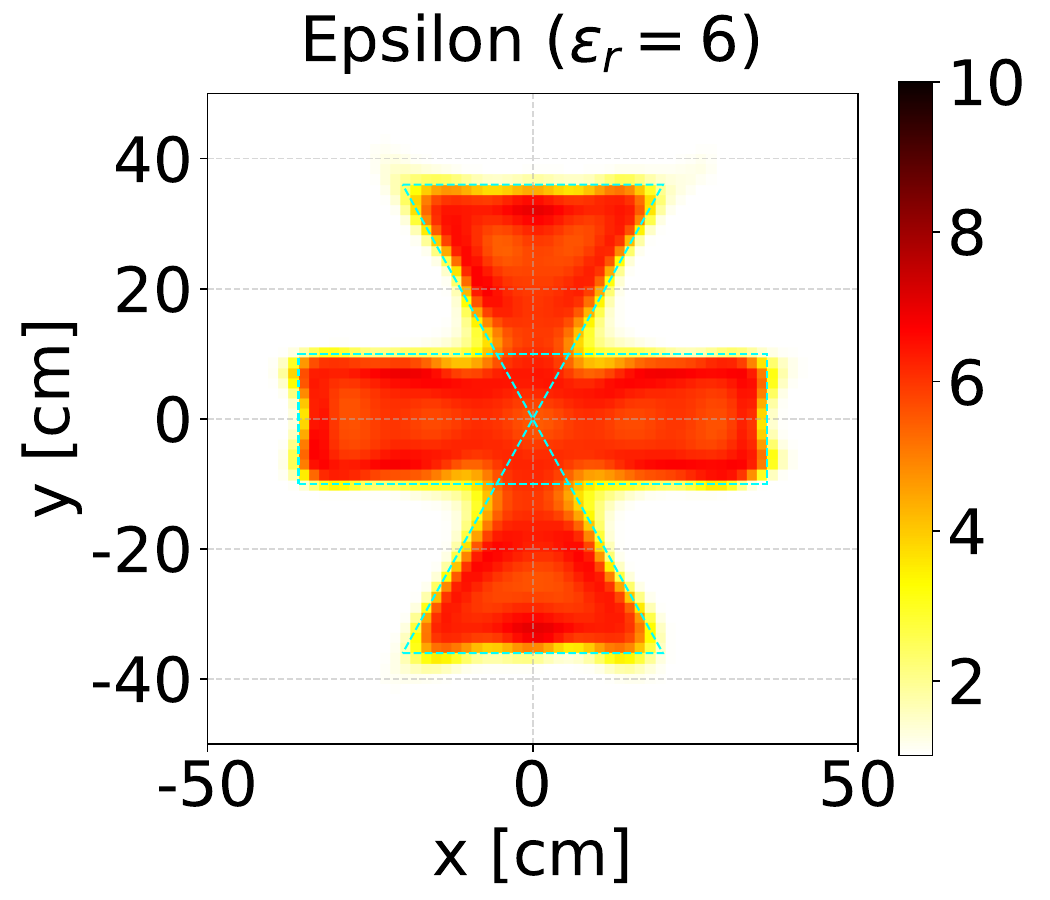}}\hfill
                \subfloat[]{\includegraphics[height=0.235\linewidth]{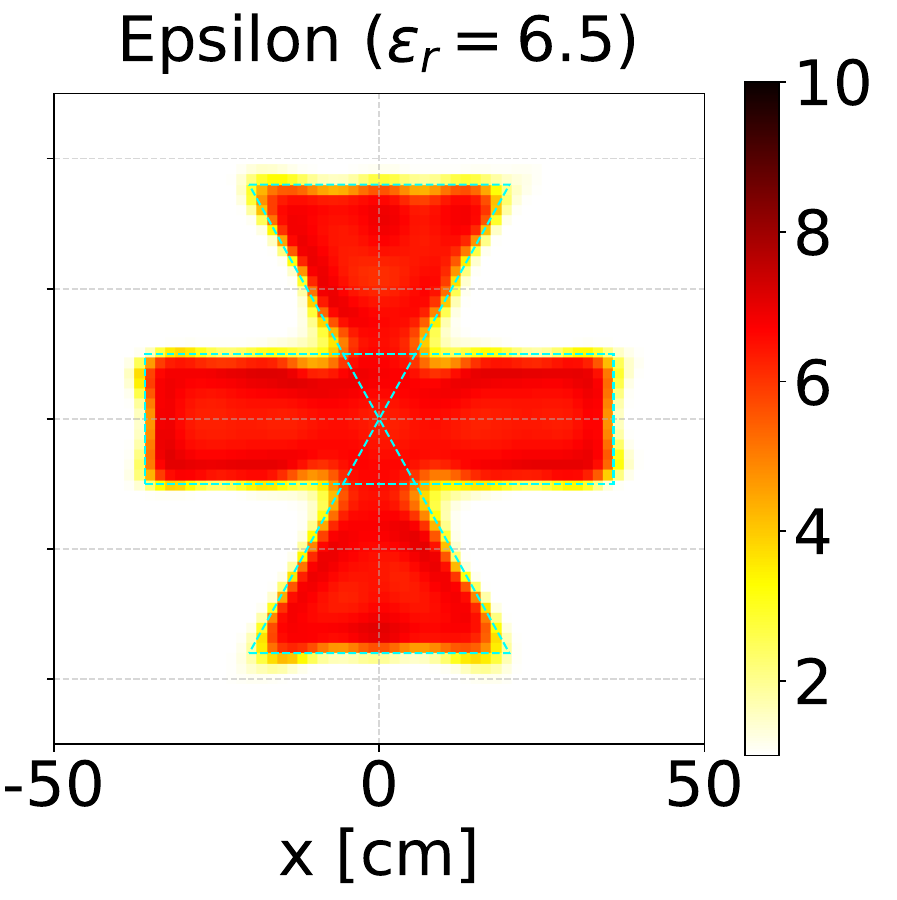}}\hfill
                \subfloat[]{\includegraphics[height=0.235\linewidth]{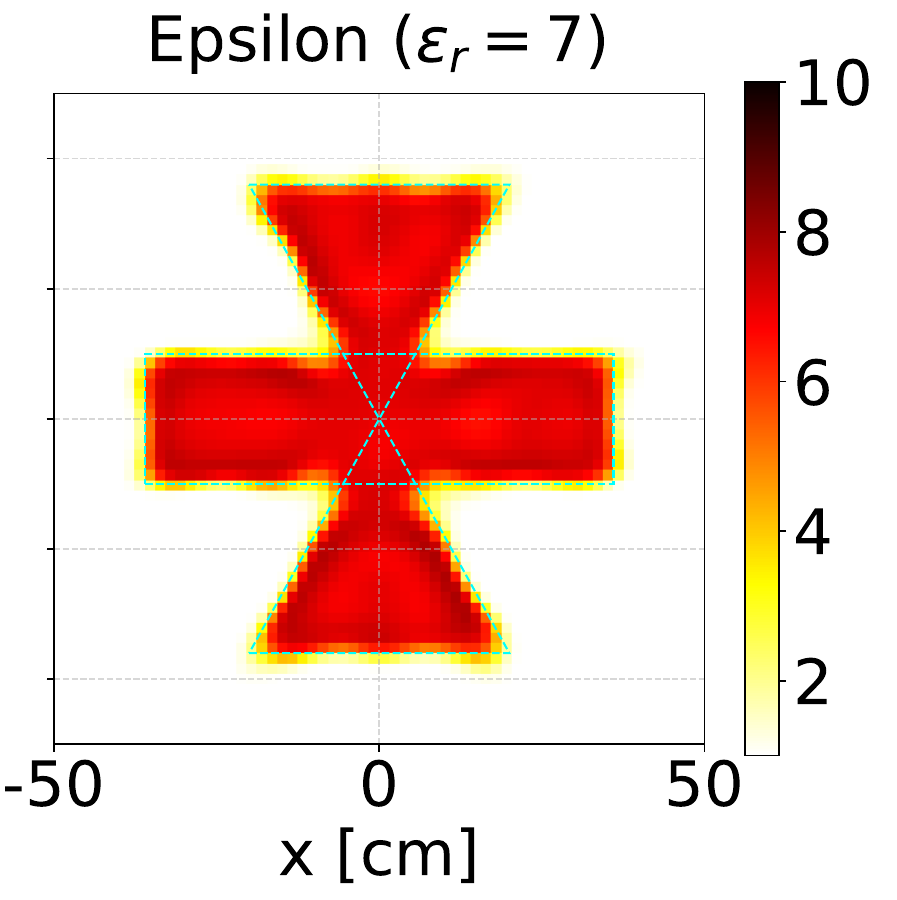}}\hfill
                \subfloat[]{\includegraphics[height=0.235\linewidth]{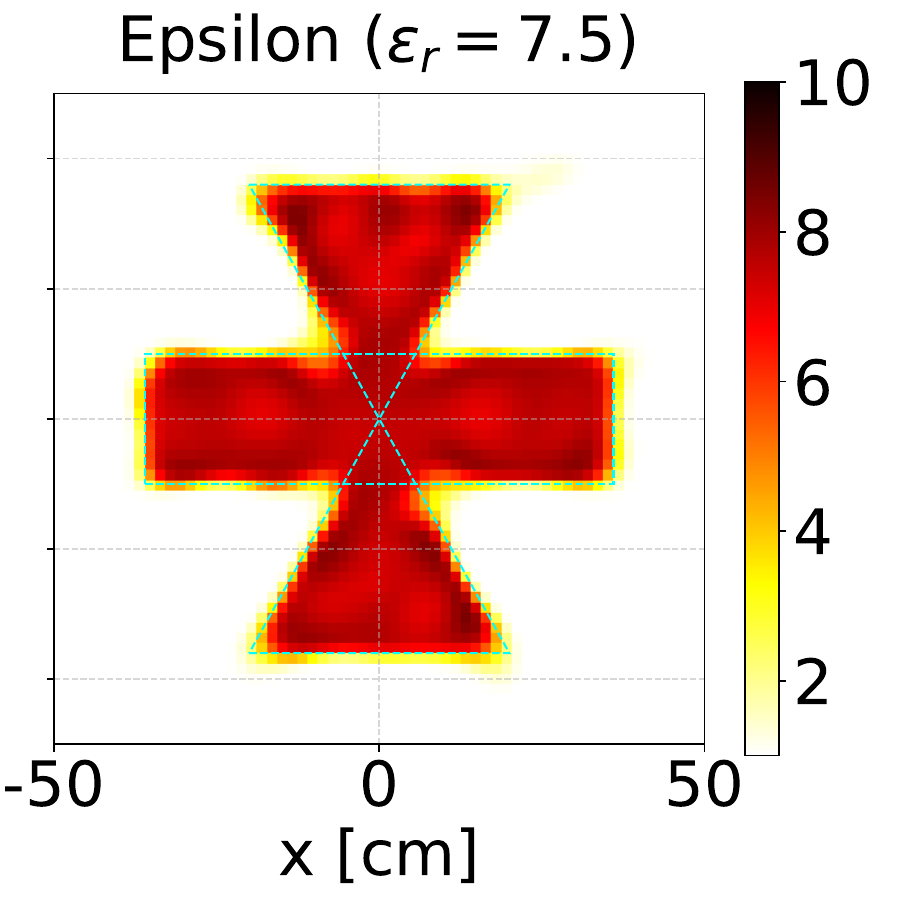}}\hspace*{\fill}

                \vspace{-1.5ex}

                \hspace*{\fill}
                \subfloat[]{\includegraphics[height=0.235\linewidth]{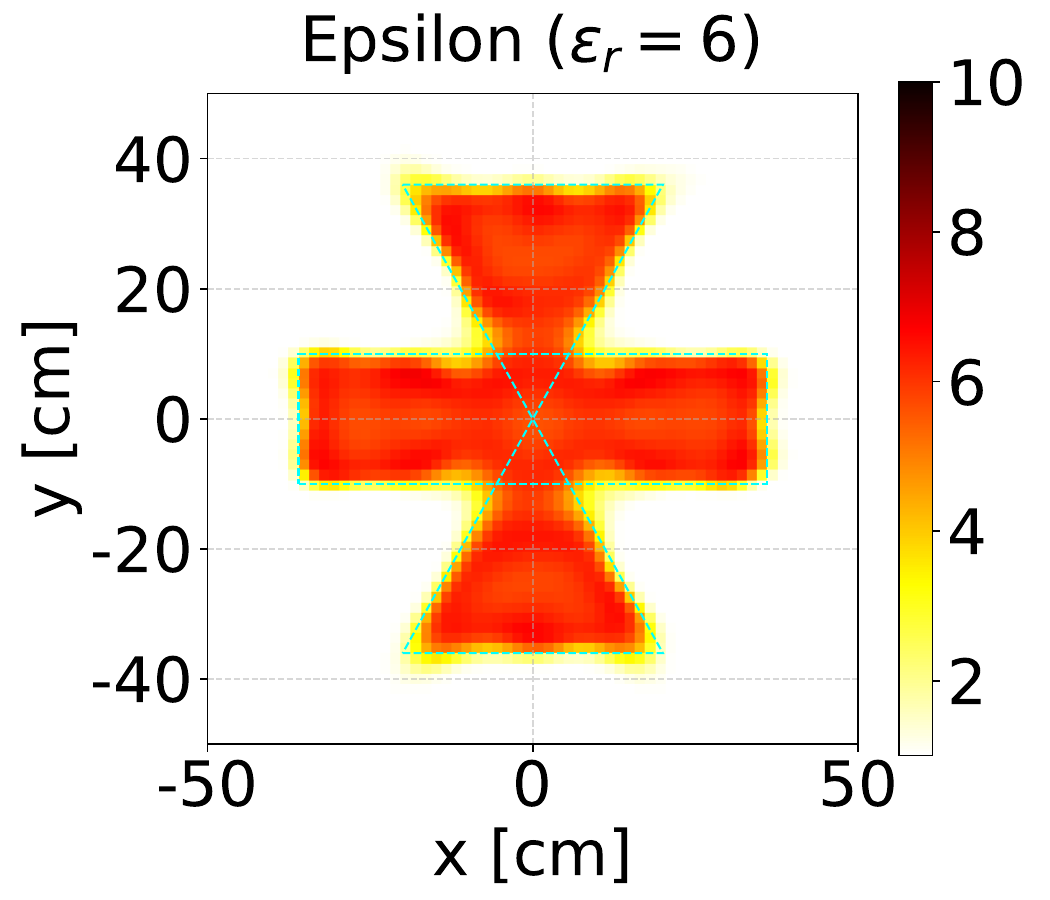}}\hfill
                \subfloat[]{\includegraphics[height=0.235\linewidth]{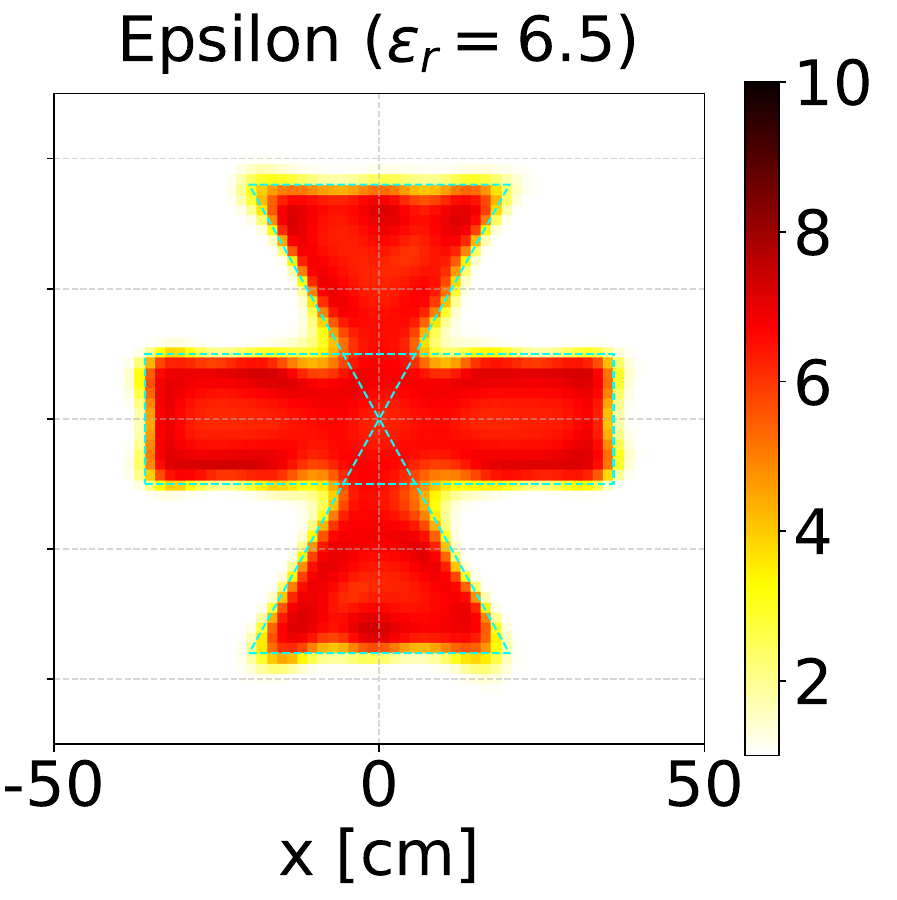}}\hfill
                \subfloat[]{\includegraphics[height=0.235\linewidth]{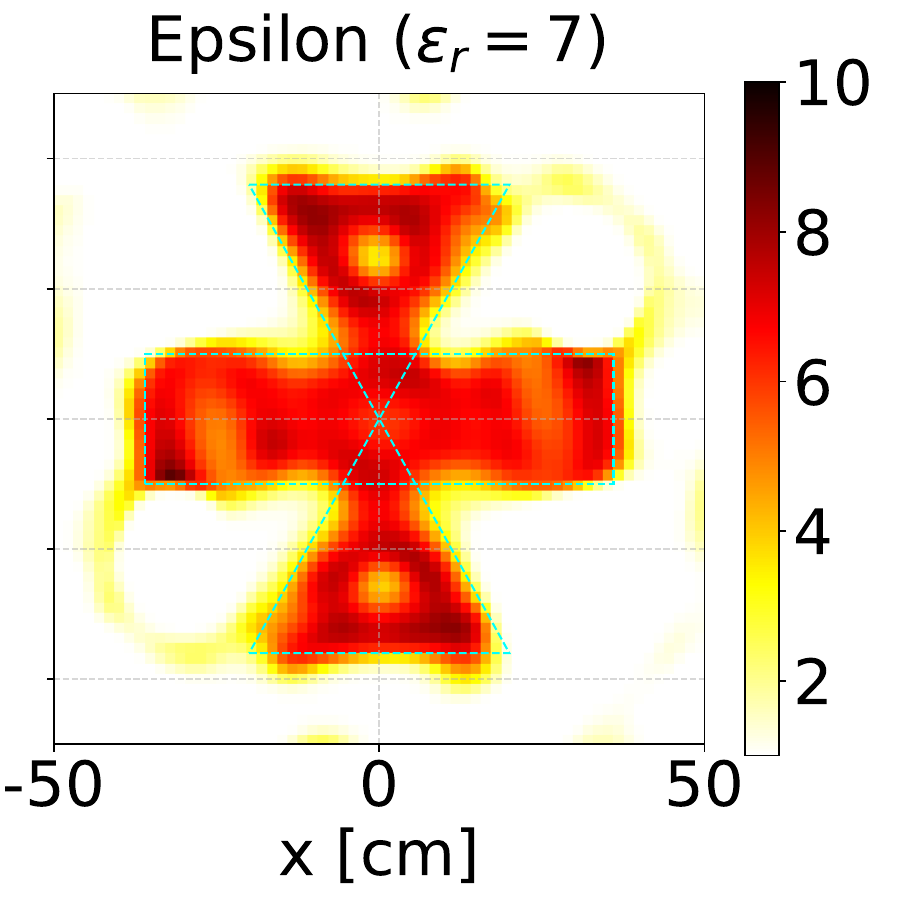}}\hfill
                \subfloat[]{\includegraphics[height=0.235\linewidth]{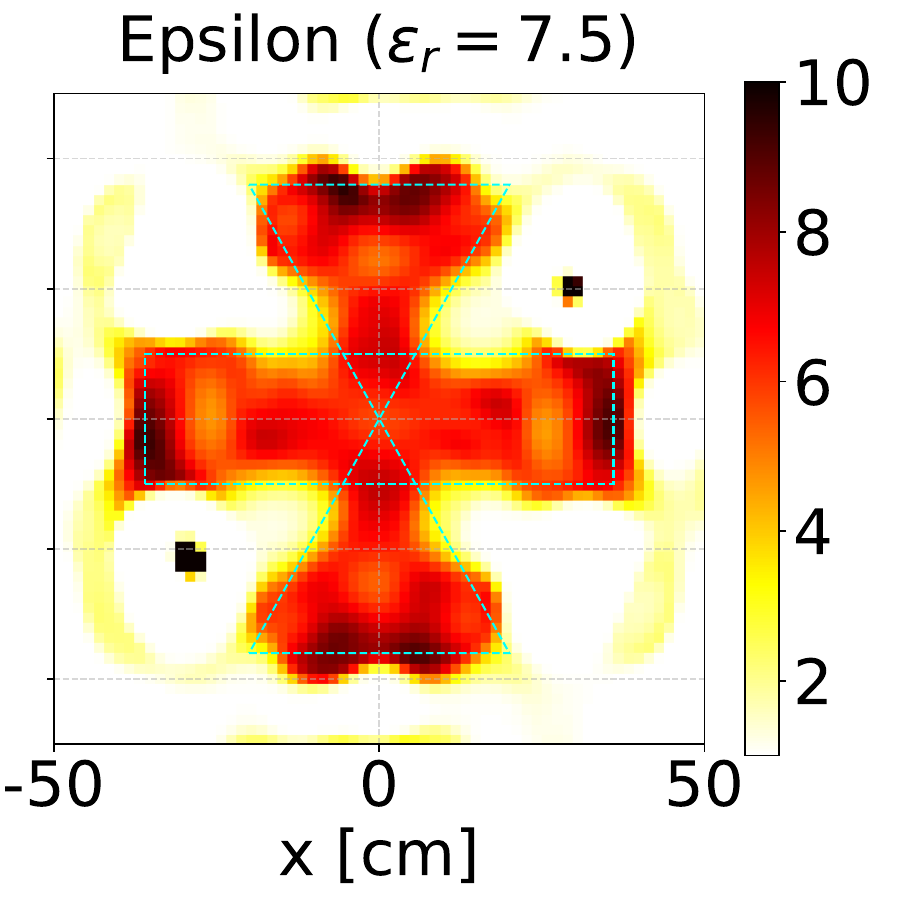}}\hspace*{\fill}
            \end{minipage}%
        }
        \caption{Final reconstructed images for CC-PINN, CC-PINN (Alt), CC-CSI, Alt-CC-INR and Alt-INR using the frequency-hopping strategy to invert ``Bowtie-Cross'' dielectric targets of different contrasts. From top to bottom: CC-PINN (a, b, c, d), CC-PINN (Alt) (e, f, g, h), CC-CSI (i, j, k, l), Alt-CC-INR (m, n, o, p) and Alt-INR (q, r, s, t). From left to right: $\varepsilon_\text{r} = 6$ (a, e, i, m, q), $\varepsilon_\text{r} = 6.5$ (b, f, j, n, r), $\varepsilon_\text{r} = 7$ (c, g, k, o, s), and $\varepsilon_\text{r} = 7.5$ (d, h, l, p, t).}
        \label{fig:eps_BowtieCrossDiel}   
    \end{figure}

    For the ``Austria'' target (where the PSNR convergence curves, final PSNR boxplots, and reconstructed images are explicitly detailed in Fig.~\ref{fig:PSNR_eps_AustriaDiel}, Fig.~\ref{fig:PSNR_Boxplots_eps_AustriaDiel}, and Fig.~\ref{fig:eps_AustriaDiel}, respectively), as the relative permittivity $\varepsilon_\text{r}$ increases from 6 to 9, the non-linearity intensifies dramatically. Alt-INR exhibits extreme instability and unpredictability: it completely fails at $\varepsilon_\text{r} = 6$ [Fig.~\ref{fig:eps_AustriaDiel}(q)], surprisingly converges at $\varepsilon_\text{r} = 7$ [Fig. \ref{fig:eps_AustriaDiel}(r)], and then collapses again at $\varepsilon_\text{r} \ge 8$. This highlights the severe fragility of optimizing without the cross-correlated constraint. CC-PINN and CC-PINN (Alt) converge well at $\varepsilon_\text{r} = 6$, but when $\varepsilon_\text{r} \ge 7$, the inefficiency of purely empirical step-size updates in navigating the rugged loss landscape leads to an early breakdown (e.g., dropping to $\sim$7--8~dB). This verifies that alternating decoupling alone is insufficient. In contrast, Alt-CC-INR maintains remarkably robust convergence across all contrasts, securing high PSNRs ($\sim$18--21~dB) and high-fidelity reconstructions. CC-CSI survives up to $\varepsilon_\text{r} = 8$ but yields significantly lower accuracy than Alt-CC-INR and suffers from severe artifacts at extreme contrasts (e.g., $\varepsilon_\text{r} = 9$).

    For the ``Bowtie-Cross'' target (Fig. \ref{fig:PSNR_eps_BowtieCrossDiel} to Fig. \ref{fig:eps_BowtieCrossDiel}), tested at $\varepsilon_\text{r} = 6$, $6.5$, $7$, and $7.5$, the performance boundaries are further exposed. Both CC-PINN and CC-PINN (Alt) succeed only at $\varepsilon_\text{r} = 6$, and collapse early as the contrast increases ($\varepsilon_\text{r} \ge 6.5$). Alt-INR performs well at lower contrasts ($\varepsilon_\text{r} \le 6.5$) but suffers from severe performance degradation and breakdown at higher contrasts ($\varepsilon_\text{r} \ge 7$, with median PSNRs plunging below 7~dB). Conversely, Alt-CC-INR robustly reconstructs the target across all contrast levels without exception (securing high median PSNRs over 20~dB). CC-CSI exhibits a strange shape-dependent fragility: it fails completely at $\varepsilon_\text{r} = 6$ and $6.5$, but interestingly converges at $\varepsilon_\text{r} = 7$ and $7.5$. This demonstrates that the alternating optimization effectively flattens the rugged non-convex landscape, fully unleashing the potential of the cross-correlated physical constraint while overcoming the optimization bottleneck of purely neural solvers.

\subsection{Validation with Measured Data}

    \begin{figure*}[!t] 
        \centering 
        
        \subfloat[]{\includegraphics[width=0.18\textwidth]{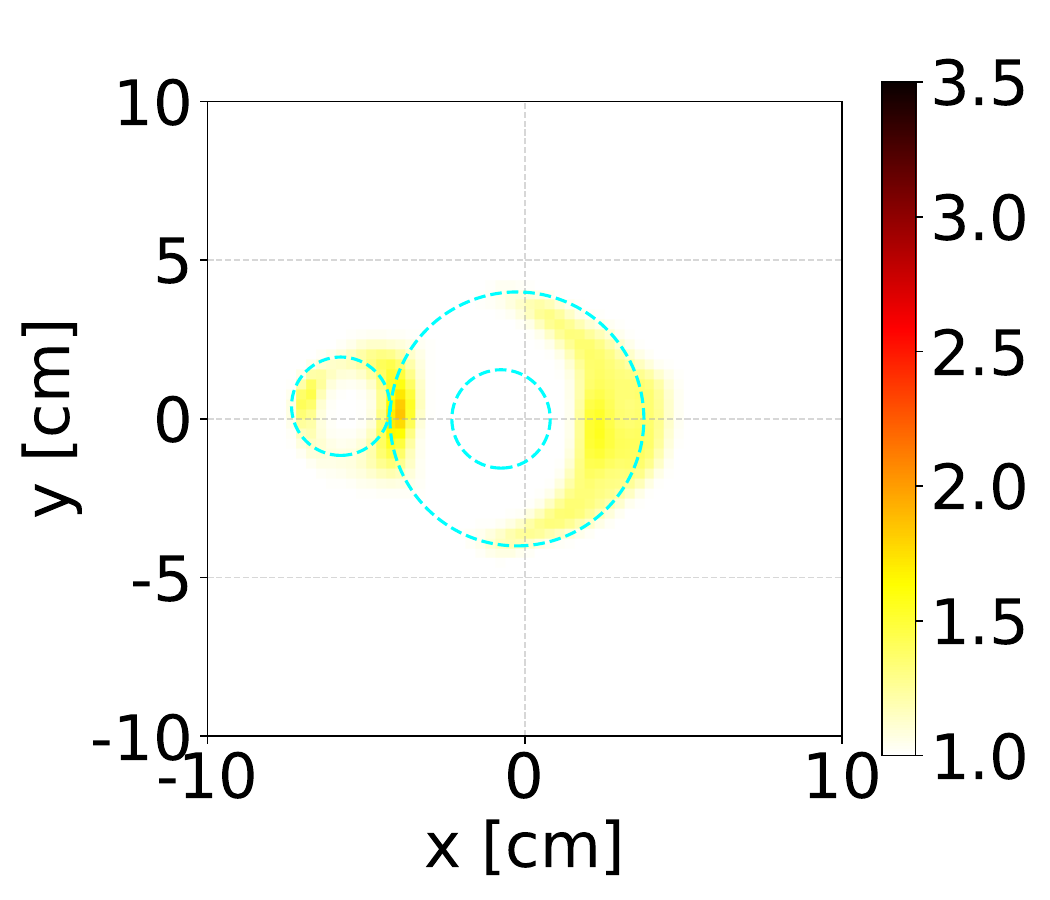}}\hfill
        \subfloat[]{\includegraphics[width=0.18\textwidth]{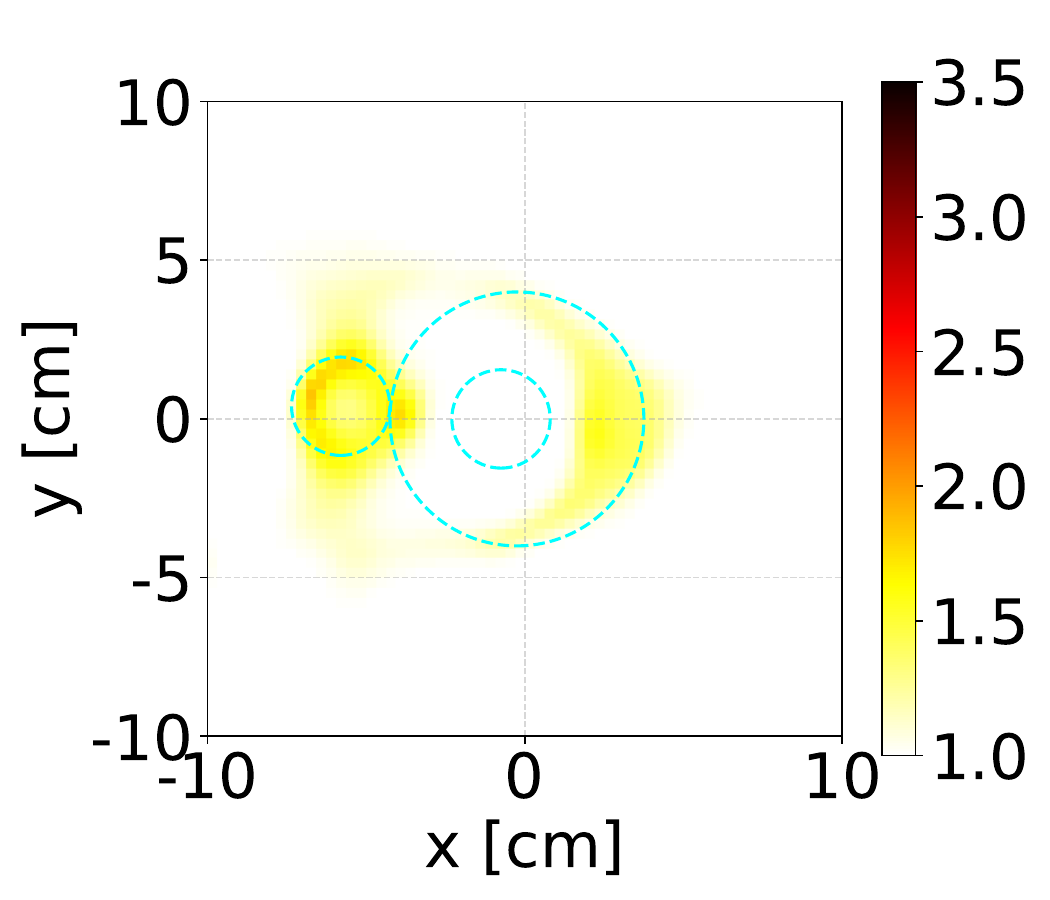}}\hfill
        \subfloat[]{\includegraphics[width=0.18\textwidth]{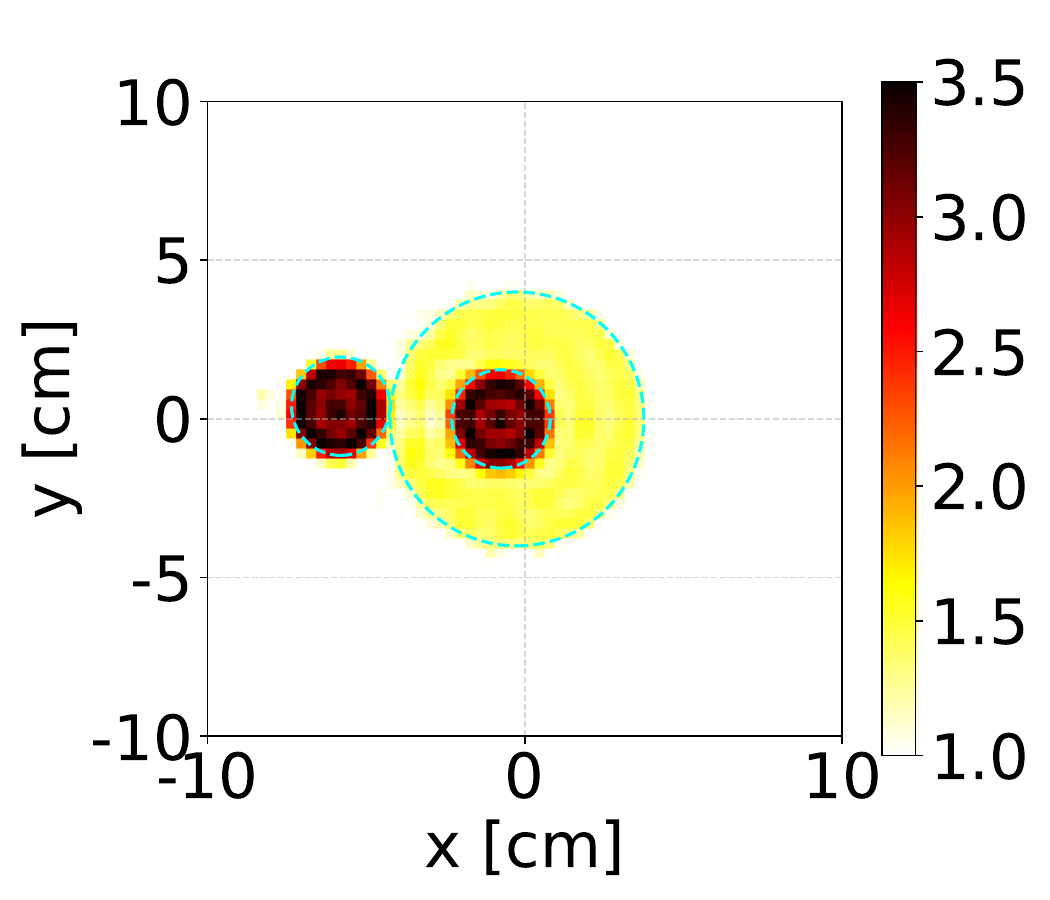}}\hfill
        \subfloat[]{\includegraphics[width=0.18\textwidth]{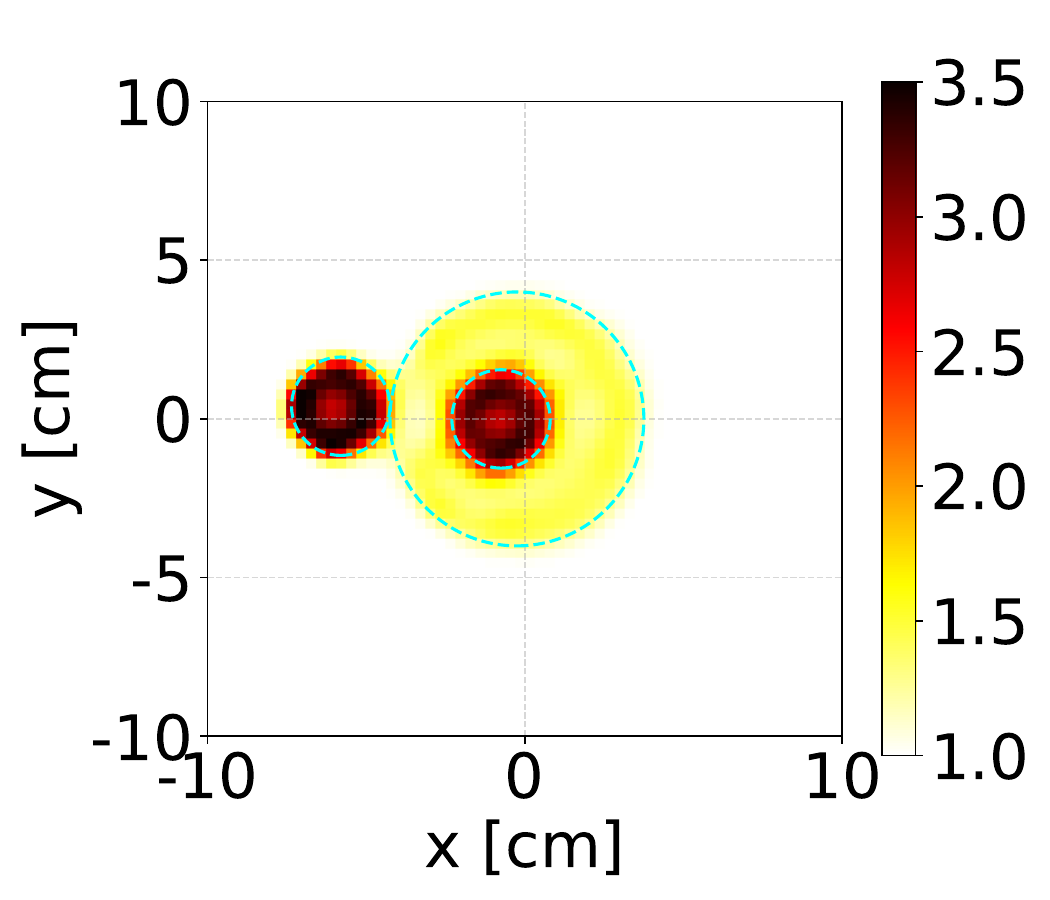}}\hfill
        \subfloat[]{\includegraphics[width=0.18\textwidth]{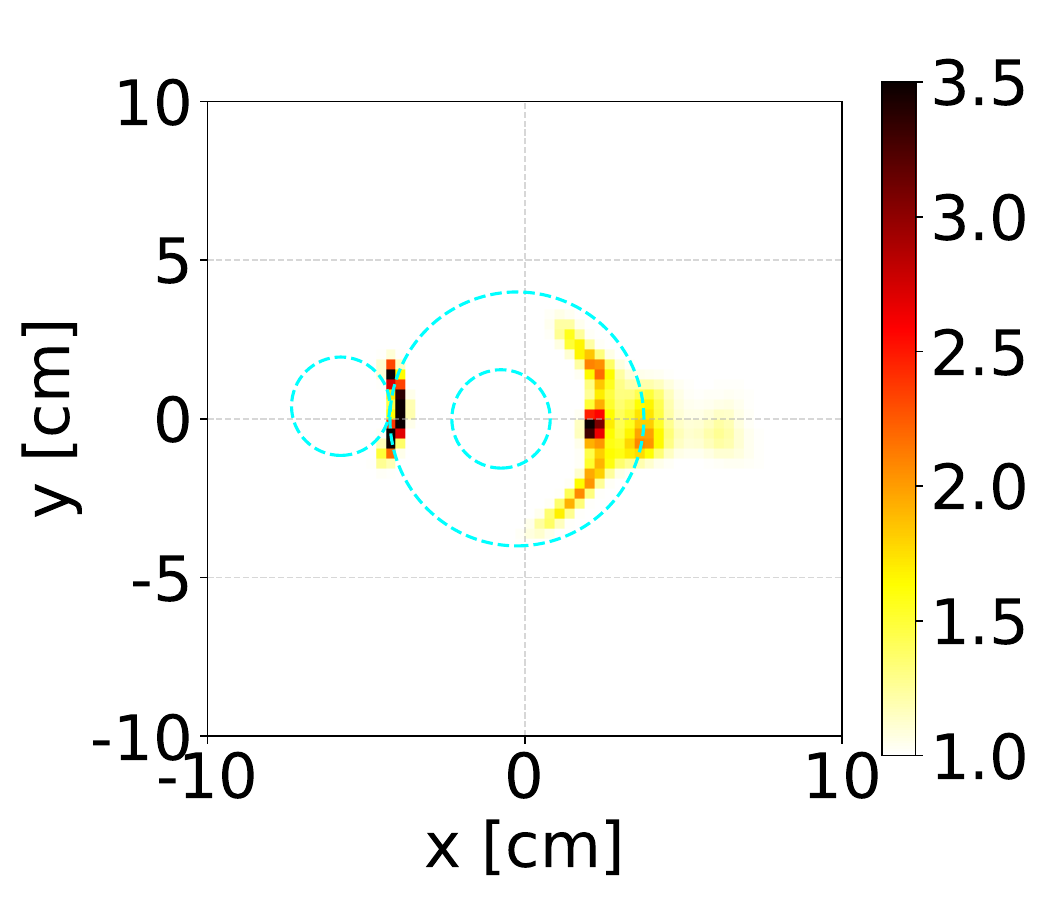}}

        \vspace{-2.0ex} 

        \subfloat[]{\includegraphics[width=0.18\textwidth]{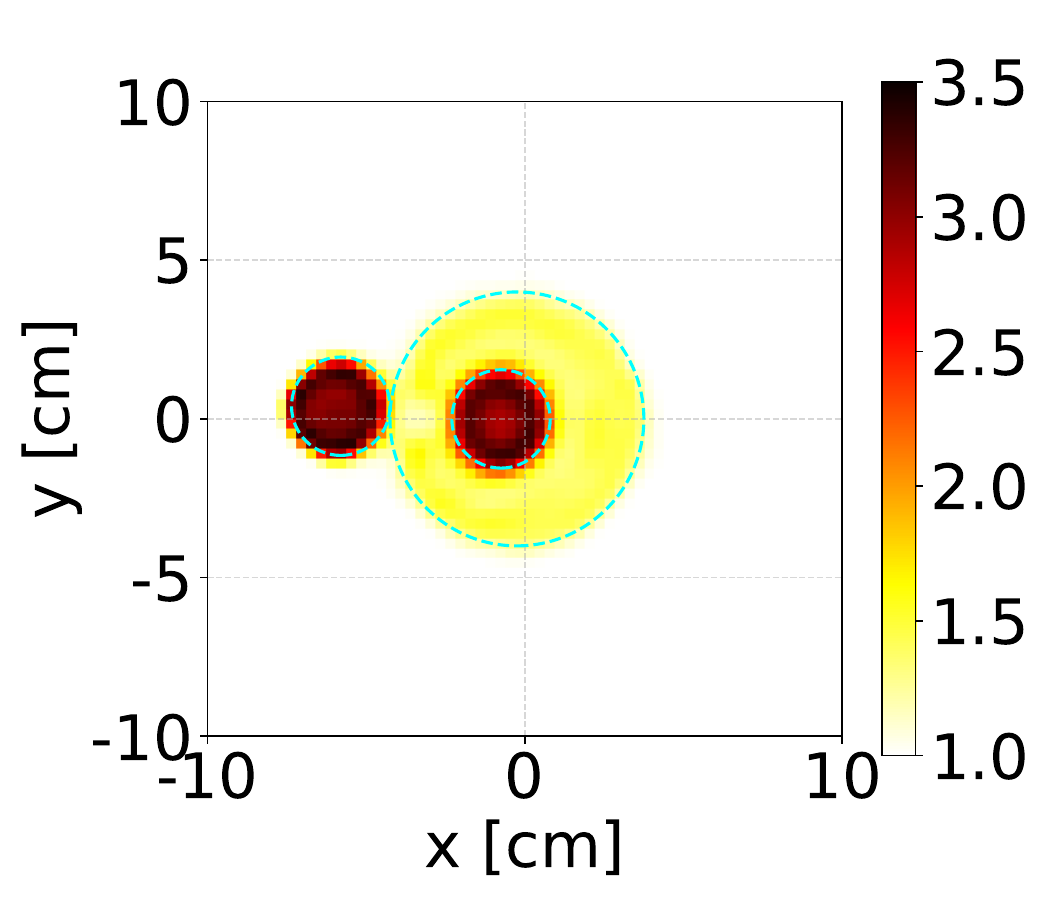}}\hfill
        \subfloat[]{\includegraphics[width=0.18\textwidth]{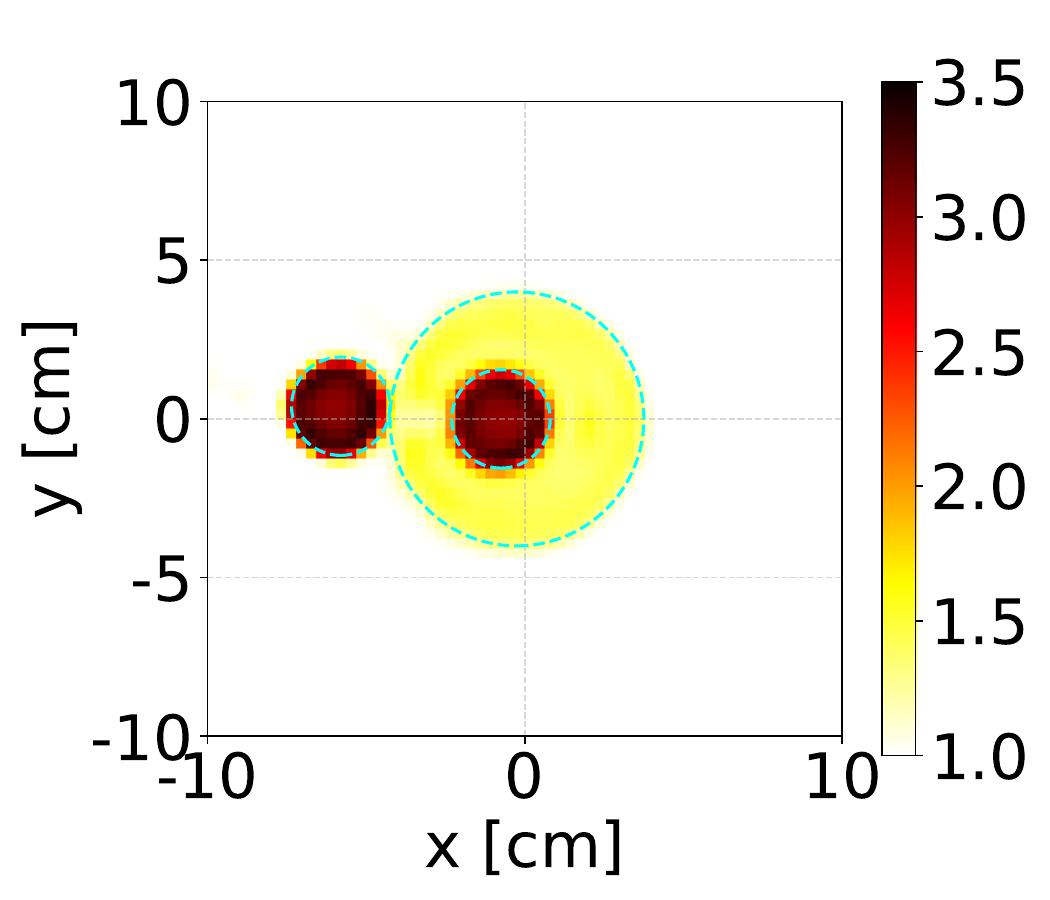}}\hfill
        \subfloat[]{\includegraphics[width=0.18\textwidth]{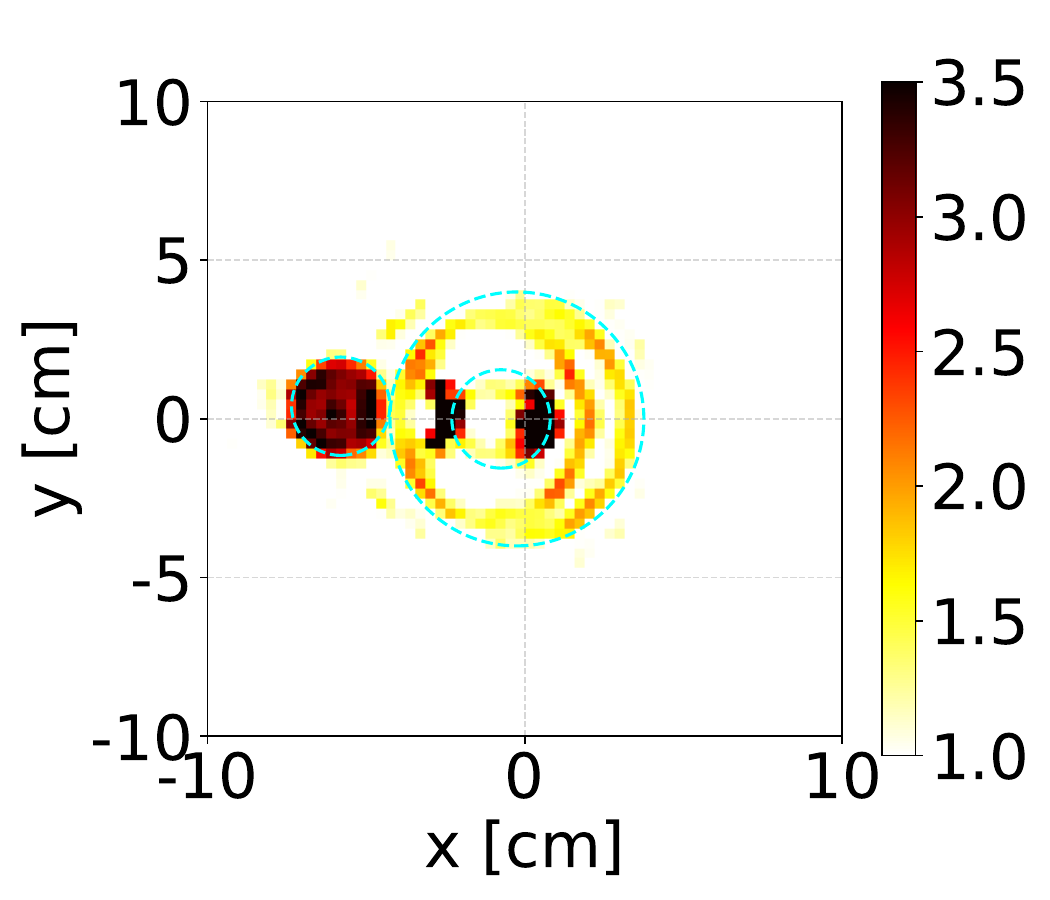}}\hfill
        \subfloat[]{\includegraphics[width=0.18\textwidth]{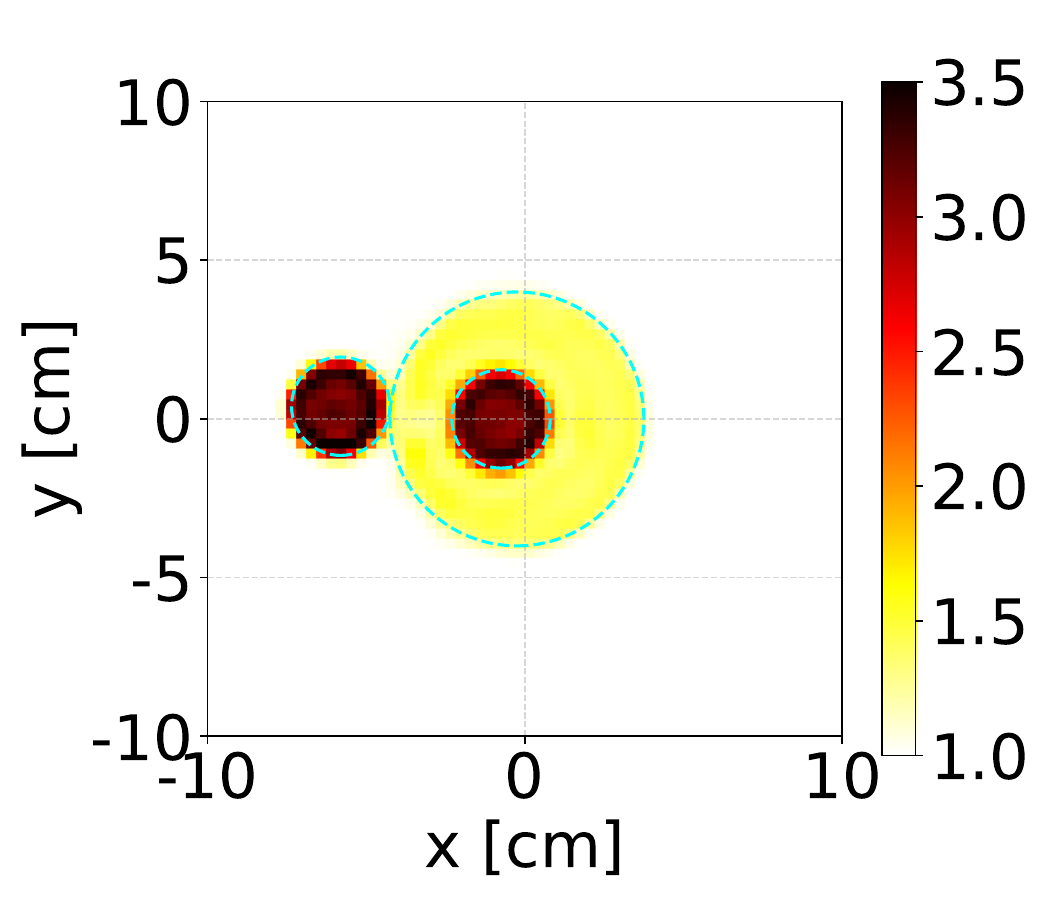}}\hfill
        \subfloat[]{\includegraphics[width=0.18\textwidth]{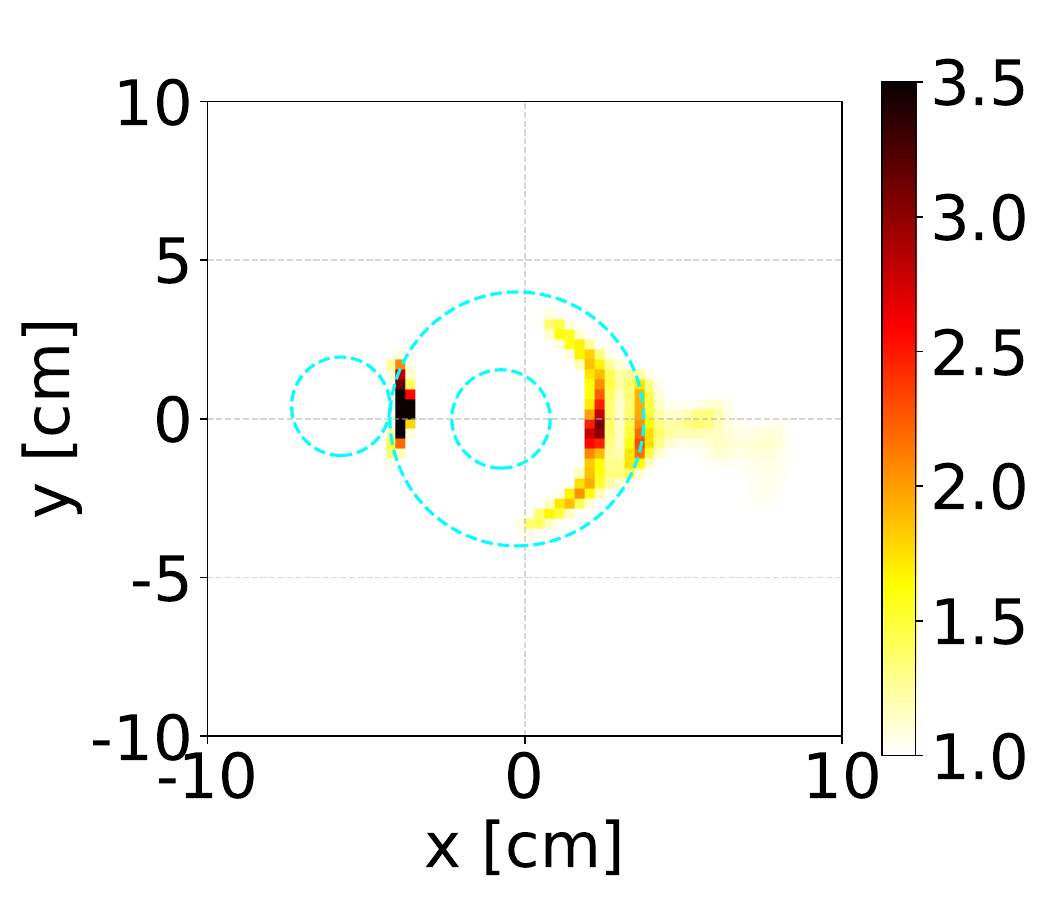}}
        
        \caption{Final reconstructed images for CC-PINN (a, f), CC-PINN (Alt) (b, g), CC-CSI (c, h), Alt-CC-INR (d, i) and Alt-INR (e, j) to invert the Fresnel dataset \textit{FoamTwinDielTM\_8910}. Top: Frequency-hopping strategy; Bottom: Simultaneous multi-frequency strategy.}
        \label{fig:eps_FoamTwinDielTM}   
    \end{figure*}

    To further validate the generalization ability in authentic physical environments, the publicly available Fresnel dataset is utilized \cite{geffrin2005free}. The original measurement system deployed 18 equally spaced transmitting antennas on an observation circle with a radius of 1.67~m, and complex scattered fields were received by 241 probes. To decrease data redundancy and accelerate computation, the spatial sampling interval on the receiving end was downsampled to $5^\circ$, yielding a compact measurement matrix of size $18 (\text{Tx}) \times 49 (\text{Rx})$. The physical domain of investigation (DOI) for inversion is defined as $[-0.1, 0.1]~\text{m} \times [-0.1, 0.1]~\text{m}$, partitioned into a discrete $64 \times 64$ grid. I select the highly challenging high-frequency subset \textit{FoamTwinDielTM\_8910} (8, 9, and 10~GHz), which contains one low-contrast foam cylinder and two high-contrast small dielectric cylinders.

    Fig. \ref{fig:eps_FoamTwinDielTM} presents the reconstructed images. Under the frequency-hopping strategy (Fig.~\ref{fig:eps_FoamTwinDielTM}, Top), unmodeled systematic errors at low frequencies easily trap conventional joint networks into pseudo-solutions. Consequently, both CC-PINN and CC-PINN (Alt) [Fig.~\ref{fig:eps_FoamTwinDielTM}(a,b)] suffer from severe artifacts. CC-CSI [Fig.~\ref{fig:eps_FoamTwinDielTM}(c)] and Alt-CC-INR [Fig.~\ref{fig:eps_FoamTwinDielTM}(d)] successfully track the target profile with a clean background. When employing the simultaneous multi-frequency strategy [Fig.~\ref{fig:eps_FoamTwinDielTM}, Bottom], the joint processing acts as an ensemble regularizer against single-frequency noise. Here, the performance of both CC-PINN and CC-PINN (Alt) improves [Fig.~\ref{fig:eps_FoamTwinDielTM}(f,g)]. However, CC-CSI [Fig.~\ref{fig:eps_FoamTwinDielTM}(h)] suffers from severe performance degradation, producing strong artifacts that obscure the target boundaries. Throughout both strategies, Alt-CC-INR [Fig.~\ref{fig:eps_FoamTwinDielTM}(d, i)] consistently achieves the highest quantitative accuracy (e.g., accurately retrieving the relative permittivity $\varepsilon_\text{r} \approx 3.0$ for the two high-contrast cylinders) and perfectly separates the disconnected domains. Alt-INR consistently fails [Fig.~\ref{fig:eps_FoamTwinDielTM}(e, j)], reaffirming that standard data-state residuals are insufficient for robust measured data inversion.

\subsection{Computational Time Comparison}

    Evaluated on an NVIDIA A6000 GPU, the execution times for a fixed number of epochs (50,000 for simulations / 20,000 for Fresnel measured data) for CC-PINN, CC-PINN (Alt), CC-CSI, Alt-CC-INR, and Alt-INR are respectively: 264.18~s / 105.11~s, 378.80~s / 147.53~s, 334.80~s / 136.39~s, 347.58~s / 137.49~s, and 310.68~s / 121.16~s. Alt-CC-INR requires moderately more time per epoch than Alt-INR due to the cross-correlated residual computation. However, compared to the traditional pure iterative method CC-CSI, Alt-CC-INR requires comparable total computation time under the same number of iterations. More importantly, within this comparable time frame, Alt-CC-INR offers distinctly superior robustness against local minima trapping for complex high-contrast distributions. All code implementations utilize GPU-based batched acceleration. To ensure full transparency and reproducibility, the complete source codes and experimental data of this work have been made publicly available at: \url{https://github.com/TUDsun/Alt-CC-INR}.

\section{Conclusion} \label{sec:conclusion}
    
    This communication proposes Alt-CC-INR, an alternating optimization framework that overcomes the fundamental optimization bottleneck of purely empirical step-size neural solvers. By combining a 2D-FFT accelerated PR-CG analytical descent with neural network updates, the mechanism successfully decouples the internal physical field evolution from the contrast parameter inference. Extensive results on both synthetic and measured data confirm that Alt-CC-INR exhibits exceptional robustness and high-fidelity reconstruction capabilities for extreme high-contrast targets, outperforming both traditional CC-CSI and joint-optimization CC-PINN. While unsupervised INRs require on-the-fly optimization unlike instantaneous supervised networks, they completely bypass the simulation-to-reality gap, providing a highly reliable framework for practical applications.

    Furthermore, when applying this method to medical inverse imaging, a non-conductive matching medium is typically utilized to further alleviate the ill-posedness, thereby enabling the robust inversion of high-permittivity biological targets. Additionally, human tissues inherently possess noticeable conductivity; the resulting attenuation of the internal electric fields substantially suppresses severe multiple scattering effects, which significantly mitigates the non-linearity and the overall difficulty of solving the inverse problem. Future work will feature ad hoc studies and rigorous experimental validations specifically tailored to these complex biomedical scenarios.

    Future extensions to 3D will inherently encounter significant data volume increases; however, this can be mitigated by upgrading to a 3D-FFT module, alongside memory-efficient distributed training to handle the VRAM demands of 3D vector fields.

\bibliographystyle{ieeetr}

\begin{thebibliography}{10}

\bibitem{amineh2020nondestructive}
R.~K. Amineh, M.~Ravan, and R.~Sharma, ``Nondestructive testing of nonmetallic pipes using wideband microwave measurements,'' \emph{IEEE Trans. Microw. Theory Techn.}, vol.~68, pp. 1763--1772, May 2020.

\bibitem{ghattas2024review}
A.~Ghattas, R.~Al-Sharawi, A.~Zakaria, and N.~Qaddoumi, ``A review on microwave non-destructive testing (NDT) of composites,'' \emph{Eng. Sci. Technol., an Int. J.}, vol.~58, p. 101848, 2024.

\bibitem{meaney2000clinical}
P.~M. Meaney, M.~W. Fanning, D.~Li, S.~P. Poplack, and K.~D. Paulsen, ``A clinical prototype for active microwave imaging of the breast,'' \emph{IEEE Trans. Microw. Theory Techn.}, vol.~48, pp. 1841--1853, Nov. 2000.

\bibitem{van1997contrast}
P.~M. van~den Berg and R.~E. Kleinman, ``A contrast source inversion method,'' \emph{Inverse Problems}, vol.~13, no.~6, pp. 1607--1620, 1997.

\bibitem{sun2017cross}
S.~Sun, B.~J. Kooij, T.~Jin, and A.~G. Yarovoy, ``Cross-correlated contrast source inversion,'' \emph{IEEE Trans. Antennas Propag.}, vol.~65, no.~5, pp. 2592--2603, 2017.

\bibitem{wang1989iterative}
Y.~Wang and W.~C. Chew, ``An iterative solution of the two-dimensional electromagnetic inverse scattering problem,'' \emph{Int. J. Imaging Syst. Technol.}, vol.~1, no.~1, pp. 100--108, 1989.

\bibitem{11106328}
S.~Sun, ``Frequency-binned cumulative hopping framework incorporating wavelength-dependent weighting strategies for inverse scattering of high-contrast objects,'' \emph{IEEE Trans. Antennas Propag.}, vol.~73, no.~10, pp. 8048--8062, 2025.

\bibitem{8565987}
L.~Li, L.~G. Wang, F.~L. Teixeira, C.~Liu, A.~Nehorai, and T.~J. Cui, ``DeepNIS: Deep neural network for nonlinear electromagnetic inverse scattering,'' \emph{IEEE Trans. Antennas Propag.}, vol.~67, no.~3, pp. 1819--1825, 2019.

\bibitem{DastfalTGRSL2025}
Z.~Dastfal, M.~Hajebi, M.~Sharifzadeh, and A.~Hoorfar, ``Learning-based profiling of buried elliptical-cylindrical objects,'' \emph{IEEE Geosci. Remote Sens. Lett.}, vol.~22, pp. 1--5, 2025.

\bibitem{9897092}
H.~Zhang, Y.~Chen, T.~J. Cui, F.~L. Teixeira, and L.~Li, ``Probabilistic deep learning solutions to electromagnetic inverse scattering problems using conditional renormalization group flow,'' \emph{IEEE Trans. Microw. Theory Techn.}, vol.~70, no.~11, pp. 4955--4965, 2022.

\bibitem{bi2024diffusionemis}
X.~Bi, Y.~Chen, and L.~Li, ``DiffusionEMIS: Diffusion model for 3-D electromagnetic inverse scattering,'' \emph{IEEE Trans. Geosci. Remote Sens.}, vol.~62, pp. 1--12, 2024.

\bibitem{du2024inhomogeneous}
N.~Du, J.~Wang, R.~Song, K.~Xu, S.~Sun, and X.~Ye, ``Inhomogeneous media inverse scattering problem assisted by {Swin} transformer network,'' \emph{IEEE Trans. Microw. Theory Techn.}, vol.~72, no.~12, pp. 6809--6820, 2024.

\bibitem{RAISSI2019686}
M.~Raissi, P.~Perdikaris, and G.~Karniadakis, ``Physics-informed neural networks: A deep learning framework for solving forward and inverse problems involving nonlinear partial differential equations,'' \emph{J. Comput. Phys.}, vol.~378, pp. 686--707, 2019.

\bibitem{du2025physics}
Y.~Du, Z.~Liu, B.~Matkerim, C.~Li, Y.~Zong, B.~Qi, and J.~Kou, ``Physics-driven neural network for solving electromagnetic inverse scattering problems,'' \emph{IEEE Trans. Antennas Propag.}, vol.~74, no.~2, pp. 1945--1956, 2026.

\bibitem{ISPNet2025}
W.~Zhang, H.~Wang, and R.~Chen, ``Physics-informed deep learning for inverse scattering of irregular targets from near-field data,'' \emph{IEEE Antennas Wirel. Propag. Lett.}, vol.~24, pp. 3734--3738, Oct. 2025.

\bibitem{sun2026CC_PINN}
S.~Sun, ``Beyond data-physics consistency: A cross-correlated physics-informed neural network for robust inverse scattering,'' \emph{arXiv:2605.01851}, 2026.

\bibitem{sitzmann2020implicit}
V.~Sitzmann, J.~Martel, A.~Bergman, D.~Lindell, and G.~Wetzstein, ``Implicit neural representations with periodic activation functions,'' \emph{Adv. Neural Inf. Process. Syst.}, vol.~33, pp. 7462--7473, 2020.

\bibitem{sun2026differentiable}
H.~Sun, D.~Liu, H.~Zhou, M.~Li, S.~Xu, and F.~Yang, ``A differentiable framework for full and phaseless data inversion using neural implicit contrast-source representation,'' \emph{IEEE Trans. Geosci. Remote Sens.}, vol.~64, Art. no. 4503514, 2026.

\bibitem{geffrin2005free}
J.-M. Geffrin, P.~Sabouroux, and C.~Eyraud, ``Free space experimental scattering database continuation: experimental set-up and measurement precision,'' \emph{Inverse Problems}, vol.~21, no.~6, pp. S117--S130, 2005.

\end{thebibliography}

\vfill

\end{document}